\documentclass[twocolumn,floatfix]{aastex63}
\usepackage{graphics,graphicx}
\hypersetup{urlcolor=blue}
\usepackage{natbib}
\usepackage[outdir=./]{epstopdf}
\usepackage{textcomp,gensymb,mathrsfs}
\usepackage{xfrac}
\usepackage{xcolor,enumitem}

\newcommand{\mps}{m\,s$^{-1}$}

\newcommand{\vsini}{$v\sin{i_*}$}

\newcommand{\kepler}{{\it Kepler}}
\newcommand{\logg}{log\,$g$}

\newcommand{\um}{$\mu$m}
\newcommand{\fbol}{$F_{\mathrm{bol}}$}

\newcommand{\teff}{\ensuremath{T_{\text{eff}}}}
\newcommand\kms{km\,s$^{-1}$}

\newcommand{\tess}{\textit{TESS}}
\newcommand{\ktwo}{\textit{K2}}
\newcommand{\gaia}{\textit{Gaia}}
\newcommand{\epscha}{$\epsilon$ Cha}%Chamaeleontis

\newcommand{\betapic}{$\beta$ Pic}
\newcommand{\starname}{TOI\,1227}
\newcommand{\planetname}{TOI\,1227\,b}

\renewcommand{\d}{\mathrm{d}}
\newcommand{\like}{\mathscr{L}}

\newcommand{\Pbad}{P_{\mathrm{b}}}
\newcommand{\Ybad}{Y_{\mathrm{b}}}
\newcommand{\Vbad}{V_{\mathrm{b}}}
\newcommand{\population}{Musca}

\shorttitle{A 11\,Myr Jovian-radius planet around a 0.17$M_\odot$ star}
\shortauthors{Mann et al.}
\graphicspath{{./}{}}

\submitjournal{The Astronomical Journal}
\received{2021 October 18}
\revised{2021 December 28}
\accepted{2022 January 20}
\published{2022 March 9}

\begin{document}

\title{TESS Hunt for Young and Maturing Exoplanets (THYME) VI: an 11 Myr Giant Planet Transiting a Very Low Mass Star in Lower Centaurus Crux}

\correspondingauthor{Andrew W. Mann}
\email{awmann@unc.edu}

\author[0000-0003-3654-1602]{Andrew W. Mann}%
\affiliation{Department of Physics and Astronomy, The University of North Carolina at Chapel Hill, Chapel Hill, NC 27599, USA} 
%affirmed

\author[0000-0001-7336-7725]{Mackenna L. Wood}%
\affiliation{Department of Physics and Astronomy, The University of North Carolina at Chapel Hill, Chapel Hill, NC 27599, USA} 
%affirmed

\author[0000-0001-8510-7365]{Stephen P. Schmidt}
\affiliation{Department of Physics and Astronomy, The University of North Carolina at Chapel Hill, Chapel Hill, NC 27599, USA} 
%affirmed

\author[0000-0002-8399-472X]{Madyson G. Barber}
\altaffiliation{UNC Chancellor’s Science Scholar}
\affiliation{Department of Physics and Astronomy, The University of North Carolina at Chapel Hill, Chapel Hill, NC 27599, USA} 
%affirmed

\author[0000-0002-4856-7837]{James E. Owen}
\affiliation{Astrophysics Group, Imperial College London, Blackett Laboratory, Prince Consort Road, London SW7 2AZ, UK}
%affirmed

\author[0000-0003-2053-0749]{Benjamin M. Tofflemire}
\altaffiliation{51 Pegasi b Fellow}
\affiliation{Department of Astronomy, The University of Texas at Austin, Austin, TX 78712, USA}
%affirmed

\author[0000-0003-4150-841X]{Elisabeth R. Newton}%
\affiliation{Department of Physics and Astronomy, Dartmouth College, Hanover, NH 03755, USA}
%affirmed

%LCC info, general advice on membership, interesting star facts, X-ray, etc. 
\author[0000-0003-2008-1488]{Eric E. Mamajek}%
\affiliation{Jet Propulsion Laboratory, California Institute of Technology, 4800 Oak Grove Dr., Pasadena, CA 91109, USA}
\affiliation{Department of Physics \& Astronomy, University of Rochester, 500 Wilson Blvd., Rochester, NY 14627, USA}
%affirmed

%Rotation analysis of target, friends, SOAR spectra, SOAR g-band transit
\author[0000-0002-9446-9250]{Jonathan L. Bush}%
\affiliation{Department of Physics and Astronomy, The University of North Carolina at Chapel Hill, Chapel Hill, NC 27599, USA} 
%affirmed

% IGRINS reduction pipeline and data
\author[0000-0001-7875-6391]{Gregory N. Mace}
\affiliation{Department of Astronomy, The University of Texas at Austin, Austin, TX 78712, USA}
%affirmed

%%THYME, FF, suggestions on follow-up
\author[0000-0001-9811-568X]{Adam L. Kraus}%
\affiliation{Department of Astronomy, The University of Texas at Austin, Austin, TX 78712, USA}
%affirmed

%%SOAR observing of r-band transit
\author[0000-0001-5729-6576]{Pa Chia Thao}%
\altaffiliation{NSF GRFP Fellow} 
\affiliation{Department of Physics and Astronomy, The University of North Carolina at Chapel Hill, Chapel Hill, NC 27599, USA} 
%affirmed

\author[0000-0001-7246-5438]{Andrew Vanderburg}%
\affiliation{Department of Physics and Kavli Institute for Astrophysics and Space Research, Massachusetts Institute of Technology, Cambridge, MA 02139, USA}
%affirmed

\author[0000-0003-4450-0368]{Joe Llama}
\affiliation{Lowell Observatory, 1400 West Mars Hill Road, Flagstaff, AZ 86001, USA}
%affirmed - waiting on comments

\author[0000-0002-8828-6386]{Christopher M. Johns--Krull}
\affiliation{Department of Physics and Astronomy, Rice University, 6100 Main Street, Houston, TX 77005, USA}
%affirmed
 
\author[0000-0001-7998-226X]{L. Prato}
\affiliation{Lowell Observatory, 1400 West Mars Hill Road, Flagstaff, AZ 86001, USA}
%affirmed - waiting on comments

\author[0000-0002-0848-6960]{Asa G. Stahl}
\affiliation{Department of Physics and Astronomy, Rice University, 6100 Main Street, Houston, TX 77005, USA}
%affirmed
 
\author[0000-0003-4247-1401]{Shih-Yun Tang}
\affiliation{Lowell Observatory, 1400 West Mars Hill Road, Flagstaff, AZ 86001, USA}
\affiliation{Department of Astronomy and Planetary Sciences, Northern Arizona University, Flagstaff, AZ 86011, USA}
%affirmed

%%inclination and his stellar parameter code
\author[0000-0002-9641-3138]{Matthew J. Fields}
\affiliation{Department of Physics and Astronomy, The University of North Carolina at Chapel Hill, Chapel Hill, NC 27599, USA} 
%affirmed

%----LCO photometry------%%%%
\author[0000-0001-6588-9574]{Karen A.\ Collins}
\affiliation{Center for Astrophysics \textbar \ Harvard \& Smithsonian, 60 Garden Street, Cambridge, MA 02138, USA}
%affirmed

%----LCO photometry------%%%%
\author[0000-0003-2781-3207]{Kevin I.\ Collins}
\affiliation{George Mason University, 4400 University Drive, Fairfax, VA, 22030 USA}
%affirmed

%----LCO photometry------%%%%
\author[0000-0002-4503-9705]{Tianjun Gan}
\affiliation{Department of Astronomy and Tsinghua Centre for Astrophysics, Tsinghua University, Beijing 100084, China}
%affirmed

%----LCO photometry------%%%%
\author[0000-0002-4625-7333]{Eric L.\ N.\ Jensen}
\affiliation{Department of Physics \& Astronomy, Swarthmore College, Swarthmore PA 19081, USA}

%----LCO photometry------%%%%
\author[0000-0003-4133-0877]{Jacob Kamler}
\affiliation{John F. Kennedy High School, 3000 Bellmore Avenue, Bellmore, NY 11710, USA}
%affirmed

%----LCO photometry------%%%%
\author[0000-0001-8227-1020]{Richard P. Schwarz}
\affiliation{Patashnick Voorheesville Observatory, Voorheesville, NY 12186, USA}
%affirmed

%------Provided Gemini-Zorro Observations ---------
\author[0000-0001-9800-6248]{Elise Furlan}
\affiliation{NASA Exoplanet Science Institute, Caltech/IPAC, Mail Code 100-22, 1200 E. California Blvd., Pasadena, CA 91125, USA}
%affirmed

\author[0000-0003-2519-6161]{Crystal~L.~Gnilka}
\affil{NASA Ames Research Center, Moffett Field, CA 94035, USA}
%affirmed

\author[0000-0002-2532-2853]{Steve~B.~Howell}
\affil{NASA Ames Research Center, Moffett Field, CA 94035, USA}
%affirmed

\author[0000-0002-9903-9911]{Kathryn V. Lester}
\affil{NASA Ames Research Center, Moffett Field, CA 94035, USA}
%affirmed

% 
\author[0000-0002-6397-6719]{Dylan A. Owens}%
\affiliation{Department of Physics and Astronomy, The University of North Carolina at Chapel Hill, Chapel Hill, NC 27599, USA}
%affirmed

%% ASTEP DATA
\author[0000-0002-3503-3617]{Olga Suarez}%
\affiliation{Universit\'e C\^ote d'Azur, Observatoire de la C\^ote d'Azur, CNRS, Laboratoire Lagrange, Bd de l'Observatoire, CS 34229, 06304 Nice cedex 4, France}
%affirmed (provided reduction text)

%% ASTEP DATA
\author[0000-0001-5000-7292]{Djamel Mekarnia}
\affiliation{Universit\'e C\^ote d'Azur, Observatoire de la C\^ote d'Azur, CNRS, Laboratoire Lagrange, Bd de l'Observatoire, CS 34229, 06304 Nice cedex 4, France}
%affirmed

%% ASTEP DATA
\author[0000-0002-7188-8428]{Tristan Guillot}%
\affiliation{Universit\'e C\^ote d'Azur, Observatoire de la C\^ote d'Azur, CNRS, Laboratoire Lagrange, Bd de l'Observatoire, CS 34229, 06304 Nice cedex 4, France}
%affirmed (provided comments)

%% ASTEP DATA
\author[0000-0002-0856-4527]{Lyu Abe}%
\affiliation{Universit\'e C\^ote d'Azur, Observatoire de la C\^ote d'Azur, CNRS, Laboratoire Lagrange, Bd de l'Observatoire, CS 34229, 06304 Nice cedex 4, France}
%affirmed

\author[0000-0002-5510-8751]{Amaury H.~M.~J. Triaud}%
\affiliation{School of Physics \& Astronomy, University of Birmingham, Edgbaston, Birmingham, B15 2TT, UK}
%affirmed

%THYME, MISTTBORN
\author[0000-0002-5099-8185]{Marshall C. Johnson}%
\affiliation{Las Cumbres Observatory, 6740 Cortona Dr., Ste. 102, Goleta, CA 93117, USA}
%affirmed

% friends of TOI1227 spectra
\author[0000-0002-1312-3590]{Reilly P. Milburn}%
\affiliation{Department of Physics and Astronomy, The University of North Carolina at Chapel Hill, Chapel Hill, NC 27599, USA} 
%affirmed

%THYME, Injection/Recovery
\author[0000-0001-9982-1332]{Aaron C. Rizzuto}%
\affiliation{Department of Astronomy, The University of Texas at Austin, Austin, TX 78712, USA}
%affirmed

%TESS Contributing Authors from the TESS Science Office; also, follow-up
\author[0000-0002-8964-8377]{Samuel N. Quinn}%
\affiliation{Center for Astrophysics \textbar \ Harvard \& Smithsonian, 60 Garden Street, Cambridge, MA 02138, USA}
%affirmed

% discussion of membership issues
\author[0000-0002-6549-9792]{Ronan Kerr}%
\affiliation{Department of Astronomy, The University of Texas at Austin, Austin, TX 78712, USA}
%affirmed

%%%% TESS architects %%%%%
\author[0000-0003-2058-6662]{George~R.~Ricker}%
\affiliation{Department of Physics and Kavli Institute for Astrophysics and Space Research, Massachusetts Institute of Technology, Cambridge, MA 02139, USA}
%affirmed

%%%% TESS architects %%%%%
\author[0000-0001-6763-6562]{Roland~Vanderspek}%
\affiliation{Department of Physics and Kavli Institute for Astrophysics and Space Research, Massachusetts Institute of Technology, Cambridge, MA 02139, USA}
%affirmed

%%%% TESS architects %%%%%
\author[0000-0001-9911-7388]{David~W.~Latham}%
\affiliation{Center for Astrophysics \textbar \ Harvard \& Smithsonian, 60 Garden Street, Cambridge, MA 02138, USA}
%affirmed

%%%% TESS architects %%%%%
\author[0000-0002-6892-6948]{Sara~Seager}%
\affiliation{Department of Physics and Kavli Institute for Astrophysics and Space Research, Massachusetts Institute of Technology, Cambridge, MA 02139, USA}
\affiliation{Department of Earth, Atmospheric and Planetary Sciences, Massachusetts Institute of Technology, Cambridge, MA 02139, USA}
\affiliation{Department of Aeronautics and Astronautics, MIT, 77 Massachusetts Avenue, Cambridge, MA 02139, USA}
%affirmed

%%%% TESS architects %%%%%
\author[0000-0002-4265-047X]{Joshua~N.~Winn}%
\affiliation{Department of Astrophysical Sciences, Princeton University, 4 Ivy Lane, Princeton, NJ 08544, USA}
%affirmed

%%%% TESS architects %%%%%
\author[0000-0002-4715-9460]{Jon M. Jenkins}%
\affiliation{NASA Ames Research Center, Moffett Field, CA, 94035, USA}
%affirmed

%TESS Contributing Authors from the TESS Science Office
\author[0000-0002-5169-9427]{Natalia~M.~Guerrero}
\affiliation{Department of Astronomy, University of Florida, Gainesville, FL 32811, USA}
%affirmed

%TESS Contributing Authors from the TESS Science Office
\author[0000-0002-1836-3120]{Avi Shporer}
\affil{Department of Physics and Kavli Institute for Astrophysics and Space Research, Massachusetts Institute of Technology, Cambridge, MA 02139, USA}
%affirmed

%TESS Contributing Authors from POC
\author[0000-0001-5347-7062]{Joshua~E.~Schlieder}
\affiliation{NASA Goddard Space Flight Center, 8800 Greenbelt Rd, Greenbelt, MD 20771, US}
%affirmed

%TESS Contributing Authors from POC
\author[0000-0002-8058-643X]{Brian~McLean}
\affiliation{Space Telescope Science Institute, 3700 San Martin Drive, Baltimore, MD, 21218, USA}
%affirmed

%TESS contributing author from SPOC
\author[0000-0002-5402-9613]{Bill Wohler}
\affiliation{SETI Institute, Mountain View, CA 94043, USA}
\affiliation{NASA Ames Research Center, Moffett Field, CA 94035, USA}
%affirmed

%Mature super-Earths and sub-Neptunes are predicted to be ~Jovian radius when younger than 10 Myr. Thus, we expect to find 5--15 Earth-radii planets around young stars even if their older counterparts harbor none. We report the discovery and validation of TOI 1227 b, a 0.85+-0.05 Jupiter-radio (9.5 Earth radii) planet transiting a very low-mass star (0.170 Solar masses) every 27.4 days. TOI 1227's kinematics and strong lithium absorption confirm it is a member of a previously discovered sub-group in the Lower Centaurus Crux OB association, which we designate the Musca group. We derive an age of 11+/-22 Myr for Musca, based on lithium, rotation, and the color-magnitude diagram of Musca members. The TESS data and ground-based follow-up show a deep (2.5%) transit. We use multi-wavelength transit observations and radial velocities from the IGRINS spectrograph to validate the signal as planetary in nature, and we obtain an upper limit on the planet mass of 0.5 Jupiter masses. Because such large planets are exceptionally rare around mature low-mass stars, we suggest that  TOI 1227 b is still contracting and will eventually turn into one of the more common <5 Earth-radii planets.

\begin{abstract}
Mature super-Earths and sub-Neptunes are predicted to be $\simeq$Jovian radius when younger than $10$\,Myr. Thus, we expect to find 5--15$R_\oplus$ planets around young stars even if their older counterparts harbor none. We report the discovery and validation of TOI 1227 b, a $0.85\pm0.05\,R_J$ (9.5$R_\oplus$) planet transiting a very low-mass star ($0.170\pm0.015\,M_\odot$) every 27.4\,days. TOI~1227's kinematics and strong lithium absorption confirm it is a member of a previously discovered sub-group in the Lower Centaurus Crux OB association, which we designate the Musca group. We derive an age of 11$\pm$2 Myr for Musca, based on lithium, rotation, and the color-magnitude diagram of \population\ members. The {\it TESS} data and ground-based follow-up show a deep (2.5\%) transit. We use multiwavelength transit observations and radial velocities from the IGRINS spectrograph to validate the signal as planetary in nature, and we obtain an upper limit on the planet mass of $\simeq$0.5\,$M_J$. Because such large planets are exceptionally rare around mature low-mass stars, we suggest that  TOI 1227 b is still contracting and will eventually turn into one of the more common $<5R_\oplus$ planets. 
\end{abstract}


\section{Introduction}\label{sec:intro}
Young planets offer a window into the early stages of planet formation and evolution. Planets younger than 100\,Myr are particularly useful for this work, as planetary systems likely evolve most rapidly in the first few hundred million years after formation \citep{2013ApJ...776....2L, 2013ApJ...775..105O}. Populations of such planets are critical to understanding planetary migration \citep{Nelson:2017aa}, photoevaporation \citep{Raymond:2008fk, Owen2018}, and atmospheric chemistry \citep[e.g., ][]{2005AsBio...5..706S, Gao_hazes}. Given the timescale for planet formation \citep[1-10\,Myr;][]{2002Natur.418..949Y, 2005A&A...434..343A}, planets aged $\lesssim$30\,Myr can even tell us about the conditions of planets right after their formation. 

The number of known young transiting planets has grown significantly in recent years \citep[e.g.,][]{Obermeier2016, David2018ab, Benatti_dstuc, THYMEIV}. This growth was primarily driven by a combination of the \ktwo\ and \tess\ missions surveying nearby young clusters and star-forming regions \citep{2014SPIE.9143E..20R, VanCleve2016}, improvements in filtering variability in young stars \citep[e.g.,][]{2016MNRAS.459.2408A, Rizzuto2017}, and more complete identification of young stellar associations from \gaia\ kinematics \citep[e.g., ][]{2018A&A...618A..93C, 2021arXiv210509338K}. Despite this progress, there are still only a handful of transiting planets at the youngest ages of $\lesssim$30\,Myr \citep{David2016b, Mann2016b, 2019ApJ...885L..12D, 2020Natur.582..497P, Rizzuto2020} and a few candidate nontransiting planets from radial velocity surveys \citep[e.g.,][]{2016ApJ...826..206J, Donati:2017aa}. 

Models predict that gas giant planets younger than $<50$\,Myr will be larger and brighter than their older counterparts \citep{2019A&A...623A..85L}. At 10--20\,Myr, progenitors of mature Jovian-mass planets are expected to be 1.2--1.6\,$R_J$ and sub-Neptunes $\sim 1 R_J$ \citep{2019A&A...623A..85L, 2020MNRAS.498.5030O}. Completeness curves from the best search pipelines \citep{Rizzuto2017}, and the discovery of much smaller planets \citep[e.g.,][]{Mann:2018, Zhou_2020} demonstrate that giant planets are readily detectable even in the presence of complex stellar variability common to young stars. Thus far, transit surveys of young stars identified only a few Jovian-radius planets in the youngest associations \citep{David2019b, Bouma_2020, Rizzuto2020}. 

As part of the \tess\ Hunt for Young and Maturing Exoplanets survey \citep[THYME;][]{Newton2019}, our team searches \tess\ data using a specialized pipeline to identify young planets missed by standard searches \citep[e.g.,][]{Rizzuto2020}, and checks previously-identified planet candidates for signs of membership in a young association \citep[e.g.,][]{Mann:2020, THYMEIV}. We identified \starname\ (2MASS J12270432-7227064, TIC 360156606) as a member of a young association, with a planet candidate (TOI1227.01) identified by the \tess\ mission, and astrometry consistent with membership in the Lower Centaurus Crux (LCC) region of the Scorpius-Centaurus (Sco-Cen) OB association \citep{2018ApJ...868...32G, Damiani2019}. The \tess-identified transit signal is $\simeq$2\% deep, suggesting a Jovian-sized planet orbiting a pre-main-sequence low-mass star. As giant planets with periods $<$100\,days are rare around older stars of similar mass \citep[$<1\%$][]{2013A&A...549A.109B}, validation of this planet would provide strong evidence of radius evolution. 

In this work, we present validation, characterization, and age estimates for \planetname, a 0.85$R_{Jup}$ planet orbiting a $\simeq$11\,Myr pre-main-sequence M5V star (0.17$M_\odot$) every 27.26\,days. In Section~\ref{sec:obs}, we detail the photometric and spectroscopic follow-up of the planet and host star. We demonstrate that the star is a member of a recently identified substructure of LCC and derive an updated age of $11\pm2$\,Myr for the parent population in Section~\ref{sec:member}. We estimate parameters for the host star in Section~\ref{sec:star}, parameters of the planet in Section~\ref{sec:transit}, and combine these with our ground-based follow-up to statistically validate the planet in Section~\ref{sec:fpp}. We place the large size of \planetname\ in context with its age and host star mass and explore its likely evolution in Section~\ref{sec:model}. We conclude in Section~\ref{sec:discussion} with a summary and discussion of follow-up and implications for future searches for young exoplanets. 
 
\section{Observations}\label{sec:obs}

\subsection{\tess\ Photometry}\label{sec:tess}

\starname\ was observed by the \tess\ mission \citep{2014SPIE.9143E..20R} from UT 22 Apr 2019 through 19 Jun 2019 (Sectors 11 and 12) and then again from 28 Apr 2021 through 26 May 2021 (Sector 38). In all three sectors, the target fell on Camera 3. The first two sectors had 2\,min cadence data as part of a search for planets around M dwarfs (G011180; PI Dressing). Sector 38 had 20\,s cadence data, as it was known to be a young TOI at that phase (G03141; PI Newton). We initially used the 2\,min cadence for all analysis for computational efficiency and \tess\ cosmic-ray mitigation\footnote{\url{https://archive.stsci.edu/files/live/sites/mast/files/home/missions-and-data/active-missions/tess/_documents/TESS_Instrument_Handbook_v0.1.pdf}} (not included in 20\,s data), but also ran a separate analysis using the Sector 38 20\,s data (see Section~\ref{sec:transit}. In both cases, we used the Pre-Search Data Conditioning Simple Aperture Photometry \citep[PDCSAP; ][]{Stumpe2012, Smith2012, Stumpe2014} \tess\ light curve produced by the Science Process Operations center \citep[SPOC; ][]{Jenkins:2016} and available through the Mikulski Archive for Space Telescopes (MAST)\footnote{\url{https://mast.stsci.edu/portal/Mashup/Clients/Mast/Portal.html}}. 

\subsection{Identification of the transit signal}

The planet signal, TOI1227.01, was first detected in a joint transit search of sectors 11 and 12 (one transit in each) with an adaptive wavelet-based detector \citep{2002ApJ...575..493J,2010SPIE.7740E..0DJ}. The candidate was fitted with a limb-darkened light curve \citep{Li:DVmodelFit2019} and passed all performed diagnostic tests \citep{DVreports1}. Although the difference image centroiding failed to converge, the difference images indicate that the transit source location was consistent with the location of the host star, \starname. A search of the residual light curve failed to identify additional transiting planet signatures. The TESS Science Office reviewed the diagnostic test results and issued an alert for this planet candidate as a TESS object of interest (TOI) on 2019 August 26 \citep{2021ApJS..254...39G}. A third transit was detected in the Sector 38 \tess\ data, consistent with the expected period and depth. 

We searched for additional planets using the Notch and LoCoR pipelines, as described in \citet{Rizzuto2017}\footnote{\url{https://github.com/arizzuto/Notch_and_LOCoR}}. This included using the significance of adding a trapezoidal Notch to the light curve detrending, as characterized by the Bayesian information criterion (BIC). The method was more effective than periodic methods for finding planets with $\lesssim$3 transits, as was the case for HIP\,67522\,b \citep{Rizzuto2020}. The transits were quite clear from the BIC test. However, no additional significant signals were detected. 

\begin{figure}[tb]
    \centering
    \includegraphics[width=0.49\textwidth]{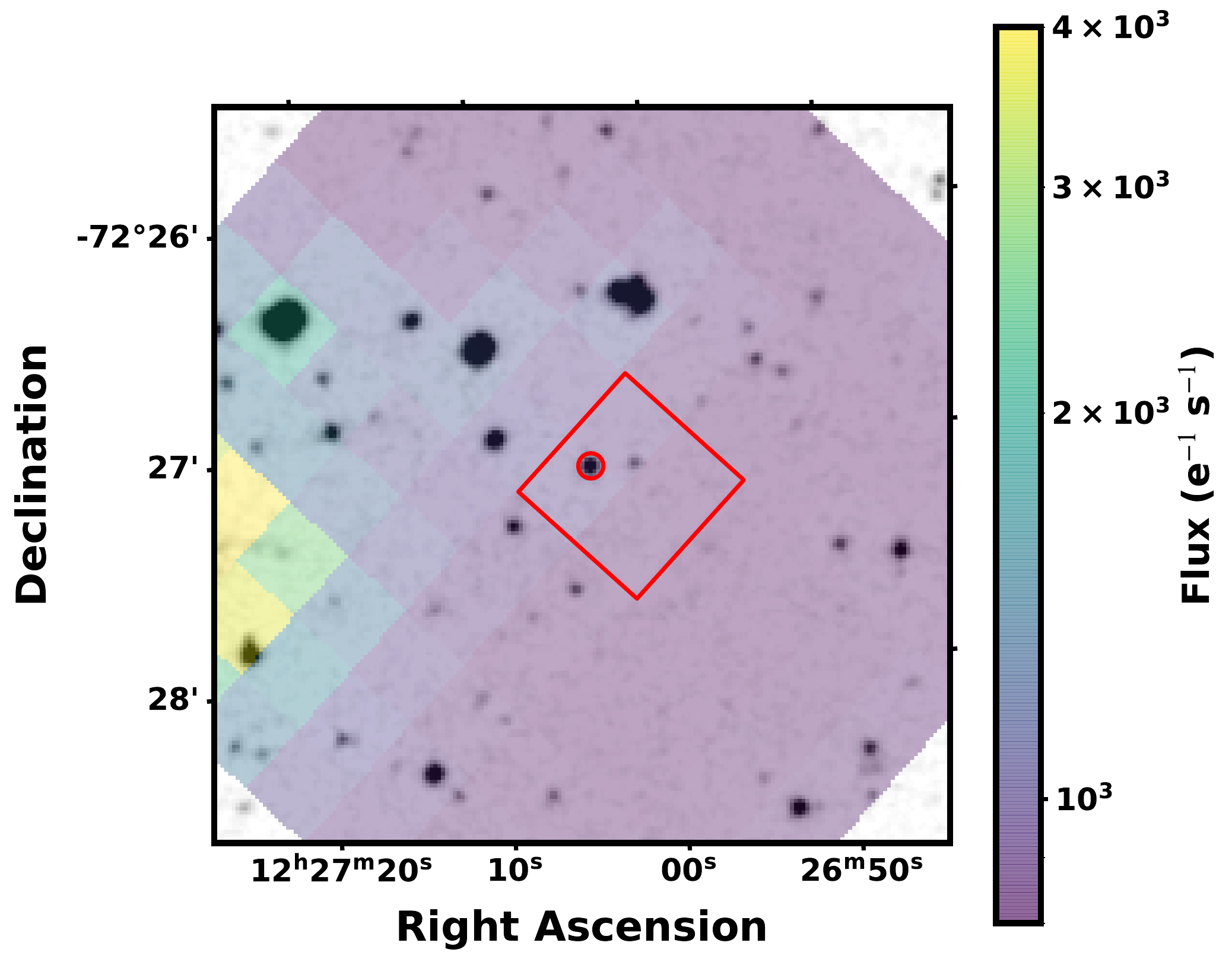}
    \caption{A \tess\ Sector 11 image colored by flux (see colorbar) on top of a DSS image (greyscale). The red circle indicates \starname\ and the red box indicates the \tess\ aperture used for Sector 11 (the aperture varied between sectors). The \tess\ PSF in this region has a full-width half-max of approximately 1.9 \tess\ pixels. The large \tess\ PSF combined with a faint source meant that there were high levels of contamination around \starname. }
    \label{fig:tess}
\end{figure}

\starname\ is a relatively faint star ($T=13.8$) in a crowded region (see Figure~\ref{fig:tess}). Within 2\arcmin\ ($\simeq$6 \tess\ pixels), there are four sources brighter than \starname, one of which is $T=8.4$ (TIC 360156594). Two sources brighter than \starname\ are within 1\arcmin. While the \tess\ aperture was small (4--6 pixels over the 3 sectors), background contamination was likely to be a problem for the \tess\ data. Correcting for contamination and confirming the planet were the major motivations for our ground-based follow-up.

%sector 11 = G011180 (Dressing: M dwarf)
%sector 12 = G011180 (Dressing: M dwarf)
%sector 38 = G03141 (Newton; young planet - 20s?)
%            G03278 (Mayo; M dwarf)
%            G03068 (Kipping; exomoons) 
%            G03052 (Doyle; fast rotators + flares)

\subsection{Ground-Based Photometry}\label{sec:phot}

The photometry is summarized in Table~\ref{tab:obslog}. We provide details by instrument below.  

\begin{deluxetable*} {lrrrrrrr} %[b!] |
\tablecaption{Log of transit observations\label{tab:obslog}}
\tablewidth{0pt}
\tablehead{
\colhead{Start} & 
\colhead{Telescope} &
\colhead{Filter} & 
\colhead{Transit \#} & 
\colhead{T$_{\rm{exp}}$} &
\colhead{Obs Duration} 
%\colhead{Coverage} 
\\[-0.2cm]
\colhead{(UT)} & 
\colhead{} &
\colhead{} & 
\colhead{\#} & 
\colhead{(s)} &
\colhead{(h)} 
}
\startdata
    2019 Apr 22\tablenotemark{a,b} & \tess\ S11 & \tess & 1 & 120 & N/A \\
    2019 Apr 22\tablenotemark{a,b} & \tess\ S12 & \tess & 2 & 120 & N/A  \\
    2020 Jan 15\tablenotemark{b} & SOAR  & $i$' & 9 & 120 & 8.1 \\ 
    2020 Jan 15 & LCO   & $i$' & 9 & 200 & 2.8 \\ %\tablenotemark{b}
    2020 May 03\tablenotemark{b} & LCO   & $i$' & 13 & 200 & 1.9 \\ 
    2020 Jul 24  & LCO  & $r$' & 16 & 200 & 2.5  \\ %\tablenotemark{b}
    2020 Jul 24\tablenotemark{b} & LCO   & $z_s$ & 16 & 200 & 2.5 \\ 
    2021 Mar 28\tablenotemark{b} & SOAR  & $g$' & 25 & 120 & 7.6 \\ 
    2021 Apr 24 & ASTEP & $Rc$' & 26 & 200 & 5.7 \\
    2021 Apr 24 & LCO   & $g$' & 26 & 300 & 6.2 \\ %\tablenotemark{b}
    2021 Apr 24\tablenotemark{b} & LCO   & $z_s$ & 26 & 210 & 6.2 \\ 
    2021 Apr 28\tablenotemark{a,b} & \tess\ S38 & \tess & 27 & 20 & N/A \\
    2021 Jun 17\tablenotemark{b} & LCO  & $g$' & 28 & 300 & 4.1 \\ 
\enddata
\tablenotetext{a}{\tess\ observed \starname\ for one transit in each of Sectors 11, 12, and 38.}
\tablenotetext{b}{Included in the global fit.}
\end{deluxetable*}

\subsubsection{Goodman/SOAR}\label{sec:goodmanphot}

We observed two transits of \planetname\ using the SOAR 4.1\,m telescope at Cerro Pach\'on, Chile, with the Goodman High Throughput Spectrograph in imaging mode \citep{Goodman_spectrograph}. For both observations, we used the red camera. In this mode, Goodman has a default 7.2\arcmin\ diameter circular field of view with a pixel scale of 0.15\arcsec\,pixel$^{-1}$. The first transit was observed on 2020 Jan 15 with the SDSS $i$' filter, and the second transit was observed on 2021 Mar 28 with the SDSS $g$' filter. Both nights had photometric conditions for the full duration of the transit observations. Although \starname\ is much fainter in $g$' than in $i$', we used a 120\,s exposure time for both filters to resolve any flares and sample the ingress. We selected a smaller region-of-interest in the readout direction (1598 pixels) to decrease readout time to 1.7\,s. Due to the small pixel scale, no defocusing was required and counts were well below half-well depth for both the target and all but one target in the field of view (HD 108342). 

About 1 hour into the 2021 Mar 28 transit observations, the SOAR guider began to fail, causing the target to shift $\simeq$5 pixels per exposure and forcing us to shift it back manually every $\sim$10 exposures. During egress, the guider completely failed and had to be reset before re-acquiring the field. We positioned the target back to its starting location and continued the observations as normal. This resulted in poorer photometric precision than normally achievable with Goodman/SOAR, and a $\simeq$15\,min gap in the data near the end of the egress.

We applied bias and flat-field corrections before extracting photometry for \starname\ using a 10-pixel radius aperture and used an annulus of 30--60 pixels to determine the local sky background. The aperture center for each exposure was the stellar centroid, calculated within a 10-pixel radius of the nominal location. We repeated this on eight nearby stars that were close in brightness to the target and showed little or no photometric variation when compared to other stars in the field. We corrected the target light curve using the weighted mean of the comparison star curves.

\subsubsection{LCO Photometry}\label{sec:lcophot}

We observed a total of seven transits with 1\,m telescopes in the Las Cumbres Observatory network \citep[LCO ;][]{Brown13}. These were all observed with Sinistro cameras, with a pixel scale of 0.389\arcsec\,pixel$^{-1}$. Two transits were observed using the SDSS $g'$ filter, one with SDSS $r'$, two with $i'$, and two with $z_s$. Exposure times are given in Table~\ref{tab:obslog}. Because of the difficulty of scheduling observations of a long-period planet ($\sim$27\,d) and weather interruptions, all LCO observations covered only part of the transit. 

The images were initially calibrated by the standard LCOGT {\tt BANZAI} pipeline \citep{McCully18}. We then performed aperture photometry on all datasets using the \texttt{AstroImageJ} package \citep[AIJ;][]{Collins17}. The aperture varied based on the seeing conditions at the observatory, but we generally used a 6--10 pixel radius circular aperture for the source and an annulus with a 15--20 pixel inner radius and a 25--30 pixel outer radius for the sky background. For all observations, we centered the apertures on the source and weighted pixels within the aperture equally. Because the event was detected on the source, the usual check of nearby sources for evidence of an eclipsing binary was not necessary. Light curves of nearby sources are available with the extracted light curves and further details on the follow-up at ExoFOP-TESS\footnote{\url{https://exofop.ipac.caltech.edu/tess/target.php?id=360156606}} \citep{EXOFOP_TESS}.

Except for a 2021 Apr 24 $g$' transit, all photometry showed the expected transit behavior. Two transits had poor coverage (no out-of-transit baseline), but still showed a shape consistent with ingress. We discuss the 2021 Apr 24 transit in more detail in Section~\ref{sec:oddtransit}.

\subsubsection{ASTEP Photometry}

On UTC 2021 Apr 24, one additional transit was observed with the Antarctica Search for Transiting ExoPlanets (ASTEP) program on the East Antarctic plateau \citep{2015AN....336..638G,2016MNRAS.463...45M}. The 0.4\,m telescope is equipped with an FLI Proline science camera with a KAF-16801E, $4096\times 4096$ front-illuminated CCD. The camera has an image scale of 0.93\arcsec\,pixel$^{-1}$ resulting in a 1$^{o}\times 1^{o}$ corrected field of view. The focal instrument dichroic plate splits the beam into a blue wavelength channel for guiding, and a non-filtered red science channel roughly matches an $R_c$ transmission curve.

Due to the extremely low data transmission rate at the Concordia Station, the data are processed on-site using an automated IDL-based pipeline described in \citet{2013A&A...553A..49A}. The calibrated light curve is reported via email and the raw light curves of about 1,000 stars are transferred to Europe on a server in Rome, Italy, and are then available for deeper analysis. These data files contain each star’s flux computed through $10$ fixed circular aperture radii so that optimal light curves can be extracted. For \starname, a 9.3-pixel radius aperture gave the best result.

\starname\ was observed under mixed conditions, with a windy clear sky, and air temperatures between $-62^o$C and $-66^o$C. The Moon was $\sim 90\%$ full and present during the observation. A strong wind ($\sim$ 6 ms$^{-1}$) led to telescope guiding issues at the beginning of the observation and prevented us from observing the ingress. The data points corresponding to these issues were removed from the analysis. However, the resulting light curve showed the expected egress.  

\subsection{Corrections for second-order extinction}

Since atmospheric extinction is strongly color-dependent, changes in airmass produce color-dependent flux losses that depend on the spectral energy distribution (SED) of the target. These color terms are weaker in redder and narrower passbands \citep{Young:1991lr}. The effect is often small, as comparison stars typically span a range of colors, and observations are timed for favorable airmass. However, \starname\ was unusual both because of its long orbital period (27.3\,d) and red color; of stars $<$5\arcmin\ from the target and differing by $<$1 magnitude in $G$, the reddest has a \gaia\ color of $BP-RP=2.1$ while \starname\ has $BP-RP=3.3$. 

Because we had few options for mitigating second-order extinction (color terms), we corrected for the effect following \citet{Mann2011}. To summarize, we estimated \teff\ for all comparison stars based on their \gaia\ colors and the tables from \citet{2013ApJS..208....9P}\footnote{\url{https://www.pas.rochester.edu/~emamajek/EEM_dwarf_UBVIJHK_colors_Teff.txt}}. We then combined the relevant BT-SETTL models (assuming Solar metallicity) with an airmass-dependent model of the atmosphere above the observing site \citep{2009JQSRT.110..533R} and the appropriate filter profile. The output was a predicted change in flux from second-order extinction alone. The trend was negligible ($<1$\,mmag) for $i$' and $z_s$ observations, so we did not apply a correction. The effect was small but non-negligible for $r$ ($>$1 mmag), and significant for $g$'-band observations (as large as 5~mmag), so we applied it to those datasets.

\subsection{Spectroscopy}

\subsubsection{Goodman/SOAR}\label{sec:goodmanspec}

We observed \starname\ with the Goodman spectrograph \citep{Goodman_spectrograph} on the Southern Astrophysical Research (SOAR) 4.1 m telescope located at Cerro Pach\'on, Chile, on two nights. We used the red camera, the 1200 l/mm grating in the M5 setup, and the 0.46\arcsec\ slit rotated to the parallactic angle. This setup gave a resolution of $R\simeq$ 5880 spanning 6150--7500\AA. 

\begin{figure}[tb]
    \centering
    \includegraphics[width=0.5\textwidth]{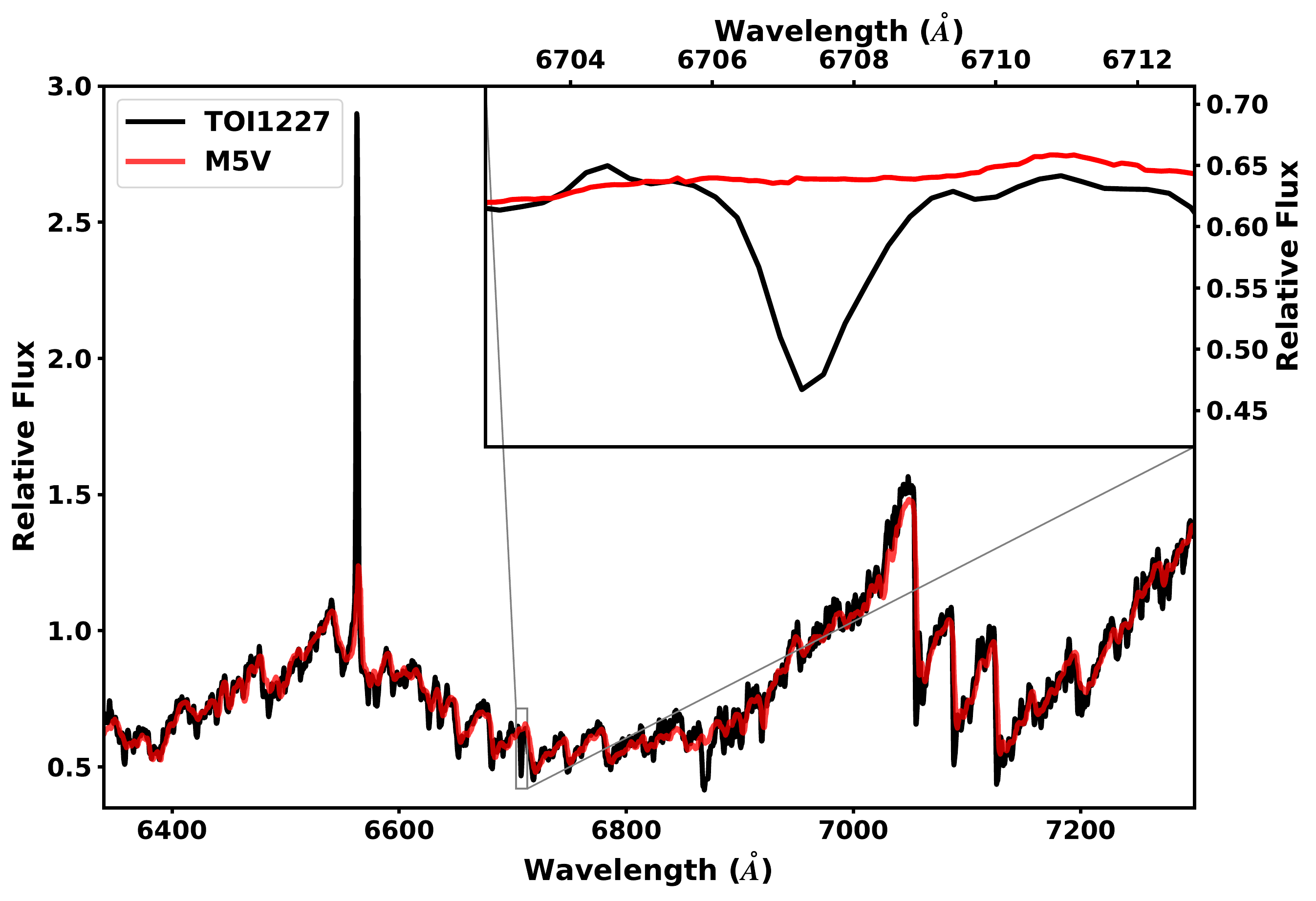}
    \caption{Goodman spectrum of \starname\ (black) and a template M5V dwarf from \citet{Kesseli2017}. Because of extinction around \starname\ and imperfect flux calibration from Goodman, we fit out a 2nd-order polynomial on the Goodman spectrum to better match the template. The Goodman spectrum is also higher resolution than the template ($R$ of 5880 vs 2000). The inset highlights the Li 6708\AA\ line, which is absent in the spectrum of older stars like the template, but strong in \starname's spectrum. }
    \label{fig:spec}
\end{figure}

We obtained the first spectrum on 2019 Dec 12 (UT) shortly after the target was alerted as a TOI, to check for signs of youth. We obtained three back-to-back exposures of \starname, each with an exposure time of 800\,s. The resulting spectrum showed the Li and H$\alpha$ expected for a young star and TiO features consistent with an M5 dwarf, as we show in Figure~\ref{fig:spec}. Once we confirmed the planetary signal (based on ground-based transits), we obtained an additional spectrum on 2021 Feb 23 (UT) under clear conditions. Our goal was a spectrum with a signal-to-noise ratio (SNR) $\gg$100 per pixel across the full wavelength range, to use for stellar characterization (Section~\ref{sec:star}). To this end, we used an exposure time of 1800\,s for five back-to-back exposures. 

To better determine the age of \starname, we also observed 13 nearby stars that are likely part of the same grouping in LCC as \starname\ (see Section~\ref{sec:member}). The 13 targets observed with Goodman were selected from the parent sample to map out lithium levels from M0--M5, which can constrain the age of the population. These targets were observed between 23 Feb 2021 and 24 Apr 2021 using an identical setup to the observations for \starname. Exposure times were set to ensure a SNR greater than 50 (per pixel) around the Li line and varied from 90\,s to 420\,s per exposure with at least 5 exposures per star for outlier (cosmic ray) removal. 

Using custom scripts, we performed bias subtraction, flat fielding, optimal extraction of the spectra, and mapping pixels to wavelengths using a 5th-order polynomial derived from the Ne lamp spectra obtained right before or after each spectrum. Where possible, we applied a small linear correction to the wavelength solution based on the sky emission or absorption lines. We stacked the individual extracted spectra using the robust weighted mean. For flux calibration, we used an archival correction based on spectrophotometric standards taken over a year. Due to variations on nightly or hourly scales, this correction is only good to $\simeq$10\%.

\subsubsection{HRS/SALT}\label{sec:salt}
We observed \starname\ with the Southern African Large Telescope \citep{2006SPIE.6267E..0ZB} High Resolution Spectrograph \citep{2014SPIE.9147E..6TC} on six nights (2020 May 08 to 2020 June 17). On each of the six visits, we obtained three back-to-back exposures. We used the high-resolution mode with an integration time of 800s per exposure. The resulting spectral resolution was $R\sim$ 46,000. The HRS data were reduced using the MIDAS pipeline \citep{MIDAS_HRS}, which performs flat fielding, bias subtraction, extraction of each spectrum, and wavelength calibration with arclamp exposures. 

We measured radial velocities from the SALT spectra by computing spectral line broadening functions \citep[BFs;][]{Rucinski1992} with the \texttt{saphires} python package\footnote{\url{https://github.com/tofflemire/saphires}} \citep{Tofflemireetal2019}. The BF results were computed from the linear inversion of an observed spectrum with a narrow-lined template. We computed the BF for 16 individual spectral orders in the SALT red arm that range from 6400 to 8900 \AA; the remaining orders had strong telluric contamination. We used a 3100 K, \logg = 4.5 PHOENIX model as our narrow-lined template \citep{2013A&A...553A...6H}. The BFs only contained one profile; we did not detect a secondary star. The BFs from each order were then combined, weighted by the SNR. We fit the combined BF with a rotationally-broadened profile to determine the stellar radial velocity and \vsini. Radial velocities from each epoch, corrected for barycentric motion, are presented in Table \ref{tab:RVs}. The mean and standard deviation of the \vsini\ measurements were 17.5$\pm$0.3 \kms, although this does not include corrections for activity/magnetic broadening or macroturbulence \citep[$\sim$1-3\kms, ][]{2020ApJ...888..116S}.

\subsubsection{IGRINS/Gemini-S}\label{sec:IGRINS}

We observed \starname\ a total of 11 times from 2020 Dec 31 to 2021 Apr 24 with the Immersion Grating Infrared Spectrometer \citep[IGRINS;][]{Park2014, Mace2016, 2018SPIE10702E..0QM} while on Gemini-South observatory (program ID GS-2020B-FT-101). IGRINS uses a silicon immersion grating \citep{2010SPIE.7735E..1MY} to achieve high resolving power ($R\simeq$ 45,000) and simultaneous coverage of both $H$ and $K$ bands (1.48--2.48\,\um) on two separate Hawaii-2RG detectors. IGRINS is stable enough to achieve RV precision of $\lesssim$40\,m\,s$^{-1}$ (or better) using telluric lines for wavelength calibration \citep{2016csss.confE..55M, 2021AJ....161..283S}.

All observations were done following commonly used strategies for point-source observations with IGRINS\footnote{\url{https://sites.google.com/site/igrinsatgemini/proposing-and-observing}}. Each target was placed at two positions along the slit (A and B), taking an exposure at each position in an ABBA pattern. Individual exposure times were between 120s and 425s, and the (total) times per epoch were between 720s and 2280s to achieve peak SNR$\gtrsim$100 per resolution element in the K-band. To help remove telluric lines, we observed A0V standards within 1 hour and 0.1 airmasses of the observation of \starname. 

We reduced IGRINS spectra using version 2.2 of the publicly available IGRINS pipeline package\footnote{\url{https://github.com/igrins/plp}} \citep{jae_joon_lee_2017_845059}, performing flat fielding, background removal, order extraction, distortion correction, wavelength calibration, and telluric correction using the A0V standards and an A-star atmospheric model. We used the spectrum right before telluric correction to improve the wavelength solution and provide a zero point for the RVs. 

We extracted the radial velocities of \starname\ using the \texttt{IGRINS RV} code\footnote{\url{https://github.com/shihyuntang/igrins_rv}} \citep{2021JOSS....6.3095T} with a 3000\,K PHOENIX model \citep{2013A&A...553A...6H} and the \texttt{TelFit} code to create a synthetic telluric spectrum \citep{Gullikson2014}.  Barycentric-corrected RVs from each epoch are listed in Table~\ref{tab:RVs}. \texttt{IGRINS RV} provided an estimate of the rotational broadening (\vsini$=16.65\pm0.24$\,\kms) and the star's systemic velocity ($13.3\pm0.3$\,\kms), the calibration of which is detailed in \citet{2021AJ....161..283S}.

\begin{deluxetable}{lllr}
\centering
\tabletypesize{\scriptsize}
\tablewidth{0pt}
\tablecaption{Radial Velocity Measurements of \starname\ \label{tab:RVs}}
\tablehead{\colhead{JD-2450000} & \colhead{$v$ (\kms)} & \colhead{$\sigma_v$ (\kms)\tablenotemark{a}} & \colhead{Instrument} }
\startdata
 8978.2812 & 12.33 &  0.14 & HRS \\ 
 8985.3302 & 11.32 &  0.32 & HRS \\ 
 9000.2566 & 13.39 &  0.24 & HRS \\ 
 9006.3086 & 12.07 &  0.29 & HRS \\ 
 9010.3192 & 11.86 &  0.37 & HRS \\ 
 9018.2586 & 14.33 &  0.42 & HRS \\ 
 9215.7831 & 13.221 & 0.034 & IGRINS \\ 
 9217.7999 & 13.278 & 0.038 & IGRINS \\ 
 9218.7799 & 13.332 & 0.044 & IGRINS \\ 
 9226.8620 & 13.252 & 0.051 & IGRINS \\ 
 9227.8498 & 13.295 & 0.043 & IGRINS \\ 
 9233.8262 & 13.229 & 0.037 & IGRINS \\ 
 9262.6350 & 13.411 & 0.043 & IGRINS \\ 
 9267.7740 & 13.474 & 0.038 & IGRINS \\ 
 9312.5170 & 13.402 & 0.060 & IGRINS \\ 
 9318.5862 & 13.592 & 0.043 & IGRINS \\ 
 9329.6748 & 13.289 & 0.034 & IGRINS \\ 
\enddata
\tablenotetext{a}{IGRINS velocity errors are {\textit relative}; the error on systemic velocity of \starname\ is $13.3\pm0.3$\,\kms and is dominated by the zero-point calibration. HRS velocities are absolute and include the zero-point calibration.}
\end{deluxetable}

\subsection{High-Contrast Imaging}\label{sec:AO}
\begin{figure}[tbh]
    \centering
    \includegraphics[width=0.5\textwidth]{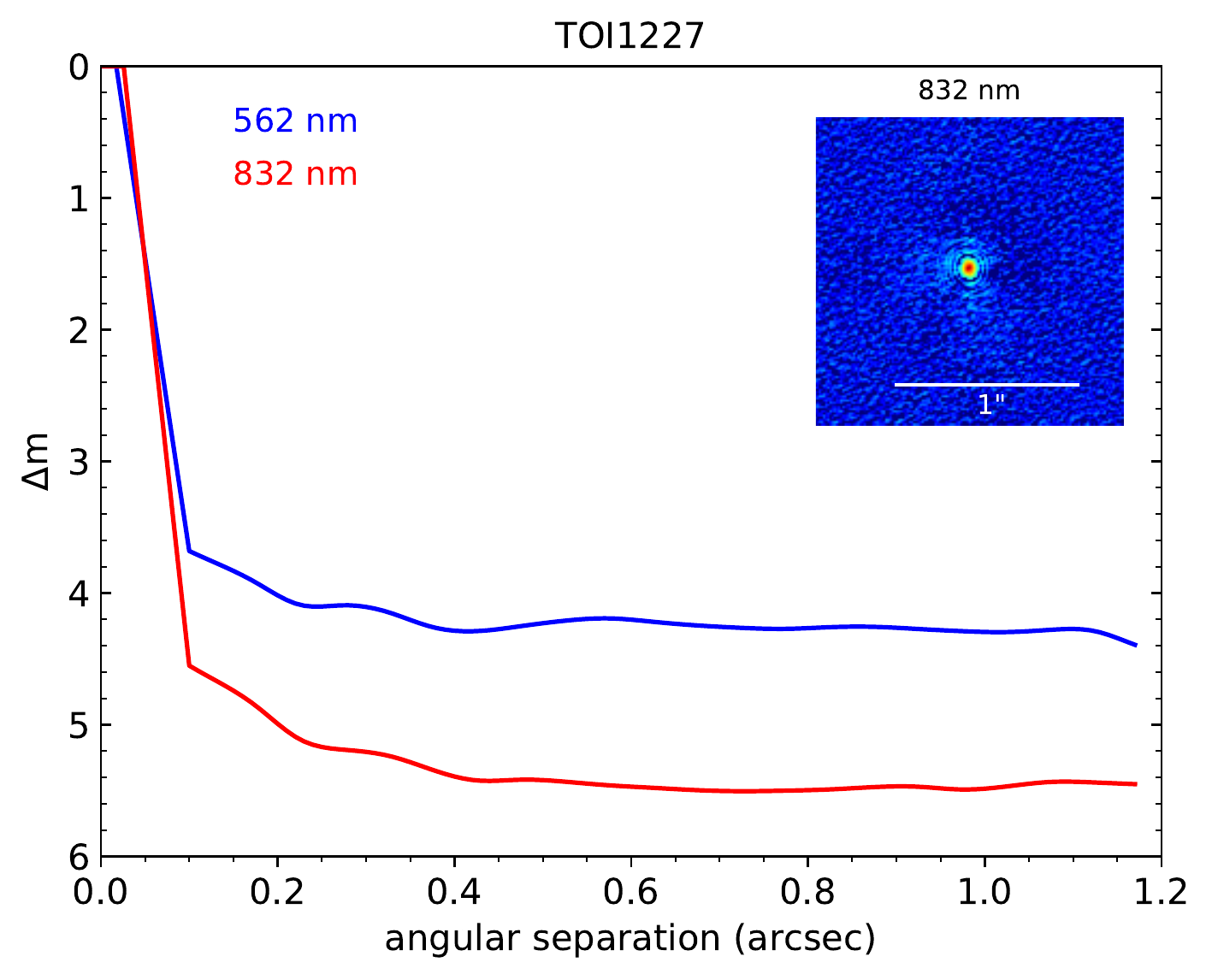}
    \caption{Detection limits from the Zorro speckle imager for \starname. Detection limits from the two narrow-band filters (5620 \AA\ and 8320 \AA) are shown in blue and red, respectively. The sub-panel in the top right shows the narrow-band 8320\AA, reconstructed image.}
    \label{fig:zorro}
\end{figure}

We observed \starname\ with the Gemini South speckle imager, Zorro \citep{Zorro}, on UT 2020 March 13 (program ID GS-2020A-Q-125). Zorro provided simultaneous two-color, diffraction-limited imaging, reaching angular resolutions of $\sim$0.02\arcsec\ in ideal conditions with a field of view of about 1.2\arcsec. \starname\ was observed in 17 sets of 1000x60\,msec exposures by the `Alopeke-Zorro visiting instrument team with the standard speckle imaging mode in the narrow-band 5620\,\AA\ and 8320\,\AA\ filters ([562] and [832]). All data were reduced with the pipeline described in \citet{2011AJ....142...19H}. No additional sources were detected within the sensitivity limits (Figure~\ref{fig:zorro}). 

The limiting contrasts for binary companions are set by the redder [832] filter, ruling out equal-brightness companions at $\rho = 0.03\arcsec$ and excluding companions fainter than the target star by $\Delta [832] < 1.5$ mag at $\rho = 0.05\arcsec$, $\Delta [832] < 4.5$ mag at $\rho = 0.10\arcsec$, and $\Delta [832] < 5.0$ mag at $\rho = 0.2\arcsec$. Converting $M_{[832]}$ to $M_G$ using the color-magnitude relations of Kraus et al. (in prep), the corresponding mass and physical scale limits implied using the 10 Myr isochrones of \citet{BHAC15} can rule out equal-mass companions at $\rho = 3$ AU, $M > 50 M_{Jup}$ at $\rho > 5$ AU, $M > 15 M_{Jup}$ at $\rho > 10$ AU, and $M > 14M_{Jup}$ AU at $\rho > 20$ AU.

\subsection{Limits on Wide Companions from Gaia EDR3 imaging}

To identify potential wide binary companions that might exert secular dynamical influences on the \starname\ planetary system, we searched the \gaia\ EDR3 catalog \citep{GaiaEDR3} for comoving and codistant neighbors. While several candidate neighbors are found at projected separations of $\rho > 500 \arcsec$ (Section~\ref{sec:pop}), none are found at $\rho < 400\arcsec$ ($\rho \lesssim 40,000$ AU), the typical outer bound for observed binary companions for Sun-like stars \citep{Raghavan2010} and well beyond the typical separation of mid-M binaries \citep{2019AJ....157..216W}.

The nominal brightness limit of Gaia is $G_{lim} < 21$ mag, which at an age of $\simeq$10\,Myr approximately corresponds to a limit of $M_{lim} \simeq 14 M_{Jup}$. \citet{Ziegler2018} and \citet{2019A&A...621A..86B} have shown that \gaia\ is sensitive to companions with $\Delta G = 6$ mag at $\rho > 2 \arcsec$ and $\Delta G = 4$ mag at $\rho > 1 \arcsec$. Thus, \gaia\ would have identified almost all companions down to the separation regime where it meets the contrast limit of the Zorro speckle imaging (Section~\ref{sec:AO}).

\subsection{Archival Spectra from ESO}
With the goal of better characterizing \starname's age, we retrieved the publicly available reduced spectra from the European Southern Observatory (ESO) archive for stars in the same population as \starname\ (Section~\ref{sec:member}). We downloaded any spectra of candidate members that included the Li line. In total, 11 objects had spectra from HARPS at the 3.6m telescope (La Silla Observatory), FEROS at the 2.2 m telescope (La Silla Observatory), and/or X-shooter at VLT (Cerro Paranal Observatory). 

\subsection{Archival Photometry, Astrometry, and Velocities}

For \starname\ and all candidate members of the parent population (Section~\ref{sec:pop}) we retrieved parallaxes, positions, proper motions, and photometry from \gaia\ Early Data Release 3 \citep[EDR3;][]{GaiaEDR3, Lindegren_2021, Riello_2021}. From EDR3 we also retrieved the Renormalised Unit Weight Error (RUWE) for all stars. RUWE value is effectively an astrometric reduced $\chi^2$ value, normalized to correct for color and brightness dependent effects\footnote{\url{https://gea.esac.esa.int/archive/documentation/GDR2/Gaia_archive/chap_datamodel/sec_dm_main_tables/ssec_dm_ruwe.html}}. RUWE should be around 1 for well-behaved sources, and higher values (RUWE$\gtrsim$1.3) suggests with the presence of a stellar companion \citep{Ziegler2020, 2021arXiv210609040W}. 

We downloaded velocities for candidate population members from a general Vizier/SIMBAD search, taking the most precise value for stars with multiple measurements. This yielded velocities for 21 stars. The full list of sources for these velocities is included in Table~\ref{tab:population}.

To aid with stellar characterization of \starname, we also pulled photometry from the Two-Micron All-Sky Survey \citep[2MASS;][]{Skrutskie2006}, the Wide-field Infrared Survey Explorer \citep[WISE; ][]{allwise}, and SkyMapper second Data Release \citep[SkyMapper DR2; ][]{Onken_2019}.

\section{The age and membership of TOI1227}\label{sec:member}

Our Goodman spectrum of \starname\ showed strong H$\alpha$ emission and lithium absorption. Combined with \starname's high position on the color-magnitude diagram (CMD), this set an upper age limit of $<30$\,Myr. However, we were able to derive a more precise estimate of the age of \starname\ from its parent population. To this end, we first identified likely members of the same association, then estimated the group age using rotation, lithium abundance, and fitting isochrones to the color-magnitude diagram.

\subsection{The parent association of TOI 1227}\label{sec:pop}

\begin{figure*}[tbh]
    \centering
    \includegraphics[width=0.95\textwidth]{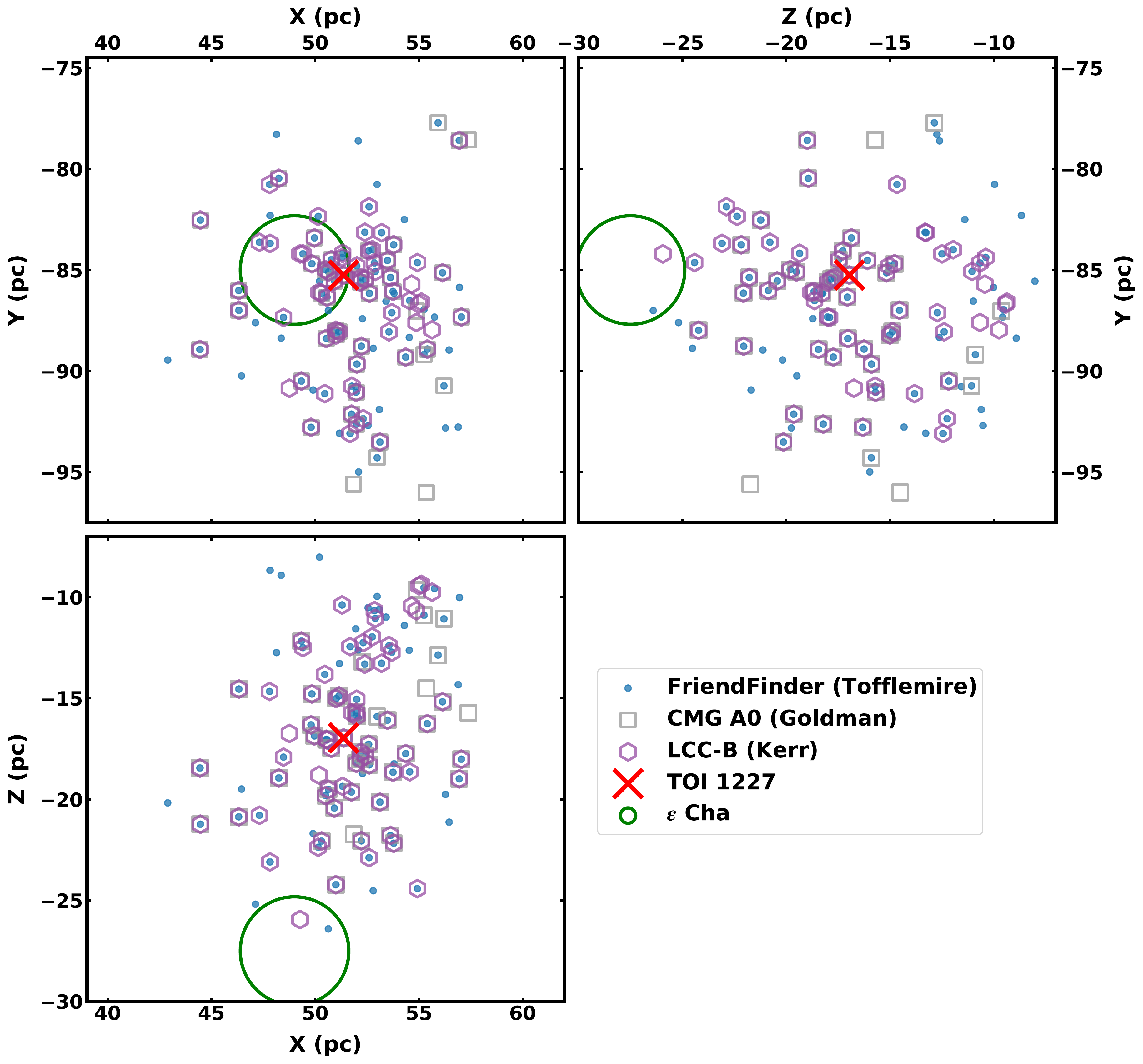}
    \caption{Galactic Heliocentric ($XYZ$) coordinates of 108 stars within the B subgroup of LCC \citep[purple hexagons,][]{2021arXiv210509338K}, the A0 population of the Crux Moving Group \citep[grey squares,][]{2018ApJ...868...32G}, or stars identified using the FriendFinder algorithm restricted to $V_{\rm{tan},\rm{off}}<1$\kms\ and within 10\,pc of \starname\ \citep[blue dots, ][]{THYMEV}. The planet host, \starname, is marked with a red X, and is in all three lists. The significant overlap between these three selections re-affirms that these are all the same population, just selected using slightly different methods. For reference, we included a green circle that includes most of the known \epscha\ members, highlighting that \epscha\ is a separate and more compact population. 
    \label{fig:xyzgroups}
    }
\end{figure*}

To identify any co-moving and coeval population to \starname, we first used the BANYAN-$\Sigma$ tool \citep{BanyanSigma}\footnote{\url{https://github.com/jgagneastro/banyan_sigma}}, providing as input the IGRINS systemic velocity and the \gaia\ EDR3 position, parallax, and proper motion. This yielded a membership probability of 99.3\% for \epscha, with 0.4\% for LCC and 0.3\% for the field. However, \starname\ is $\simeq$25\,pc away from the core of \epscha, while nearly all known members of \epscha\ are packed in a sphere $\lesssim$5\,pc across \citep[][also see Figure~\ref{fig:xyzgroups}]{2013MNRAS.435.1325M}. The \texttt{BANYAN} algorithm preferred placing \epscha\ over LCC likely because of better agreement in $UVW$ despite poor $XYZ$ agreement. However, the current \texttt{BANYAN} model did not account for more recent findings of significant velocity substructure in LCC \citep{2018ApJ...868...32G, 2021arXiv210509338K}, which changes the $UVW$ model for LCC and hence the membership probabilities. 

\starname\ was more recently identified as a member of the ``B'' sub-population of LCC by \citet{2021arXiv210509338K} and the A0 sub-group of the Crux Moving Group (CMG) by \citet{2018ApJ...868...32G}. These two groups are effectively the same, but use different selection methods and treatment of sub-groups. \citet{2021arXiv210509338K} considered this group (LCC-B) part of the LCC substructure, while \citet{2018ApJ...868...32G} considered the CMG a moving group associated with LCC with multiple sub-groups (A0, A, B, and C). The $XYZ$ positions, proper motions, and available radial velocities of LCC-B/CMG-A0 members are similar to \starname\ (better so than \epscha; Figure~\ref{fig:xyzgroups}), so we adopted the LCC-B/CMG-A0 as the initial parent population. 

We looked for additional members using the \texttt{FriendFinder} algorithm\footnote{\url{https://github.com/adamkraus/Comove}} \citep{THYMEV}. \texttt{FriendFinder} uses the \gaia\ EDR3 astrometry and radial velocity of a source (we used our IGRINS systemic velocity) to compute Galactic $UVW$ and $XYZ$, then projects the $UVW$ motion into the tangential velocities that would be expected for nearby stars if they were to share the same space motion. We selected stars with separations $<$10\,pc from \starname\ and tangential velocities $<$1\,\kms\ from the expected values calculated by \texttt{FriendFinder}. These cuts were based on the fact that 1\,\kms$\simeq$1\,pc\,Myr$^{-1}$ and an estimated age for the population of $\simeq$10\,Myr. We estimated these cuts would yield $\lesssim$2 field interlopers based on the background population of \gaia\ stars. 

\begin{figure}[tbh]
    \centering
    \includegraphics[trim=0 0 65 0,clip,width=0.49\textwidth]{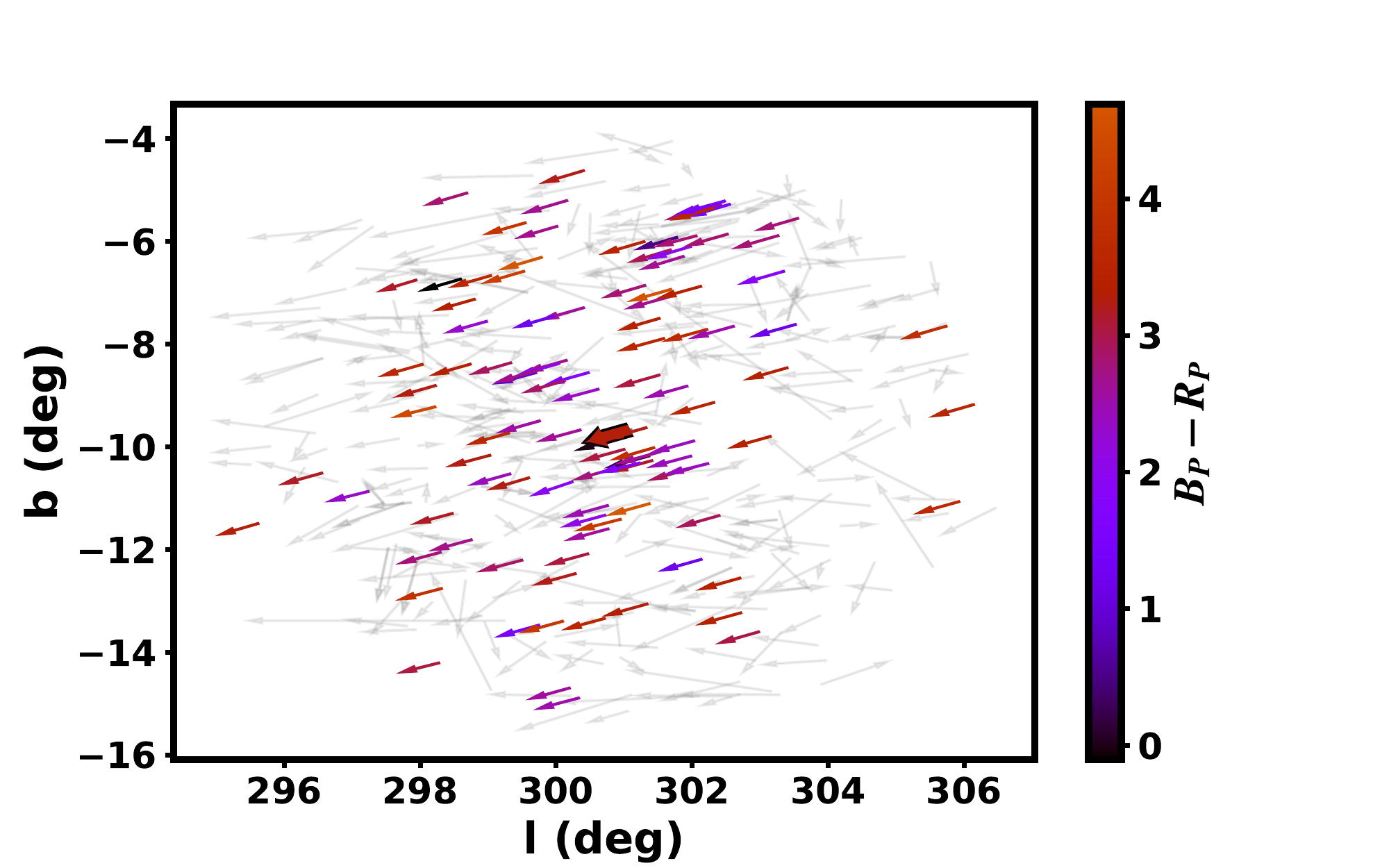}
    \caption{Galactic coordinates ($l$, $b$) of stars around \starname, with arrows indicating the direction and magnitude of the proper motion for each star. Likely members of \population\ (Section~\ref{sec:pop}) are colored by their $B_P-R_P$ color. All star within 10\,pc of \starname\ (excluding candidate members) are shown as grey arrows. \starname\ is thicker with a black outline.
    \label{fig:lb_pm}
    }
\end{figure}

The three candidate membership lists--those from \citet{2018ApJ...868...32G}, \citet{2021arXiv210509338K}, and this work--have significant overlap in both Galactic position and proper motion (Figures~\ref{fig:xyzgroups} and \ref{fig:lb_pm}). Of the 108 stars in any of the three lists, 40 were in all three, and 27 were in two of the three. Further, most of the stars missing from \citet{2018ApJ...868...32G} were missing precise parallaxes in \gaia\ DR2 (the former used DR2 while the other two selections used EDR3). Many of the stars in the \texttt{FriendFinder} list but not in \citet{2021arXiv210509338K} were removed by one of the quality cuts imposed by the latter work (e.g., $BP/RP$ flux excess). The \texttt{FriendFinder} list was also missing a small number of objects slightly further away from \starname\ because of our separation cut.

To be inclusive, we adopted the combination of all three lists as the membership list for \starname's association; these 108 stars are listed in Table~\ref{tab:population}.

The resulting population has two (contradictory) names in the literature, both of which are difficult to remember. Given the presence of a young transiting planet, it deserves a more memorable name than `A0' or 'B'. Most of the members land within the Musca constellation, so we refer to the merged group of stars as the \population\ group (or just \population). 

\begin{figure}[tbh]
    \centering
    \includegraphics[width=0.47\textwidth]{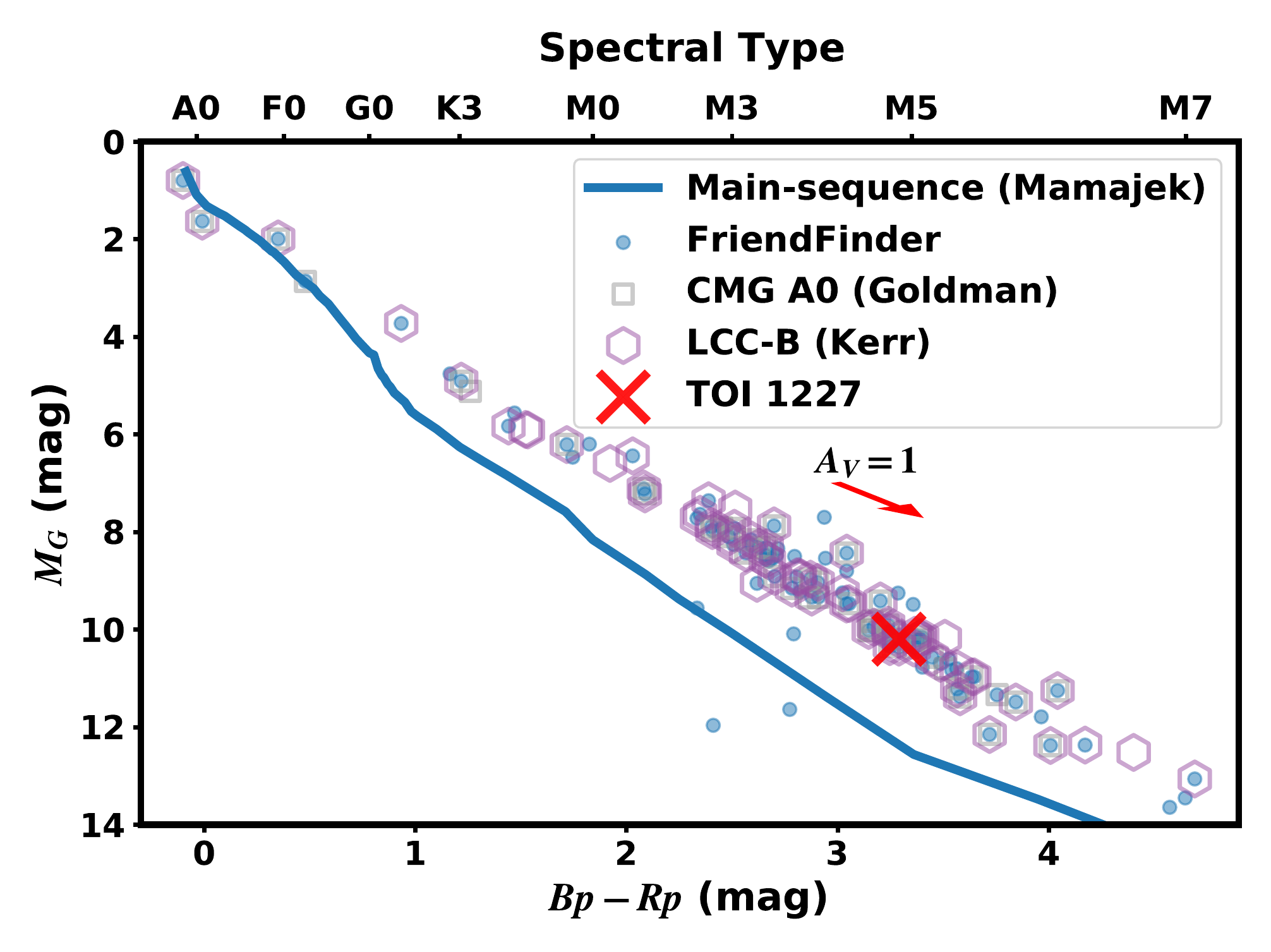}
    \caption{Color-magnitude diagram of stars in \population\ (stars associated with \starname), with symbols by the selection method or reference. We removed stars with poor astrometric fits from \gaia\ \citep[RUWE$>1.5$;][]{GaiaEDR3}. \starname\ is marked as an red "X", and lands within the tight sequence. The blue line indicates an empirical main-sequence for reference \citep{2013ApJS..208....9P}. 
    \label{fig:cmd}
    }
\end{figure}

\starname\ resides at the center of \population, both based on Galactic position (Figure~\ref{fig:xyzgroups}) and proper motion (Figure~\ref{fig:lb_pm}). The mean velocity of the members with literature RVs (12.8\kms) is also in good agreement with the value for \starname. A randomly selected star that matches \population\ in $XYZ$ has a $\ll$1\% probability of matching \population\ in proper motion and radial velocity by chance. The CMD also indicates a common age for stars in the population (Figure~\ref{fig:cmd}), with \starname\ matching the sequence. Combined with the presence of lithium in its spectrum, \starname's membership in \population\ is unambiguous.

\subsection{Rotation}\label{sec:rotation}

To better measure the age of \population\ (and hence \starname), we measured rotation periods for all candidate members using \tess\ Full-Frame Images (FFIs) downloaded from the Mikulski Archive for Space Telescopes (MAST). First, we created initial light curves from the FFI cutouts centered on each candidate member. After background subtraction, we used the \texttt{unpopular} package \citep{2021arXiv210615063H} to generate a Causal Pixel Model (CPM) of the telescope systematics for each star, which we subtracted from the initial light curve. We used the CPM curves because it does a better job preserving long-period signals than PDCSAP \citep{2016ApJ...832..133S, 2017arXiv171002428W}.

We searched the resulting single-sector light curves for rotation periods between 0.1--30 days using the Lomb-Scargle algorithm \citep{LombScargle}, repeating for each available sector. We selected the rotation period from the sector that returned the largest Lomb-Scargle power. As a check, we phase-folded the resulting light curves along that rotation period to inspect each measurement by eye. A few stars had multiple peaks in their Lomb-Scargle periodograms, which might indicate an unresolved binary. In such cases, we took the stronger signal (which was always the shorter period). Each rotation measurement was assigned a quality score during the visual inspection following \citet{2021arXiv210613250R}, with Q0 indicating an obvious rotation signal, Q1 a questionable signal, Q2 a spurious detection, and Q3 a light curve dominated by noise. In total, we measured usable rotation periods (Q0 or Q1) for 90 stars out of a sample of 108. Of the remaining 18, 11 were too faint to retrieve a reliable light curve, one showed a clear dipper pattern \citep[and was identified as such by ][]{2020arXiv200912830T}, three had high flux contamination from nearby stars, and the remaining three had no significant period detection. Rotation period measurements are included in Table~\ref{tab:population}. The high detection rate is consistent with a young population and suggests a low rate of field-star interlopers in our selection.

\begin{figure}[t]
    \centering
    \includegraphics[width=0.47\textwidth]{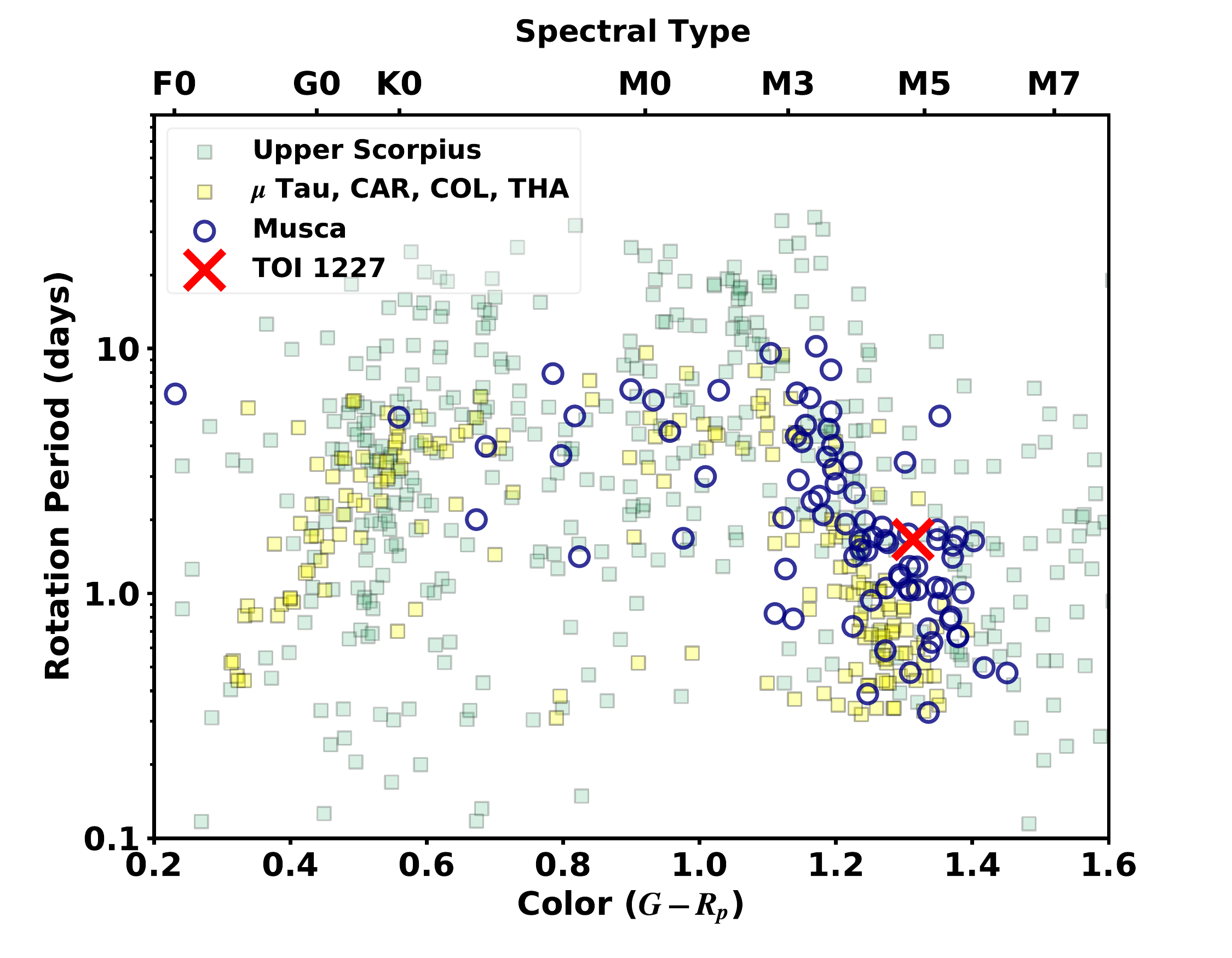}
    \caption{Rotation periods for candidate members of \population\ (dark circles). Only stars with high-quality rotation periods (Q0 or Q1) are shown. For reference, we show rotation periods from $\simeq$10\,Myr Upper Scorpius from \citet{Rebull2018} and 40-60\,Myr ($\mu$ Tau, Carina, Columba, and Tucana-Horologium) from \citet{2020ApJ...903...96G}. 
    \label{fig:rotation}
    }
\end{figure}

The rotation period distribution (Figure~\ref{fig:rotation}) is consistent with the spread seen in the 10\,Myr Upper Scorpius association \citep{Rebull2018}, and marginally consistent with the tighter rotation sequence at 40--60\,Myr from \citet{2020ApJ...903...96G}. \starname's rotation period (1.65\,days) was consistent with the \population\ sequence. This effectively sets a limit of $\lesssim60$\,Myr for the group and further confirms \starname\ as a member. 

\subsection{Lithium }\label{sec:lithium}
\begin{figure*}[tbh]
    \centering
    \includegraphics[width=0.47\textwidth]{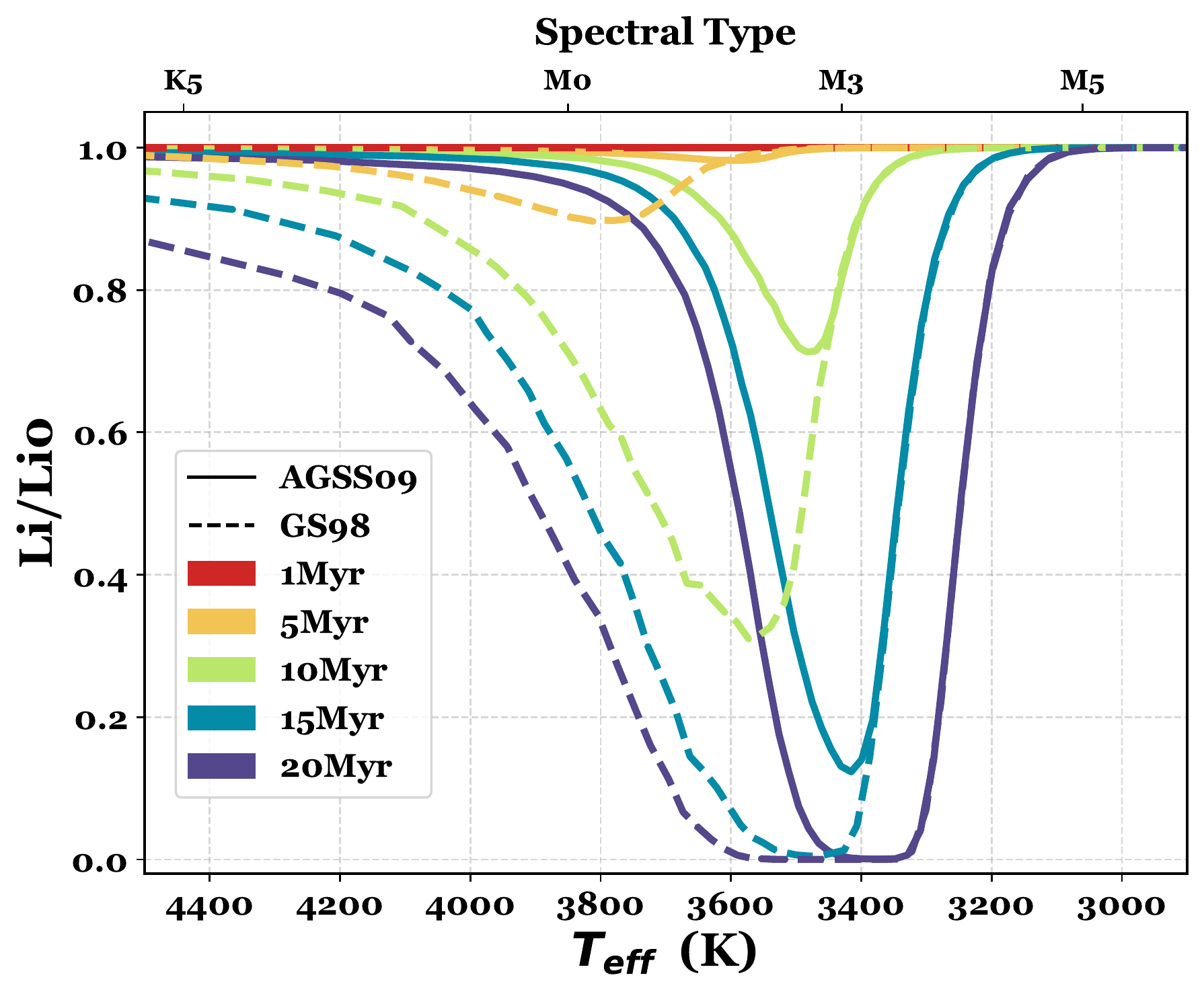}
    \includegraphics[width=0.51\textwidth]{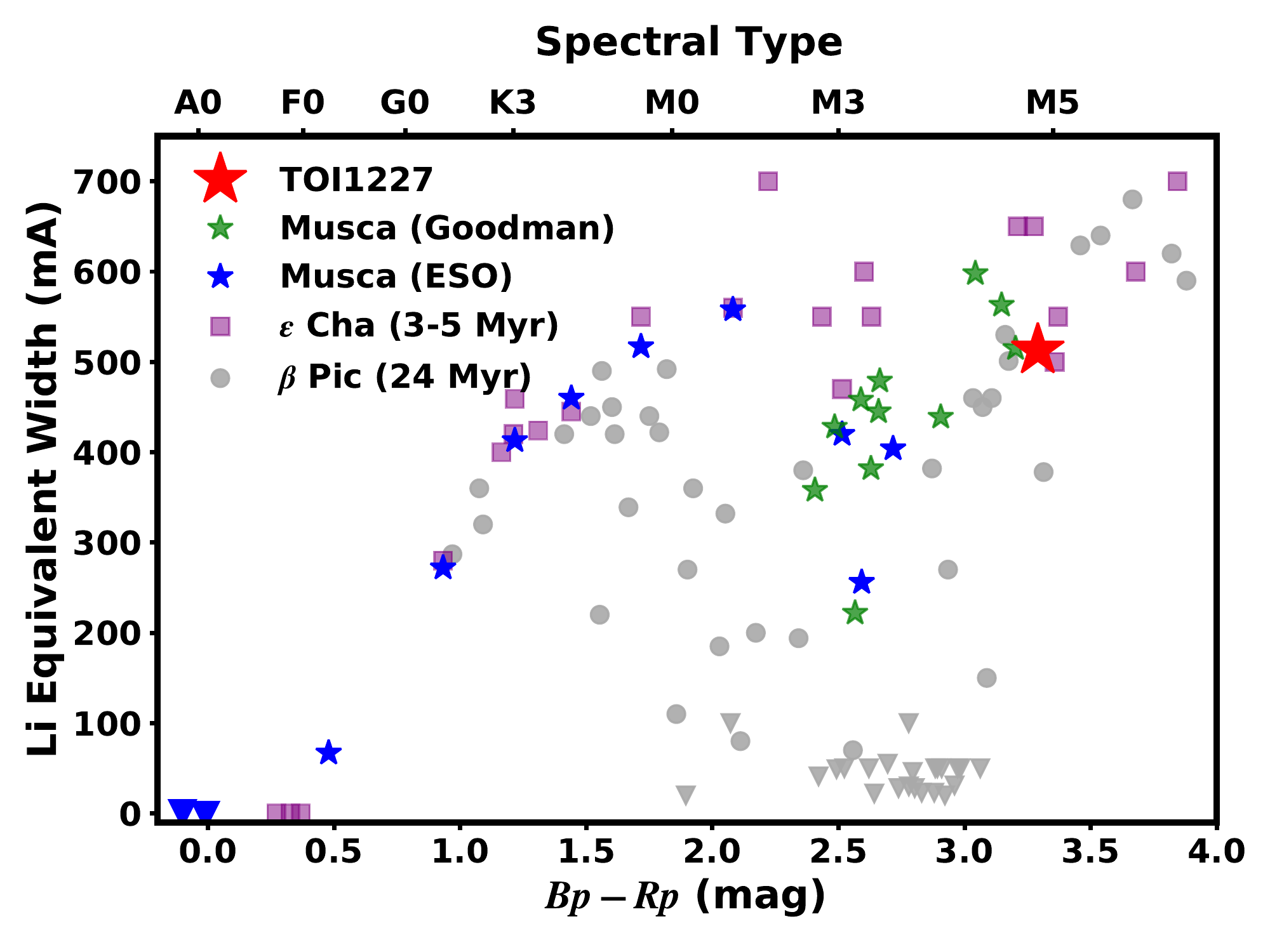}
    \caption{Left: Li abundance relative to the initial level predicted by the DSEP magnetic models for two different Solar abundance scales (dashed and solid lines) with colors corresponding to different times. The existence of a significant drop in the Li levels without a full depletion limits an association's age to 8--14\,Myr, depending on the assumed abundance scale \citep{GS98, AGSS09}. Right: lithium equivalent width of \starname\ (red) and its kinematic neighbors measured from Goodman (green star) or high-resolution ESO archival spectra (blue star). We compare this to the sequence from \epscha\ \citep[purple; ][]{2013MNRAS.435.1325M}, and $\beta$\,Pic \citep[grey; ][]{2017AJ....154...69S}. Arrows indicate upper limits.  \epscha\ shows a clear sequence, while $\beta$\,Pic shows full Li depletion around M2-M4. The stars associated with \starname\ show a dip around M3, but not a full depletion, bracketing the age between the two groups (5\,Myr and 24\,Myr). The top axis of both plots shows estimated spectral types. 
    \label{fig:lithum}
    }
\end{figure*}

Lithium (Li) is a powerful indicator of age in young stars \citep{1996ApJ...459L..91C}. M dwarfs deplete their lithium over 10--200\,Myr at a rate that depends on their spectral type; this forms a region where stars have no surface lithium (i.e., the lithium depletion boundary or LDB). The location and size of the LDB are strongly sensitive to the association's age, and LDB ages are largely independent of the isochronal age \citep{2014prpl.conf..219S}. At the youngest ages ($<$20\,Myr), lithium is only partially depleted, leading to a `dip' in the Li levels short of a full boundary \citep[Figure~\ref{fig:lithum}; also see][]{rizzuto2015}. The location and depth of the lithium dip also depend strongly on age. 

For our Li determinations, we measured the equivalent width of the Li 6708\,\AA\ line for 22 stars using our high-resolution ESO archival (11 targets) and Goodman (13 targets) spectra (two stars overlap). To account for the variations in resolution, \vsini, and velocity between targets and the instrument used, we first fit nearby atomic lines with a Gaussian profile (e.g., iron lines for warmer stars and potassium lines for the M dwarfs). We used the width from these fits to define the bounds of the Li line. To estimate the pseudo-continuum, we iteratively fit the 6990--6720\AA\ region excluding the Li line, each time removing regions $>$4$\sigma$ below the fit (there were no emission lines in this region). We did not attempt to correct for contamination from the Fe line at 6707.44\,\AA\ or broad molecular contamination in the cooler stars, which likely set a limit on the precision of our equivalent widths at the $\simeq$10\% level.

A single star, TIC 359357695, had two clear sets of nearly equal-depth lines (an SB2). Interestingly, this star is a known TOI (1880), indicating a roughly equal-mass eclipsing binary. For the Li equivalent width, we measured each line individually with a manually-applied offset. We then combined the two equivalent widths. 

In Figure~\ref{fig:lithum}b we compare the Li sequence for \population\ to that from \betapic\ \citep[$\simeq$24\,Myr;][]{2017AJ....154...69S} and \epscha\ \citep[3-5\,Myr;][]{2013MNRAS.435.1325M}. The \epscha\ cluster has high Li levels over the full sequence, while \betapic\ showed a full depletion around M3--M5. \population\ resides between these two, with a dip in Li levels around M3 but not full depletion; this effectively bounds the age of \population\ between the two groups. Based on Li predictions from the Dartmouth Stellar Evolution Program \citep[DSEP; ][]{Dotter2008} with magnetic enhancement \citep{Feiden2012b}, we estimated the Li age to be 8--14\,Myr (Figure~\ref{fig:lithum}). 

\subsection{Comparison to theoretical isochrones}\label{sec:isochrones}

We compared the candidate members of \population\ to the PARSECv1.2S stellar isochrones \citep{PARSEC}. To handle contamination from binaries and nonmember interlopers, we used a mixture model described in detail in Appendix~\ref{sec:mixture}. We also removed stars with RUWE$>1.3$ (likely binaries) and stars outside the colors covered by the isochrones (i.e., stars with $M_G>11.8$ and $G-R_P>1.45$). Many tight pairs were resolved in \gaia, but not in 2MASS or similar ground-based surveys. The mixture model is not set up to handle these cases; a data point has a single outlier probability independent of the band. Thus, we performed this comparison using \gaia\ magnitudes only. An inspection of the sequence using 2MASS magnitudes suggested that this would not change the derived age in any significant way. 

\begin{figure}[tbh]
    \centering
    \includegraphics[width=0.47\textwidth]{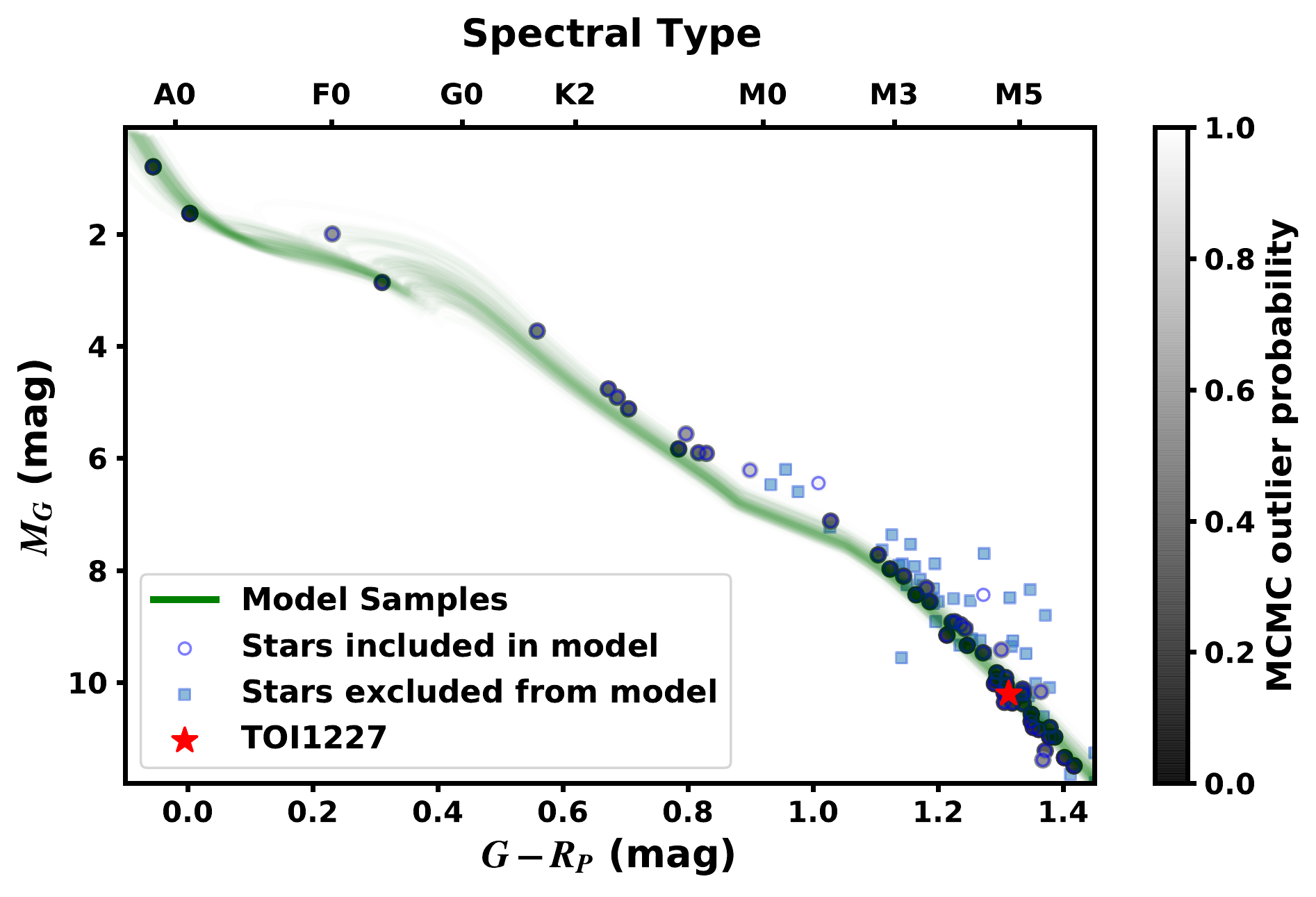}
    \caption{PARSEC isochrone fit to \population\ members using a mixture model. The model comparison is done in color, but approximate spectral classes are given on the top axis. Stars included in the fit (blue circles) are filled based on their outlier probability. Excluded points (mostly due to a high RUWE) are shown as blue squares. The green lines are 200 random samples from the MCMC posterior. Models predict slightly different ages at different mass ranges; the mixture model takes the majority age by down-weighting stars in poorly matched regions (increasing their outlier probability). 
    \label{fig:isochrone}
    }
\end{figure} 

The fit, shown in Figure~\ref{fig:isochrone}, yielded an age of 11.6$\pm$0.5\,Myr, consistent with 11.8\,Myr from \citet{2018ApJ...868...32G}, 13.0$\pm$1.4\,Myr from \citet{2021arXiv210509338K}, and $\simeq$12\,Myr for the lower end of LCC from \citet{2016MNRAS.461..794P}. The age errors were likely underestimated, as our fit did not fully account for model systematics \citep[often $\simeq$1-3\,Myr, e.g.,][]{Bell2015}. Indeed, the fit slightly underpredicted the luminosity of K2--K5 dwarfs and overpredicted the value for M4 and later. As an additional test, we repeated the fit using magnetic DSEP, which yielded a consistent age of 11.1$\pm$1.4\,Myr, suggesting model differences are comparable to our measurement errors. 

\subsection{The age of Musca}

Given the constraints from the isochrones, lithium, rotation, and the literature, we adopt an age of 11\,Myr with a conservative uncertainty of $\pm$2\,Myr. We used this age for \planetname\ as well, as any age spreads are likely to be similar to, or smaller than, the adopted uncertainties. 

\section{Host Star Analysis}\label{sec:star}

We summarize properties of \starname\ in Table~\ref{tab:prop}, the details of which we provide below. 

\begin{deluxetable}{lcc}
\centering
\tabletypesize{\scriptsize}
\tablewidth{0pt}
\tablecaption{Properties of the host star \starname. \label{tab:prop}}
\tablehead{\colhead{Parameter} & \colhead{Value} & \colhead{Source} }
\startdata
\multicolumn{3}{c}{Identifiers}\\
\emph{Gaia} EDR3 & 5842480953772012928\\
TIC & 360156606\\
2MASS & 12270432-7227064 \\
WISE & J122704.22-722706.5	\\
Skymapper & 397425267 \\
\hline
\multicolumn{3}{c}{Astrometry}\\
\hline
$\alpha$ (J2016.0)  &  186.767344 &  \emph{Gaia} EDR3\\
$\delta$ (J2016.0) & -72.451852  & \emph{Gaia} EDR3 \\
$\mu_\alpha \cos{\delta}$ (mas\,yr$^{-1}$)& -40.294$\pm$0.026 & \emph{Gaia} EDR3\\
$\mu_\delta$  (mas\,yr$^{-1}$) & -10.808$\pm$0.030 & \emph{Gaia} EDR3\\
$\pi$ (mas) & 9.9046$\pm$0.0242 & \emph{Gaia} EDR3\\
\hline
\multicolumn{3}{c}{Photometry}\\
\hline
G$_{Gaia}$ (mag) & 15.218$\pm$0.003 & \emph{Gaia} EDR3\\
BP$_{Gaia}$ (mag) & 17.195$\pm$0.006 & \emph{Gaia} EDR3\\
RP$_{Gaia}$ (mag) & 13.905$\pm$0.004 & \emph{Gaia} EDR3\\
V (mag) & 17.00$\pm$1.13 & TICv8.0\\
r' & 16.346$\pm$0.013 & Skymapper DR2\\
i' & 14.333$\pm$0.009 & Skymapper DR2\\
zs & 13.523$\pm$0.007 & Skymapper DR2\\
J (mag) & $11.890 \pm 0.024$ &  2MASS\\
H (mag) & $11.312 \pm 0.022$  & 2MASS\\	
Ks (mag) & $11.034\pm 0.021$ & 2MASS\\
W1 (mag) & $10.887 \pm 0.023$ & ALLWISE\\
W2 (mag)& $10.649 \pm 0.021 $ & ALLWISE\\
W3 (mag)& $10.516 \pm 0.062$ & ALLWISE\\ 
\hline
\multicolumn{3}{c}{Kinematics \& Position}\\
\hline
RV$_{\rm{Bary}}$ (km\, s$^{-1}$) & 13.3$\pm$0.3 & This work\\
U (km\, s$^{-1}$) & -9.85$\pm$0.16 & This work\\%-9
V (km\, s$^{-1}$) & -19.88$\pm$0.25 & This work\\%-19.8
W (km\, s$^{-1}$) & -9.13$\pm$0.06 & This work\\%-8.2
X (pc) & 51.34$\pm$0.16 & This work\\
Y (pc) & -85.22$\pm$0.27 & This work\\
Z (pc) & -16.95$\pm$0.05 & This work\\
\hline
\multicolumn{3}{c}{Physical Properties}\\
\hline
$P_{\rm{rot}}$ (days) & $1.65\pm0.04$ & This work \\
\vsini (km\, s$^{-1}$) & $ 16.65\pm0.24 $ & This work\\
$i_\star$ ($^\circ$) & $ >73$ & This work\\
\fbol\,(erg\,cm$^{-2}$\,s$^{-1}$)& ($7.87\pm0.53)\times10^{-11}$ & This work\\ 
T$_{\mathrm{eff}}$ (K) & $3072 \pm 74$ & This work\\
SpT & M4.5V--M5V & This work \\
M$_\star$ (M$_\odot$) & $ 0.170\pm0.015 $ & This work \\
R$_\star$ (R$_\odot$) &  $0.56 \pm 0.03$ & This work \\
L$_\star$ (L$_\odot$) & $(2.51\pm0.17)\times10^{-3}$ & This work \\
$A_V$ & $0.21^{+0.11}_{-0.09}$ & This work \\
$\rho_\star$ ($\rho_\odot$) & $0.94\pm0.18$ & This work \\
Age (Myr) & 11 $\pm$ 2 & This work \\
L$_X$ (erg\,s$^{-1}$) & 10$^{28.32}$ & This work\\
log($L_X/L_{\rm{bol}}$) (dex) & -2.66 & This work\\
\enddata
\end{deluxetable}

\subsection{Effective temperature, luminosity, radius, and mass}

\begin{figure}[tb]
    \centering
    \includegraphics[width=0.49\textwidth]{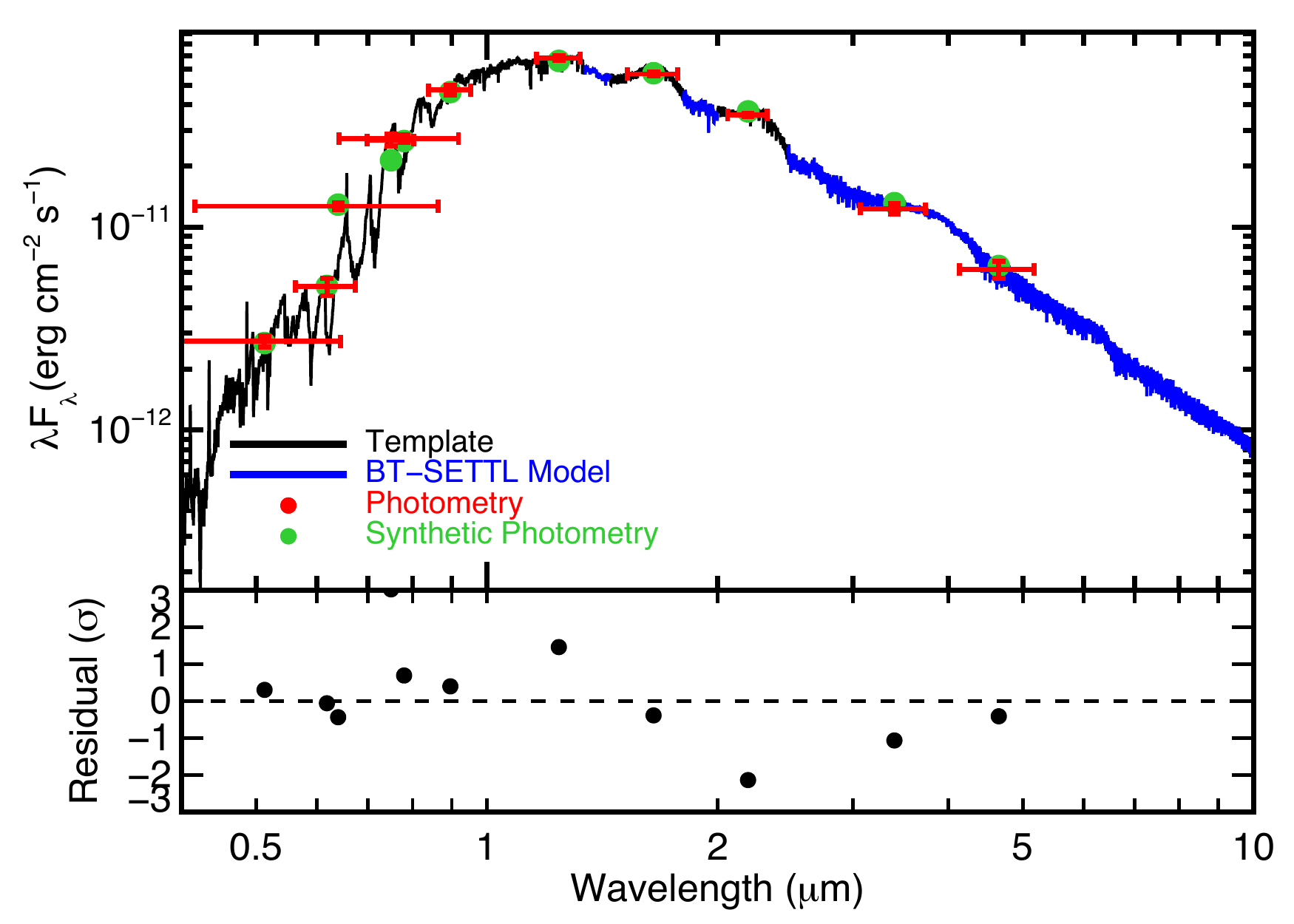}
    \caption{Example fit from comparing BT-SETTL models and young un-reddened templates to the observed photometry and Goodman spectrum. The red points are the observed photometry converted to fluxes using the appropriate zero-point and uncertainty, with the vertical errors denoting the uncertainties and the horizontal errors the approximate filter width. Green points are synthetic photometry derived from combining the spectrum with the relevant filter profile. The bottom panel shows the residual in standard deviations. \starname\ was not detected in WISE $W4$ and the $W3$ point is not shown for clarity (but is used in the fit). The template/model shown here (black) is an M4.5-M5 star in the $\beta$~Pic association and a 3075\,K model (blue). }
    \label{fig:sed}
\end{figure}

We first fit the spectral-energy-distribution (SED) following the methodology from \citet{Mann2016b} and highlighted in Figure~\ref{fig:sed}. To briefly summarize, we compared the observed photometry and Goodman spectrum to a grid of template spectra drawn from nearby (likely reddening-free) young moving groups. The observed data was simultaneously fit with Solar-metallicity BT-SETTL CIFIST models \citep{BHAC15}. The atmospheric models were used both to estimate \teff\ and fill in gaps in the template spectra (e.g., beyond 2.3\,\um). We combined this full SED with the \gaia\ EDR3 parallax to estimate both the bolometric flux (\fbol) and the total luminosity ($L_*$). With \teff\ and $L_*$, we calculated the stellar radius ($R_*$) using the Stefan-Boltzmann relation. The fit included six total free parameters: the choice of template, $A_V$, \teff, and three parameters that account for flux and wavelength calibration offsets between the Goodman spectra, stellar templates, and model spectra. To account for variability in the star, we added (in quadrature) 0.03 mags to the errors of all optical photometry. The resulting fit yielded \teff$=3072\pm84$\,K, \fbol$=(7.87\pm0.53) \times 10^{-11}$\,erg\,cm$^{-2}$\,s$^{-1}$, $L_*=(2.51\pm0.17) \times 10^{-3}\,L_\odot$, $A_V=0.21^{+0.11}_{-0.09}$\,mag, and $R_*=0.56\pm0.03\,R_\odot$. Swapping to the PHOENIX model grid from \citet{2013A&A...553A...6H} yielded nearly identical \fbol\ and a higher (but consistent) \teff\ of 3145$\pm$67\,K.

During the fit, the surface-brightness predictions of the atmospheric models were scaled to match the observed photometry. The scale factor is equivalent to $R_*^2/D^2$, which combined with the \gaia\ EDR3 parallax, provided an additional estimate of $R_*$. The technique is similar to the infrared flux method \citep[IRFM; ][]{1977MNRAS.180..177B}. IRFM-based radii of low-mass stars have not consistently agreed with empirical measurements \citep{Casagrande2008, Boyajian2012}, but more recent efforts have been more successful \citep{MorellNaylor2019}, and our resulting $R_*$ estimate ($0.57\pm0.03\,R_\odot$) was consistent with our estimate using the Stefan-Boltzmann relation. 

We also derive \teff\ using our IGRINS spectra. While we consider the SED-based estimate to be reliable, we were concerned about significant spot coverage impacting our derived \teff\ (and hence $R_*$). Spot coverage would manifest as a cooler temperature at redder wavelengths, where the (cooler) spots have a larger impact on the total flux. Hot spots would have the opposite effect, which would also be visible as a wavelength-dependent \teff. The IGRINS data offers another advantage here; \teff\ determined from the low-resolution Goodman spectra was driven by the molecular bands, while the high-resolution $H$- and $K$-band data from IGRINS are less sensitive to missing or erroneous molecular opacities. Thus, while the two fits use the same BT-SETTL models, the fits using IGRINS data were sensitive to different systematics in those models \citep{2018A&A...610A..19R}. 

We fit the highest SNR IGRINS spectra obtained. We fit each order separately, but only included orders with low telluric contamination (determined by eye) and SNR $>$80. For each fit, we explored six free parameters: \teff, $\log~g$, radial velocity, a broadening parameter (including instrumental and rotational effects), the coefficients of the first two Chebyshev polynomials (used to correct for flux calibration offsets), and a factor that describes missing uncertainties in the data as a fraction of the observed spectrum. We fixed [Fe/H] to Solar, consistent with measurements of young regions around Sco-Cen \citep{2006A&A...446..971J}. We fit all free parameters using the Monte Carlo Markov Chain code \texttt{emcee} \citep{Foreman-Mackey2013}. Each order was run with 30 walkers for 50,000 steps after an initial burn-in of 10,000 steps and with uniform priors bounded only by the model grid or physical limits. 

\begin{figure*}[tb]
    \centering
    \includegraphics[width=0.49\textwidth]{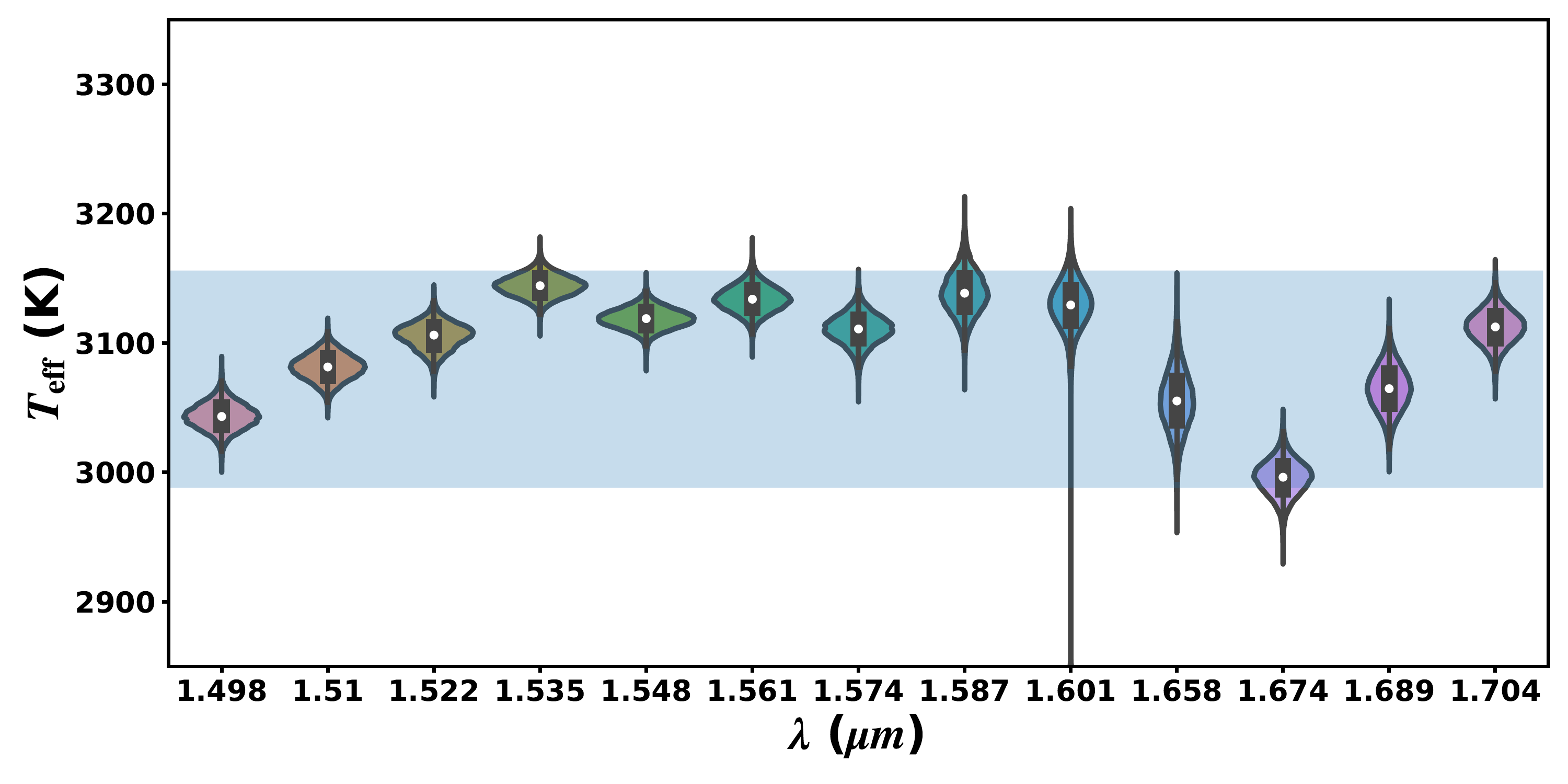}
    \includegraphics[width=0.49\textwidth]{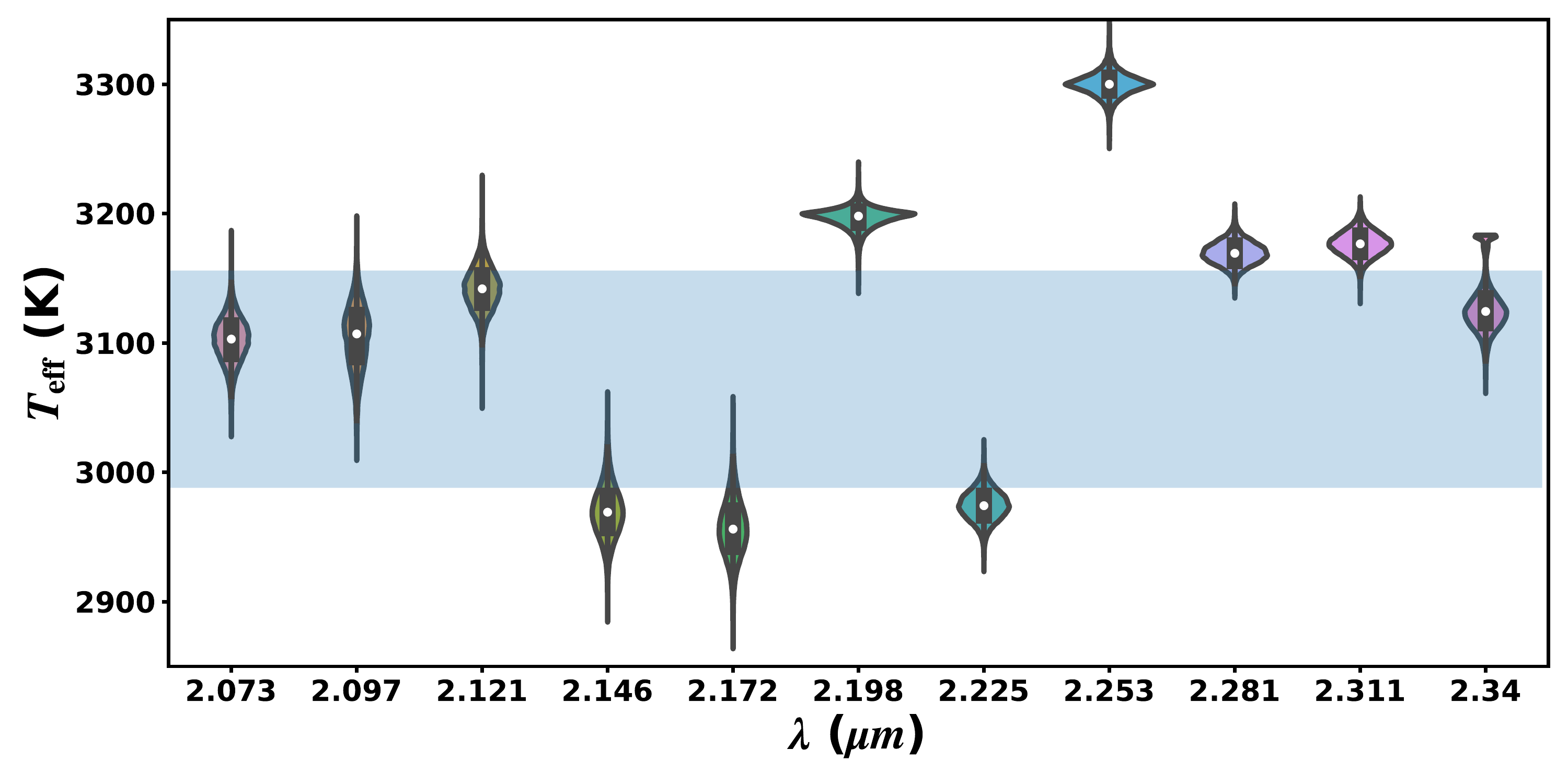}
    \caption{Marginal probability distributions on \teff\ \citep['violin' plot;][]{2020zndo....592845W} for 24 IGRINS orders fit independently over the $H$- (left) and $K$-band (right). Each violin has an equal area distributed according to the MCMC posterior, so narrow/tall violins represent larger uncertainties than wider/shorter ones. The blue bar represents the 1$\sigma$ distribution from our optical spectrum and SED fit. 
    }
    \label{fig:igrins_teff}
\end{figure*}

Parameters besides \teff\ were effectively nuisance parameters in this analysis, as most other parameters (e.g., \vsini) were better determined with more empirical methods. We summarize the \teff\ results in Figure~\ref{fig:igrins_teff}. \teff\ measurements from a given order were often precise (typical errors of 15--50\,K), but varied between orders by 50--100\,K; we considered the latter a more accurate reflection of the true errors, as systematics in the models change with wavelength. The \teff\ for each order was generally consistent with our fit to the SED and optical spectrum, indicating that our derived \teff\ was reliable and spots (while still present) did not significantly impact our derived \teff.

We repeated our fit to the IGRINS data after co-adding all 11 spectra. Because the spectra were taken across the rotational phase, these likely sample different spot coverage patterns. However, the resulting spectrum was significantly higher SNR ($>200$). The results were broadly consistent, with smaller errors in each order (10--30\,K) but a similar variation between orders (50--100\,K). The variation between orders was not consistent with the effect of spots, further suggesting that the spread in inferred \teff\ was driven by systematics in the models, rather than observational noise or surface inhomogeneities on the host. 

To determine $M_*$ and as an additional check on our stellar parameters above, we compared the observed photometry to Solar-metallicity magnetic DSEP evolution models. We compared \gaia{} and 2MASS absolute magnitudes to the model predictions with an MCMC framework using \texttt{emcee}. We fit for age, $A_V$, and $M_*$. An additional parameter ($f$, in magnitudes) captured underestimated uncertainties in the data or models. We used a hybrid interpolation method that found the nearest age in the model grid and then performed linear one-dimensional interpolation in mass to obtain stellar parameters and synthetic model photometry. To account for errors from our nearest-age approach and ensure uniform sampling in age, we pre-interpolated bilinearly in age and mass using the \texttt{isochrones} package \citep{2015ascl.soft03010M}. The interpolated grid was far denser (0.1\,Myr and 0.01$M_\odot$) than expected errors. To redden the model photometry, we used \texttt{synphot} \citep{pey_lian_lim_2020_3971036}, following the extinction law from \citet{1989ApJ...345..245C}. We placed Gaussian priors for age (11$\pm$2\,Myr; derived from the population), $A_V$ (0.21$\pm$0.1\,mag; from the SED fit), and \teff\ (3072$\pm$74\,K; from the SED fit). All other parameters evolved under uniform priors. From the best-fit posteriors, we were able to interpolate additional posteriors on other stellar parameters from the evolutionary model, including $R_*$ and \teff.

The fit yielded $\text{age}=11.5\pm1.0$\,Myr, $M_*=0.165\pm0.010\,M_\odot$, $A_V=0.148\pm0.074$\,mag, $R_*=0.554\pm0.007\,R_{\odot}$, and $T_{\text{eff}}=3050\pm20$\,K. Changing the \teff\ prior to uniform yielded consistent, but less precise parameters (e.g., $M_*=0.170\pm0.015M_\odot$). As an additional test, we repeated the analysis using the PARSEC models, which yielded parameters somewhat discrepant with our SED and population analysis ($\text{age}=14.9\pm2.1$\,Myr, $T_{\text{eff}}=2900\pm30$\,K, and $M_*=0.20\pm0.01\,M_{\odot}$). PARSEC did not reproduce the observed colors of the coolest stars in \population, which manifested as systematically high luminosities for a fixed color past $\simeq$M4 (see Figure~\ref{fig:isochrone}). Thus, we prefer the magnetic DSEP fit. However, we adopted the more conservative mass ($M_*=0.170\pm0.015\,M_\odot$), which was $2\sigma$ consistent with the PARSEC value. 

\subsection{Rotation period, rotational broadening, and stellar inclination}
\citet{Canto2020} reported a rotation period ($P_{\rm{rot}}$) of 1.663$\pm$0.028\,days for \starname\ using \tess\ data. Our analysis of rotation periods in Section~\ref{sec:member} yielded a consistent 1.65$\pm0.04$\,days, which we adopted for our analysis. We computed \vsini\ as part of extracting radial velocities from IGRINS/Gemini and HRS/SALT spectra. The mean value from the IGRINS data was 16.65$\pm$0.24\,\kms, while the SALT data yielded a marginally inconsistent value of 17.8$\pm$0.3\,\kms. We adopted the former value, as the IGRINS data are significantly higher SNR and less impacted by spots or molecular line contamination.

We used the combination of \vsini, $P_{\rm{rot}}$, and $R_*$ to estimate the stellar inclination ($i_*$), and hence test whether the stellar spin and planetary orbit are consistent with alignment. A basic version of this calculation can be done by estimating the $V$ term in \vsini\ using $V=2\pi R_*/P_{\rm{rot}}$, although in practice it requires additional statistical corrections, including the fact that we can only measure alignment projected onto the sky. To this end, we followed the formalism from \citet{2020AJ....159...81M}. The resulting stellar inclination was consistent with alignment with the planet, yielding a limit on inclination of $i_*>73^\circ$ at 95\% confidence, and $i_*>77^\circ$ at 68\% confidence.

\subsection{X-Ray Luminosity}

\starname\ was detected in X-rays in a {\it ROSAT} pointing of the globular cluster NGC 4372 in 1993 \citep{Johnston1996}. The X-ray source was listed as \#6 in the NGC 4372 (IDed in SIMBAD as [JVH96] NGC 4372 6), however, the quoted values were not usable here, as they quote an X-ray flux assuming that the source is at the distance to NGC 4372 and only gave upper limits on the hardness ratios. The X-ray detection was reanalyzed by \citet{Voges2000} in the 2nd {\it ROSAT} PSPC Catalog, which IDs the X-ray source as 2RXP J122703.8-722702, situated 4.9\arcsec\ away from \starname\ with a positional error of $\pm$9\arcsec\, (i.e. consistent with \starname\ being the source of the X-rays).  
They list a soft X-ray count rate of $f_X$ = (1.967\,$\pm$\,0.626) $\times$ 10$^{-3}$ ct\,s$^{-1}$ with hardness ratios HR1 = 0.06\,$\pm$\,0.32 and HR2 = 0.58\,$\pm$\,0.32 with an effective exposure time of 7399\,s observed over UT 4-6 September 1993. Using the energy conversion factor equation of \citet{Fleming1995}, this count rate and HR1 hardness ratio translates to a soft X-ray energy flux of $f_X$ = 1.70\,$\pm$\,10$^{-14}$ erg\,s$^{-1}$\,cm$^{-2}$, which at the distance to \starname\, translates to a soft X-ray luminosity of L$_X$ = 10$^{28.32}$ erg\,s$^{-1}$. 

Given our estimate of the star's bolometric luminosity (Table~\ref{tab:prop}), this translates to a fractional X-ray luminosity of log(L$_X$/L$_{bol}$) = $-$2.66. This is within the range of activity levels stars in the saturated regime display. However, X-ray levels tell us little about \starname's age; M5V stars remain in the saturated well into field ages \citep{West2015, 2017ApJ...834...85N}.

\section{Transit Analysis}\label{sec:transit}
We fit the \tess, SOAR, and LCO photometry simultaneously using the \texttt{MISTTBORN} (MCMC Interface for Synthesis of Transits, Tomography, Binaries, and Others of a Relevant Nature) fitting code\footnote{\url{https://github.com/captain-exoplanet/misttborn}} first described in \citet{Mann2016a} and expanded upon in \citet{MISTTBORN}. \texttt{MISTTBORN} uses \texttt{BATMAN} \citep{Kreidberg2015} to generate model light curves and \texttt{emcee} \citep{Foreman-Mackey2013} to explore the transit parameter space using an affine-invariant Markov chain Monte Carlo (MCMC) algorithm. 

We used \texttt{MISTTBORN} to fit for the five regular transit parameters: time of inferior conjunction ($T_0$), orbital period of the planet ($P$), planet-to-star radius ratio ($R_p/R_\star$), impact parameter ($b$), and stellar density ($\rho_\star$). For each wavelength observed, we included two linear and quadratic limb-darkening coefficients ($q_1$, $q_2$) following the triangular sampling prescription of \citet{Kipping2013}. We included data from four unique bands: \textit{TESS}, $g$', $i$', and $z_s$, requiring eight limb-darkening parameters in total ($r$' was not used for reasons detailed below). Gas drag and gravitational interactions are expected to dampen out eccentricities and inclinations of extremely young planets like \planetname\ \citep{2004ApJ...602..388T}, so we locked the eccentricity to zero (although we check this assumption later).  

To model stellar variations, \texttt{MISTTBORN} included a Gaussian Process (GP) regression module, utilizing the \texttt{celerite} code \citep{celerite}. We initially followed the procedure in \citet{celerite}, using a mixture of two stochastically driven damped simple harmonic oscillators (SHOs) at periods $P_{GP}$ (primary) and $0.5P_{GP}$ (secondary). However, we found the second signal was poorly constrained, suggesting a single SHO was sufficient. Instead, we adopted a fit that included three GP terms: the dominant period ($\ln{P_{GP}}$), amplitude ($\ln{\rm{Amp}}$), and the decay timescale for the variability (quality factor, $\ln{Q}$). 

Stellar variation from spots is wavelength-dependent. This raised a problem when fitting multiple transits over such a wide wavelength range with a single GP amplitude. The most robust solution would be to fit the data using multiple GPs with a common period \cite[e.g., GPFlow;][]{GPflow2017}. However, most of the ground-based transits were partials and many lacked significant out-of-transit baseline, so the GP is poorly constrained from an individual transit. Further, our GP kernel was able to fit nonsimultaneous data of multiple wavelengths, because the rotation signal can evolve during the 27 days between transits. In cases where we had simultaneous data from multiple filters (3 transits), we used only the more precise dataset. This cut excluded all $r$' data but still left us with nine observed transits (6 from the ground). We used the excluded observations for our false-positive analysis (specifically the test of chromaticity; Section~\ref{sec:fpp}).

When handling crowded regions and faint sources like \starname\ (see Figure~\ref{fig:tess}), the SPOC reduction of \tess\ data has been known to over subtract the sky background \citep{2020AJ....160..153B}, leading to a deeper transit depth. This was a potential issue for data collected before Sector 27, after which SPOC applied a correction to their reduction.\footnote{See \href{https://archive.stsci.edu/tess/tess_drn.html}{Data Release Note 38} for more information. Note that Sectors 1-13 have been reprocessed with SPOC R5.0, which corrected the sky background bias issue for these sectors.} An initial fit of our data yielded \tess\ depths somewhat deeper than the ground-based transits, and the Sector 11-12 data deeper than the Sector 38 data. To correct for this, we re-analyzed the target pixel files and estimate that the PCDSAP light curves required an additive correction of 37e$^{-1}$\,pixel$^{-1}$ for Sector 11 and 75e$^{-1}$\,pixel$^{-1}$ for Section 12. No significant bias was seen for the Sector 38 data. We applied this correction to the light curve before running \texttt{MISTTBORN}, and to account for uncertainties we fit for two dilution terms ($C_{S11}$ and $C_{S12}$) applied to the \tess\ Sector 11 and 12 data following \citet{Newton2019}:
\begin{equation}
    LC_{\rm{diluted}} = \frac{LC_{\rm{undiluted}}+C}{1+C},
\end{equation}
where $LC_{\rm{undiluted}}$ was the model light curve generated by \texttt{MISTTBORN}. A negative dilution would correspond to ovsersubtraction of the background (i.e., if the corrections above were underestimated).

We applied Gaussian priors on the limb-darkening coefficients based on the values from the \texttt{LDTK} toolkit \citep{2015MNRAS.453.3821P}, with errors accounting for errors in stellar parameters and the difference between models used (which differ by 0.04-0.08). All other parameters were sampled uniformly with physically motivated boundaries (e.g., $|b|<1+R_P/R_*$, $0<R_P/R_*<1$, and $\rho_*>0$).

We ran the MCMC using 50 walkers for 500,000 steps including a burn-in of 50,000 steps. This run was more than 50 times the autocorrelation time for all parameters, indicating it was more than sufficient for convergence. 

\begin{figure*}[tbp]
    \centering
    \includegraphics[width=0.94\textwidth]{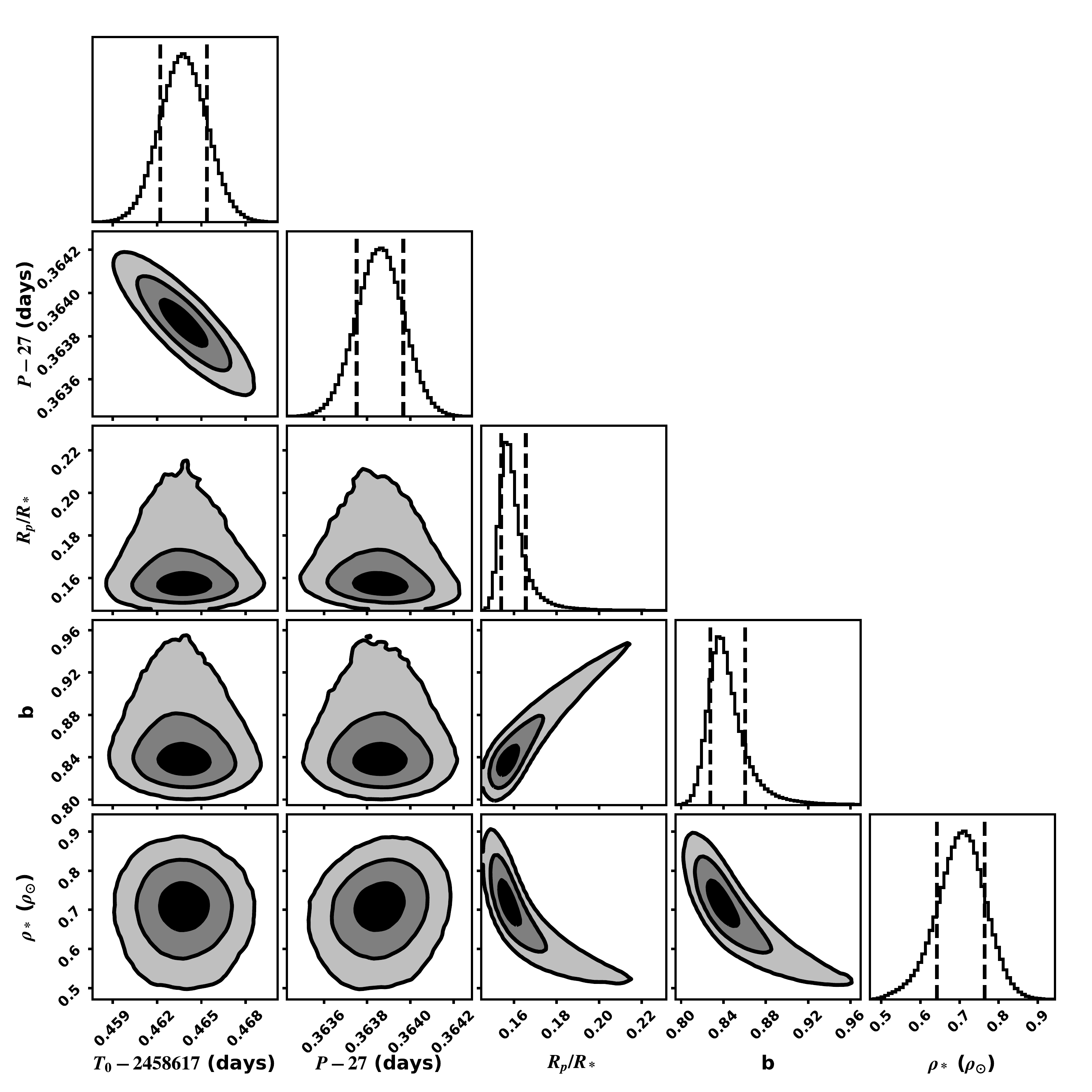}
    \caption{Corner plot of the major transit parameters (GP and limb-darkening parameters excluded for clarity). Most parameters are roughly Gaussian although $R_P/R_*$ and $b$ have a long tail corresponding to a high impact parameter and larger planet. At the extreme ($R_P/R_*=0.22$), this would still correspond to a planetary radius (1.2$R_J$). 
    }
    \label{fig:transit_corner}
\end{figure*}

All output parameters from the \texttt{MISTTBORN} analysis are listed in Table~\ref{tab:transfit}, with a subset of the parameter correlations in Figure~\ref{fig:transit_corner}. Combining the derived $R_P/R_{\star}$ with our estimated stellar parameters from Section~\ref{sec:star} yielded a planet radius of $R_P=0.854^{+0.067}_{-0.052}R_J$.

\begin{deluxetable}{lccc}
\tabletypesize{\scriptsize}
\tablewidth{0pt}
\tablecaption{Global Transit-Fit Parameters. \label{tab:transfit} }
\tablehead{
\colhead{Parameter} & \colhead{with GP} & \colhead{No GP\tablenotemark{a}}}
\startdata
\multicolumn{3}{c}{Transit Fit Parameters} \\
\hline
$T_0$ (BJD-2457000) & $1617.4621 \pm 0.0016$ & $1617.4627 \pm 0.001$ \\ 
$P$ (days) & $27.36397 \pm 0.00011$ &$27.363916 \pm 7.9\times10^{-5}$ \\ 
$R_P/R_{\star}$ & $0.1568^{+0.009}_{-0.0047}$ &  $0.1546^{+0.0035}_{-0.0024}$\\ 
$b$ & $0.849^{+0.024}_{-0.015}$&  $0.836^{+0.014}_{-0.012}$\\ 
$\rho_{\star}$ ($\rho_{\odot}$) & $0.755^{+0.062}_{-0.072}$ & $0.735 \pm 0.049$\\ 
$q_{1,TESS}$ & $0.32 \pm 0.12$&  $0.21\pm0.10$\\ 
$q_{2,TESS}$ & $0.293^{+0.082}_{-0.083}$ & $0.275 \pm 0.083$\\ 
$q_{1,g}$ & $0.74 \pm 0.12$& $0.80\pm0.10$ \\ 
$q_{2,g}$ & $0.384^{+0.057}_{-0.059}$& $0397\pm0.056$ \\ 
$q_{1,i}$ & $0.56 \pm 0.11$& $0.619\pm0.090$ \\ 
$q_{2,i}$ & $0.336^{+0.071}_{-0.073}$& $0.358\pm0.068$ \\ 
$q_{1,z}$ & $0.27 \pm 0.12$ & $0.299^{+0.093}_{-0.088}$\\ 
$q_{2,z}$ & $0.283^{+0.08}_{-0.081}$& $0.287\pm0.080$ \\ 
$C_{S11}$ & $-0.081^{+0.11}_{-0.092}$ & $-0.06\pm0.08$\\ 
$C_{S12}$ & $0.04^{+0.12}_{-0.1}$& $0.03\pm0.09$  \\ 
$\ln(P_{GP})$& $0.532^{+0.051}_{-0.037}$&  \ldots \\ 
$\ln(Amp)$ & $-10.39^{+0.15}_{-0.14}$&    \ldots\\  
$\ln(Q)$ & $1.14^{+0.23}_{-0.20}$ &   \ldots \\ 
\hline 
\multicolumn{3}{c}{Derived Parameters\tablenotemark{b}} \\ 
\hline 
$a/R_{\star}$ & $34.01^{+0.97}_{-1.0}$ &   $34.48^{+0.75}_{-0.79}$\\ 
$i$ ($^{\circ}$) & $88.571^{+0.062}_{-0.093}$ &  $88.611^{+0.047}_{-0.054}$\\ 
$T_{14}$ (days) & $0.2013^{+0.0049}_{-0.0043}$ &   $0.2013^{+0.0029}_{-0.0027}$\\ 
$R_P$ ($R_J$) & $0.854^{+0.067}_{-0.052}$ &  $0.842^{+0.049}_{-0.047}$\\ 
$a$ (AU) & $0.0886^{+0.0054}_{-0.0057}$&  $0.0898 \pm 0.0052$ \\ 
$S$ ($S_{\oplus})$ & $3.21^{+0.38}_{-0.36}$&   $3.12 \pm 0.33$ \\ 
\hline 
\enddata
\tablenotetext{b}{The fit including the GP used the \tess\ 2\,m data, while the second fit used the 20\,s cadence data from Sector 38. Although the output values are consistent, the fit including the GP is preferred.}
\tablenotetext{b}{Derived parameters calculated using stellar parameters in Table~\ref{tab:prop}.}
\end{deluxetable} 

\begin{figure*}[tbh]
    \centering
    \includegraphics[width=0.98\textwidth]{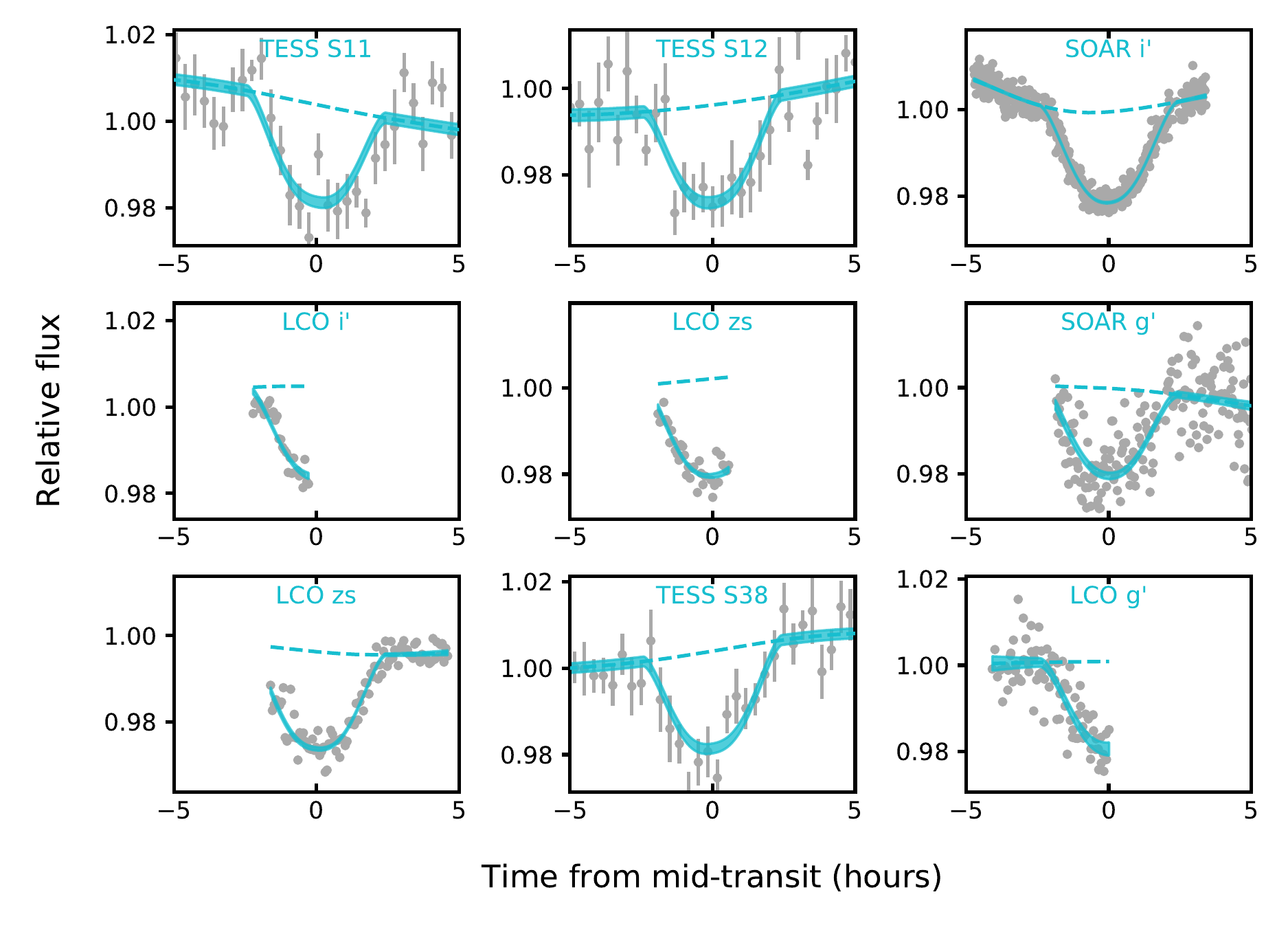}
    \caption{All nine transits from \tess\ (3 transits) and ground-based follow-up (6 transits) that were used in our MCMC global fit, shown in chronological order. The cyan shaded region shows the combined GP and transit model. The dashed line shows the GP fit absent the transit model. The scales of all panels match. For clarity, we have binned the \tess\ data and show the scatter in each bin as error bars; for all other light curves, one point represents one exposure. 
    }
    \label{fig:gp_transit}
\end{figure*} 

\begin{figure*}[tbh]
    \centering
    \includegraphics[trim=30 20 0 0,clip,width=0.64\textwidth]{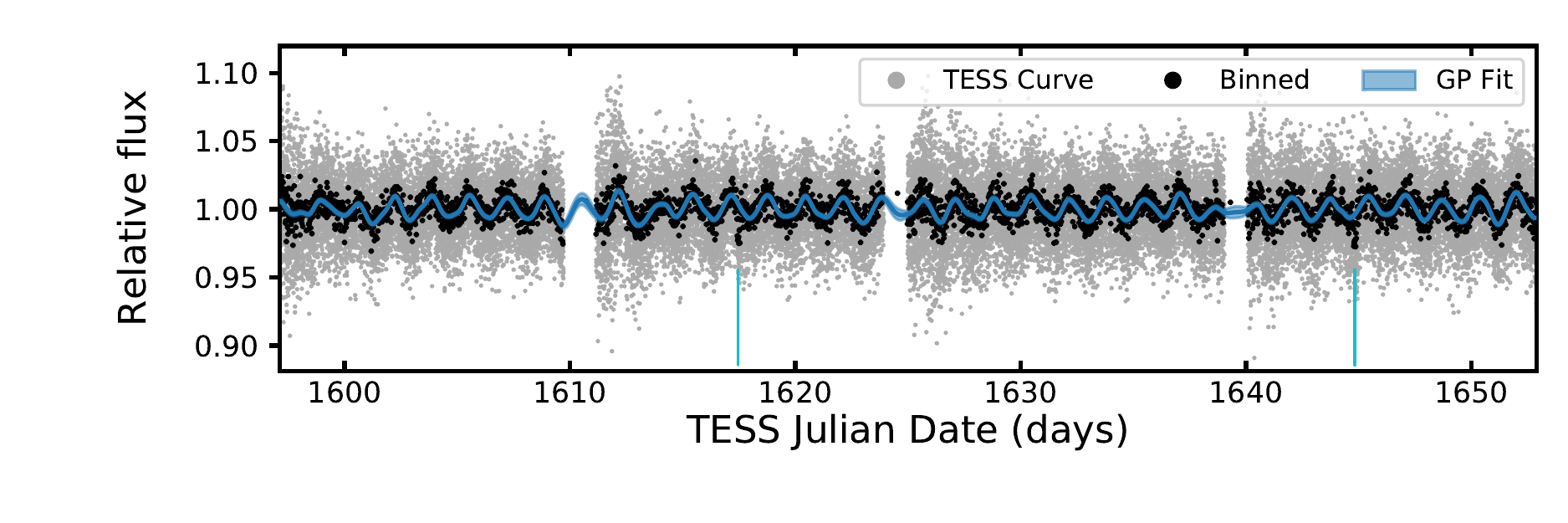}
    \includegraphics[trim=0 20 0 0,clip,width=0.35\textwidth]{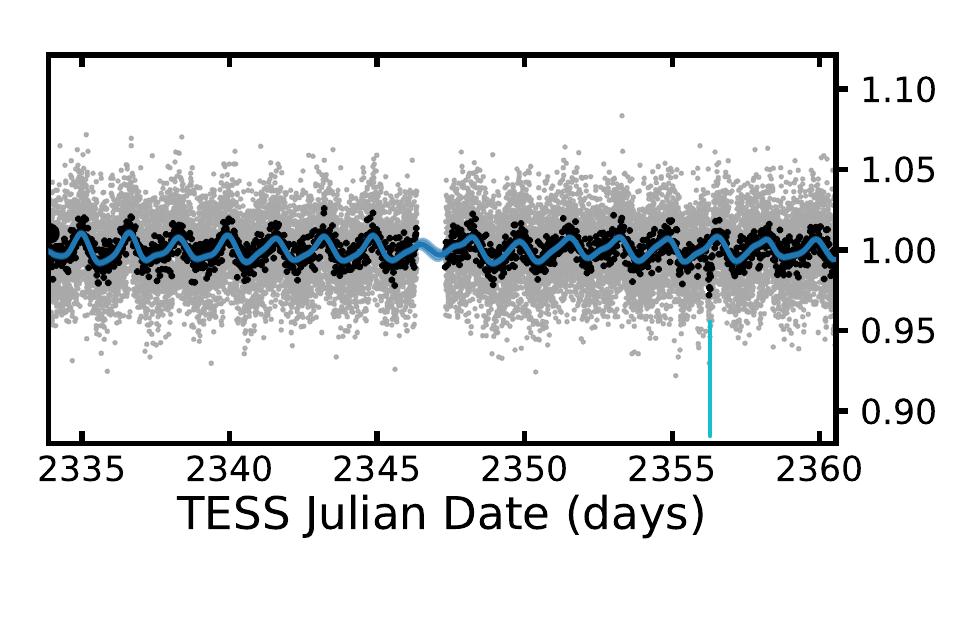}
    \caption{\tess\ light curve from Sectors 11 and 12 (left) and 38 (right). PDCSAP light curve in grey with binned points in black. The GP model fit is shown as a blue shaded region (indicating 1$\sigma$ range); outside of gaps in coverage, errors on the GP fit are usually too small to see. The three cyan vertical lines mark the location of \planetname\ transits. 
    }
    \label{fig:gp_transit_full}
\end{figure*}

The resulting fit from \texttt{MISTTBORN} reproduced the individual transits (Figure~\ref{fig:gp_transit}) as well as the overall light curve variations seen in \tess\ (Figure~\ref{fig:gp_transit_full}). The dilution parameters were both consistent with zero, suggesting our corrections to the PDCSAP curves were reasonable. The transit depth and stellar radius gave a planet radius of $0.859^{+0.065}_{-0.052}\,R_J$ ($9.6\pm0.7\,R_\oplus$). The posterior showed a tail at larger radii, corresponding to a higher impact parameter (Figure~\ref{fig:transit_corner}). At $99.7$\% confidence \planetname\ is $<$1.2\,$R_J$ ($13.4\,R_\oplus$). 

Including the GP was computationally expensive, which led us to use the (binned) 120\,s \tess\ data from Sector 38 (although 20\,s data was available). As discussed above, a single GP fitting chromatic variability over multiple wavelengths and a long time period may be misleading. As a test, we refit the transit without the GP, removing the stellar variability before running the MCMC fit. For the \tess\ data, we fit the shape of the transits and the low-frequency variability simultaneously following  \citet{2020AJ....159..243P}, then divided out the variability model. For the ground-based transits, we fit each transit individually including the GP model as above (such that each transit can have a unique GP amplitude) and divided out the best-fit GP model from the light curve. Because most of the ground-based photometry had a limited out-of-transit baseline, only four of the transits were used.

We ran the GP-free fit through \texttt{MISTTBORN} for 200,000 steps after a burn-in of 20,000 steps. Priors were the same as from our earlier fit. 

\begin{figure}[tbh]
    \centering
    \includegraphics[width=0.48\textwidth]{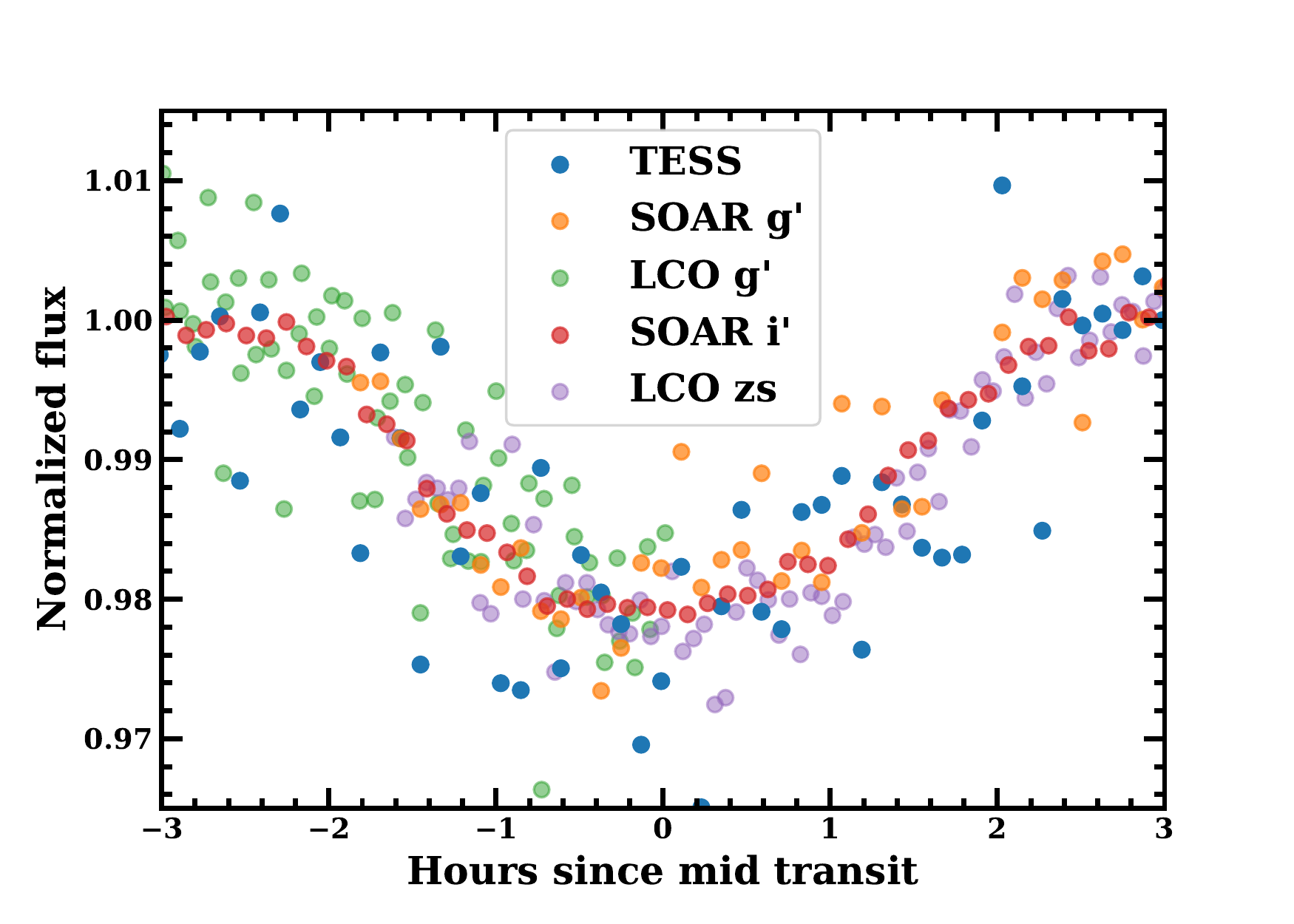}
    \caption{Phase-folded transit data from \tess\ and four ground-based transits after removing stellar variability. The \tess\ and SOAR $i'$ data have been binned for clarity. Dilution on \tess\ data has been removed. The transit shape varies slightly due to differences in limb darkening, but the transit parameters generally agree. 
    }
    \label{fig:nogpTransit}
\end{figure}

The GP-free fit was largely consistent with our GP fit, but with smaller uncertainties.  The resulting fit parameters are listed in Table~\ref{tab:transfit}. We also show the phase-folded transit data in Figure~\ref{fig:nogpTransit}.

We assumed zero eccentricity for both fits, but any nonzero eccentricity might (depending on the argument of periastron) manifest difference between our stellar density derived in Section~\ref{sec:star} and the value from the fit to the transit. The transit-fit density agreed with the spectroscopic value. However, the spectroscopic and isochronal parameters are far less precise and eccentricities inferred this way are degenerate with the argument of periastron. Thus, we can only say that a low eccentricity ($<0.2$) is preferred. 

The transit-fit density is more than a factor of two more precise than the value estimated from our SED fit and stellar models (Section~\ref{sec:star}). If the $e=0$ assumption is valid, the transit-fit density could be used to improve the estimated $R_*$ \citep[e.g., ][]{Seager:2003lr, 2007ApJ...664.1190S}. However, this provided only a marginal improvement on the $R_*$ uncertainty (0.0250$R_\odot$ vs 0.030$R_\odot$) due to a large error on $M_*$. Since the $M_*$ estimate is model-dependent, we opted to keep the stellar parameters from Section~\ref{sec:star}.

% \begin{figure}[tbh]
%     \centering
%     \includegraphics[width=0.45\textwidth]{density.pdf}
%     \caption{Posterior of the stellar density ($\rho_*$) derived from our SED fit and model interpolation (orange, Section~\ref{sec:star}) compared to that from the transit fit (blue). Disagreement between these two posteriors would suggest a non-zero eccentricity and/or errors with our derived stellar parameters. 
%     }
%     \label{fig:density}
% \end{figure} 

\section{False-positive Analysis}\label{sec:fpp}
We considered the three most common false positive (non-planetary) scenarios to explain the transit signal: an eclipsing binary, a background eclipsing binary, and a hierarchical (bound) eclipsing binary. Other scenarios that may apply specifically to young stars (e.g., stellar variability) were quickly dismissed due to the depth, duration, and shape of the transit, as well its consistency over more than a year.

\subsection{Eclipsing Binary}
We compared the radial velocity (RV) data from IGRINS (Table~\ref{tab:RVs}) to the predicted velocity curve of a planet or binary at the orbital period of the outer planet (27.4\,d). We did not use HRS data for this analysis. There may be zero-point offsets between velocities from IGRINS and HRS. HRS velocities also showed significantly higher velocity scatter ($>$500\,\mps), likely due to higher stellar variability in the optical.

\begin{figure*}[htb]
    \centering
    \includegraphics[width=0.96\textwidth]{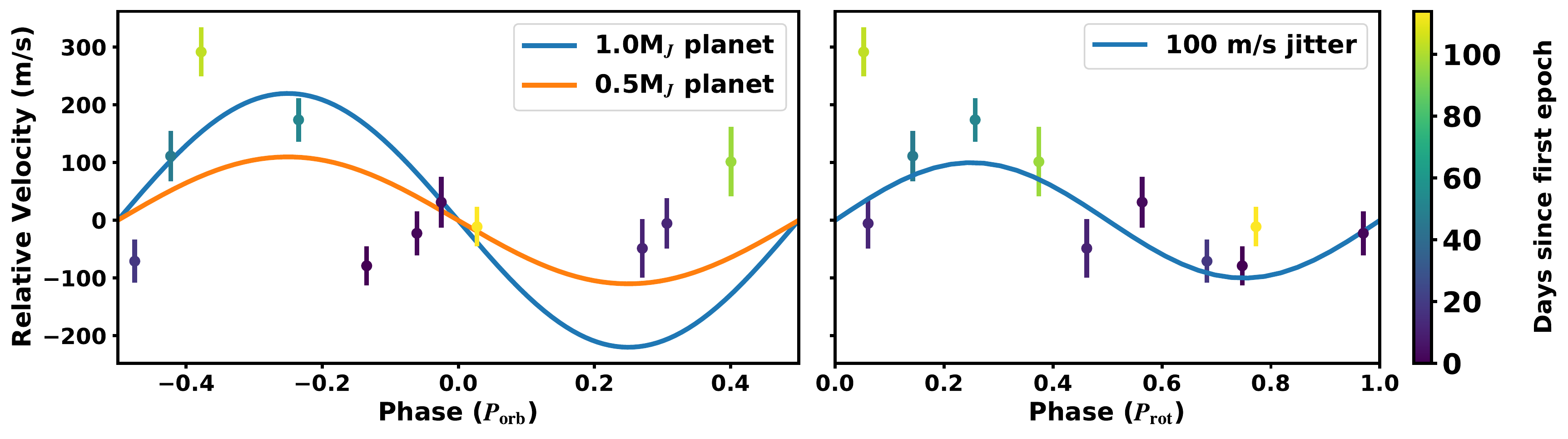}
    \caption{Radial velocities of \starname\ from IGRINS (Section~\ref{sec:IGRINS}) phased to the period of the planet (left) or the rotation period (right). Points are color-coded by time since the first epoch. For reference, we show the expected variations for 0.5$M_J$ ($\simeq110$\mps, orange) and 1$M_J$ planets ($\simeq$220\mps, blue) in then left figure and a simple 100\mps\ jitter (blue) from stellar rotation (right). The velocities are consistent with a $\simeq$100\mps\ jitter from rotation. HRS velocities are less precise and there may be zero-point differences between the two instruments, so they are not shown here.
    }
    \label{fig:rvs}
\end{figure*} 

A Jovian-mass planet would induce an RV signal of $\simeq$220\,\mps, while the scatter in the IGRINS velocities was $\simeq$130\,\mps\ (Figure~\ref{fig:rvs}), ruling out even a highly eccentric brown dwarf. There was a marginal detection of a $\sim$0.5\,$M_J$ planet, particularly if we allow the planet's orbit to be slightly eccentric. However, a modest stellar jitter expected from rotation and long-term variation in the spot pattern could explain all of this variation. Denser sampling and a full radial velocity model including stellar noise would be more likely to yield a detection, but we can still place an upper limit of $M_P<1.7\,M_J$ at $>$95\% confidence. This rules out any possible brown dwarf companion.

\subsection{Background Eclipsing Binary}

Given a maximum eclipse depth of 50\% and an observed transit depth of 2.5\%, the star producing the observed signal must be within $\Delta m<3.26$ mags of \starname. Since the depth is consistent across all filters, this constraint applies to all $griz$ magnitudes. The multi-band speckle data (Section~\ref{sec:AO}) ruled out such a companion or background star down to 90\,mas. \citet{Vanderburg2019} detail a similar metric based on the ratio of the ingress time ($T_{12}$) to the time between first and third contact ($T_{13}$). However, because \planetname\ has a high impact parameter, this metric gives marginally worse constraints than the transit depth alone. 

 The proper motion of \starname\ was large enough to see `behind' the source using digitized images (DSS) from the Palomar Observatory Sky Survey (POSS-I, POSS-II) and the Southern ESO Schmidt (SERC) Survey \citep[patient imaging;][]{Muirhead2012,2018AJ....155...72R}. Between the blue POSS images (1976.3) and the most recent SOAR transit data (2021.2), \starname\ moved more than 1.8\arcsec. This motion was much larger than the resolution of our SOAR transit imaging ($\simeq0.8\arcsec$) and speckle imaging (0.09\arcsec), and somewhat larger than the resolution of the POSS plates (1.7\arcsec). No source was present in the DSS images at the modern location of \starname. After subtracting the source using a nonparametric model of the PSF from \texttt{StarFinder} \citep{2000A&AS..147..335D}, we were able to rule out any source $<$3 magnitudes fainter than \starname. Additionally, both USNO-A2.0 and USNO-B reported no other nearby sources data \citep{Monet:2003fj}, but were sensitive to stars $>$3 magnitudes fainter than \starname\ (in $R$).
 
\subsection{Hierarchical Eclipsing Binary}
 
To check for a bound companion eclipsing binary, we first ran the \texttt{MOLUSC} code \citep{2021arXiv210609040W}. \texttt{MOLUSC} simulated binary companions following an empirically-motivated random distribution in binary parameters. We limited the mass of the synthetic companions to between 10\,$M_J$ to 0.17\,$M_\odot$, as the goal was only to identify any unseen stellar or brown dwarf companion. For each synthetic binary, \texttt{MOLUSC} computed the corresponding velocity curve, brightness, and (sky-projected) separation, as well as enhancement to the noise that would be measured by \gaia. The results were compared directly to our observed high-resolution images, radial velocities, and the \gaia\ astrometry. When running \texttt{MOLUSC}, we included all radial velocity data, high-resolution imaging, \gaia\ astrometry, and limits implied by \starname's CMD position compared to the population (the latter effectively rules out companions $<$0.3\,mag fainter than \starname\ at \gaia\ bandpass and resolution). In total, \texttt{MOLUSC} generated 5 million synthetic companions and determined which could not be ruled out by the data (survivors). 

\begin{figure}[htb]
    \centering
    \includegraphics[width=0.48\textwidth]{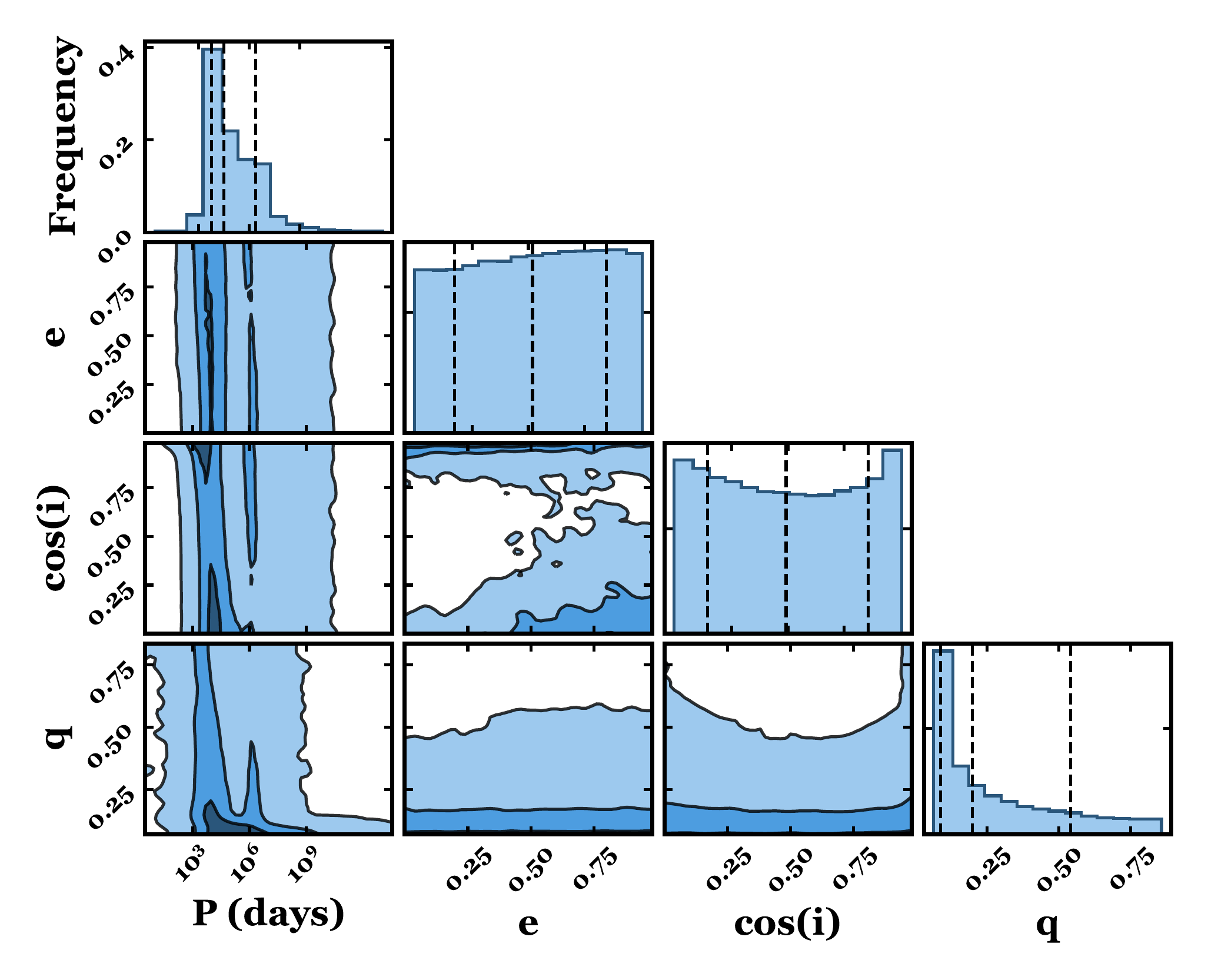}
    \caption{Distribution of surviving companions from \texttt{MOLUSC} in period ($P$), eccentricity ($e$), cosine of the inclination ($\cos(i)$) and mass ratio ($q$). The distribution represents only 9\% of the total generated companions, the rest of which were ruled out by the observational data. 
    }
    \label{fig:molusc}
\end{figure}

\texttt{MOLUSC} found that only 9\% of the generated companions survived (Figure~\ref{fig:molusc}). Most of the survivors were faint or stars on wider orbits that happen to have orbital parameters that put them behind or in front of \starname\ during all observations. Furthermore, when we restricted the survivors to those that could reproduce the observed transit (unresolved in imaging and sufficiently bright for the observed transit), only 1\% of the possible companions remained. 

We also used the transit depths measured independently in each of the observed filters to measure the color of any potential undetected companions. At redder wavelengths, the contrast ratio of the primary and companion approaches unity, so a transit or eclipse from the redder companion would appear deeper in the longer-wavelength data. \citet{Desert2015} and \citet{Tofflemireetal2019} showed how to convert the ratio of the transit depths for any two bands ($\delta_2/\delta_{1}$) into the difference in colors between the target and any potential companion:
\begin{equation}\label{eqn:CST2}
    m_{1,c}-m_{2,c} < m_{1,p} - m_{2,p} + 2.5\rm{log} \left(\frac{\delta_2}{\delta_2}\right),
\end{equation}
where $m_{1,x}-m_{2,x}$ is the color of the primary ($p$) or companion ($c$). 

We excluded \tess\ from this comparison because of the ambiguity introduced by the dilution term (Section~\ref{sec:transit}). We also excluded the LCO $g'$ data observed on 2021 Apr 24 (see Section~\ref{sec:oddtransit} for more details). We split the data into four datasets corresponding to the four observed filters; we combined the ASTEP $R_c$ data with the SDSS $r$' data for simplicity, although our coverage in this is poor and the result did not have a significant impact when compared to the effect of other observations taken at other wavelengths. We then fit the dataset for each filter independent of the others. Our fit followed the same method outlined in Section~\ref{sec:transit}, but we placed priors on the wavelength-independent parameters ($\rho$, $b$, $P$, $T_0$, and $\ln(P_{GP})$) drawn from our global fit. 

\begin{figure}[htb]
    \centering
    \includegraphics[width=0.48\textwidth]{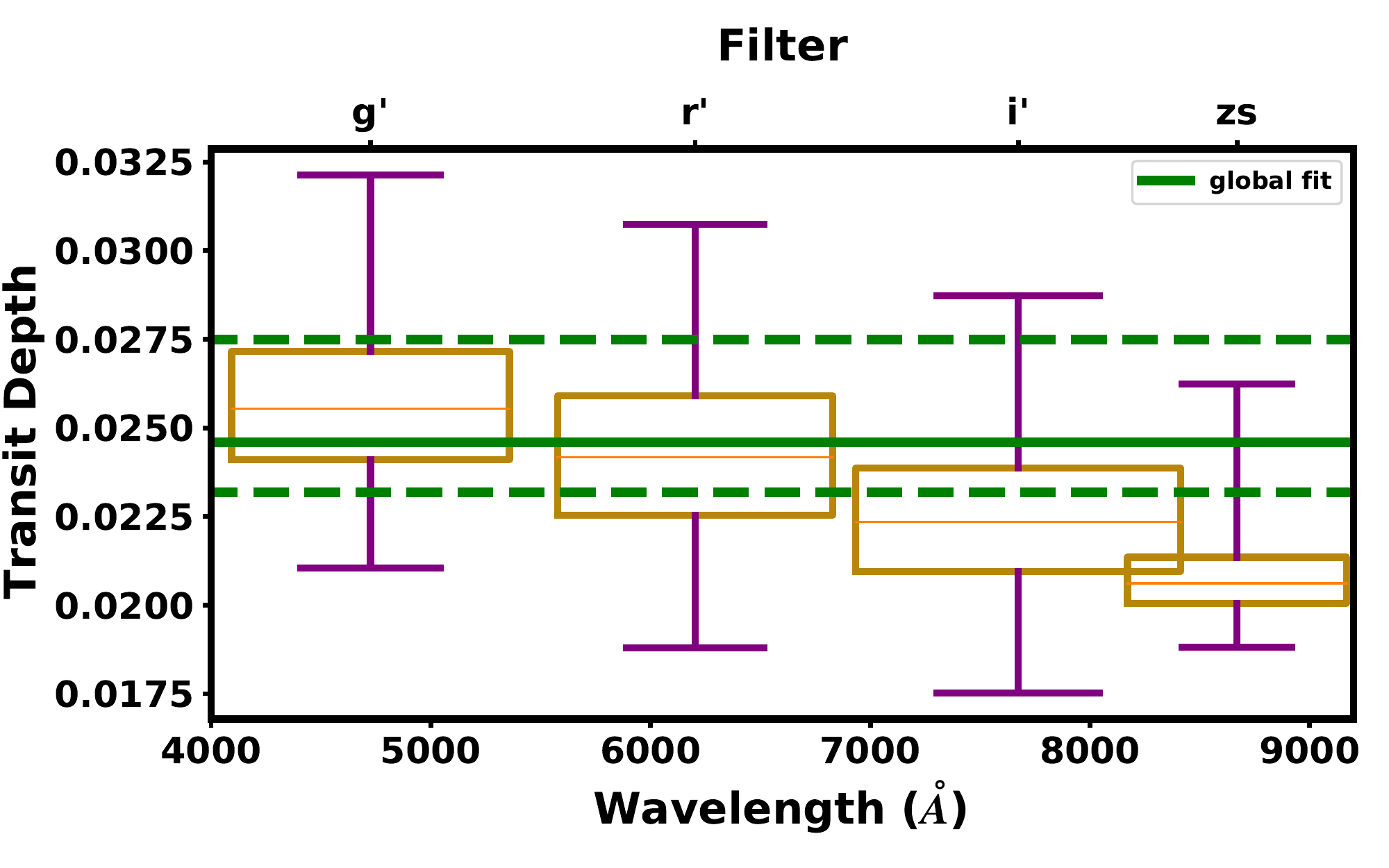}
    \caption{Box and whisker plot of the ground-based transit depths. The orange line indicates the median of each fit, with boxes indicating the 1$\sigma$ range and whiskers the 2$\sigma$ range. The global fit is shown as a green line with dashed lines corresponding to the 1$\sigma$ range. The width of each box roughly corresponds to the filter width. If the transit signal were due to an eclipse around a redder companion (or background star), the transit should be deeper in the red where the flux ratio is closer to unity. The opposite was seen; transits were deeper in $g$ than in $z_s$ (although all transit depths were consistent). 
    }
    \label{fig:transitcolor}
\end{figure}

As we show in Figure~\ref{fig:transitcolor}, all transits yielded consistent depths. The deepest transits were the two bluest: $g'$ and $r'$ (although $r$' depth was very uncertain). A bound eclipsing binary would have yielded a {\it shallower} transit at the bluest wavelengths. Taking the 95\% confidence of the depth ratio, any companion harboring the transit/eclipse signal must be $<$0.06\,mag redder than \starname\ in $g-z$. To satisfy these color criteria, the companion would have to be approximately equal mass to \starname, making the two appear brighter by $>$0.3\,mag in $G$. However, \starname\ sat within $0.2$\,mag of the group CMD (Figure~\ref{fig:isochrone}). More quantitatively, the resulting tight color constraints eliminated {\it all} surviving companions from the \texttt{MOLUSC} analysis, ruling out this scenario and validating the signal as planetary. 

% g= 17.659
% r= 16.346
% i= 14.333
% Tess=13.802
% z= 13.523

\subsection{The unusual April 2021 transit}\label{sec:oddtransit}
\begin{figure}[tbp]
    \centering
    \includegraphics[width=0.45\textwidth]{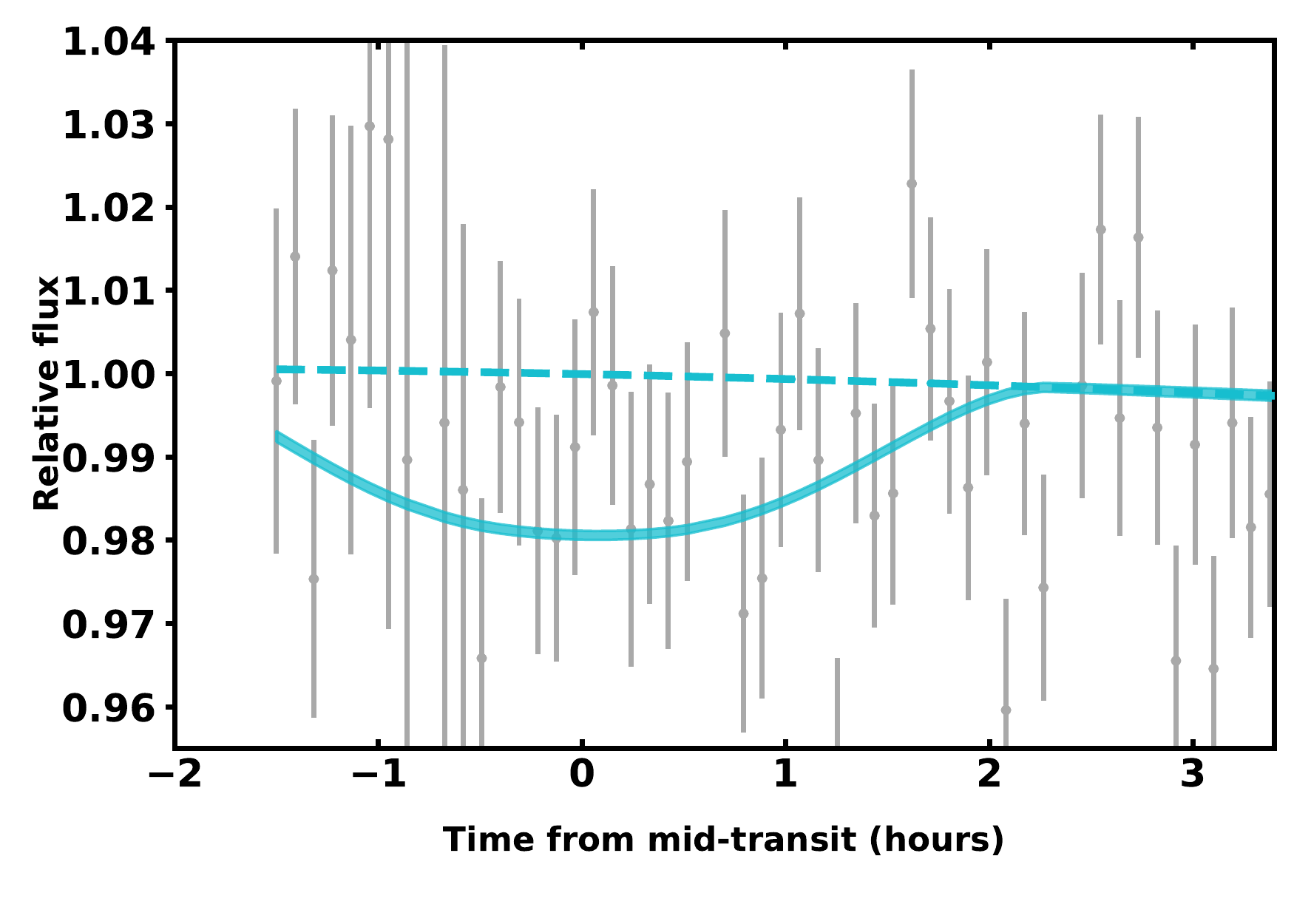}
    \caption{The Apr 2021 $g$' transit data (grey points) was consistent with a non-detection or at least a significantly weaker transit. The model derived from the fit to all $g$'-band data are shown as a teal shaded region encompassing the uncertainties, and the GP without the transit is shown as a dashed line. The large uncertainties meant that this particular transit had a small weight. 
    }
    \label{fig:gtransit}
\end{figure}

We excluded the $g'$ LCO transit from 2021 Apr 24 in our above analysis. The observations missed ingress but covered most of the transit, all of the egress, and about 1.5\,h post-transit. Before color corrections, no clear transit was detected, and even with variability and color corrections the transit was weaker than expected (Figure~\ref{fig:gtransit}). While the scatter in the data was large (1.5\%), the data suggested a maximum transit depth of $\simeq$1\%. That same transit was observed simultaneously in $z_s$, which showed the expected shape; and in $R_c$, which showed an egress as expected. Although the large errors and partial coverage in the in $R_c$ data could not rule out a weaker transit similar to that seen in $g$'. 

Taken alone, the chromatic transit seen in the three data sets from 2021 Apr 24 suggests the transit-like signal is from a background or bound eclipsing binary where the eclipsing system is much redder than the source. However, this conclusion was strongly contradicted by our much more precise $g$', $i$', and $z_s$ transits from multiple other nights, which we used in our false-positive analysis (Section~\ref{sec:fpp}). No false-positive scenario can explain both a non-detection in $g$' on Apr 24, and the clear detections on other nights, particularly since we have clear detections on both even and odd transits (see Table~\ref{tab:obslog}). A more likely explanation is that the 2021 Apr 24 transit was impacted by spots or flares. 

Flares from the $\simeq$24\,Myr M dwarf AU Mic have been known to distort, weaken, and nearly erase transits of the two planets, even at \tess\ wavelengths \citep{2021A&A...649A.177M, 2021arXiv210903924G}. These effects should be much larger in $g'$, which could explain why $z_s$ was not impacted significantly. 

If the planet crosses a heavily spotted region, it will block less light than when crossing the warmer region. For such cool stars, small changes in \teff\ have a large effect on the blue end of the spectrum and a relatively small effect where the SED peaks (in $z_s)$; the result is that modest spots can make a $g$' transit extremely weak without impacting the $z_s$ transit. Such spot crossings can easily happen in one transit and not others, as the rotation period and orbital period are not harmonics of each other and spots on young M dwarfs are known to appear and disappear on month timescales \citep{2021arXiv210613250R}.  

To test this, we fit each of the three transits from 2021 Apr 24 individually, placing priors on the period, $T_0$, $b$, $\rho$, and limb-darkening. As we explain below, the GP kernel can fit out much of the discrepancy (i.e., a flexible GP can get the transits to match), so we did not use a GP model in this test. Instead, we fit a linear trend to the out-of-transit data. This is a conservative assumption, as it will make the errors on the transit depth smaller and hence harder to explain. Fitting for a spot in the transit would be challenging with the data, so we instead fit for the possible range of spots that could explain these depths following the basic methodology of \citet{2018ApJ...853..122R}, but forcing the transit to cross the spotted region rather than a pristine transit cord and a spotted star. There were three free parameters: the spot temperature ($T_{\rm{spot}}$), the fraction of the star that is spotted ($f_S$), and the transit depth had there been no spots ($D$). We restricted $5\%<f_S<80\%$, because smaller fractions would not be sufficient to fill the transit cord and larger would be noticeable as a change in \teff\ with wavelength (see Section~\ref{sec:star} and Figure~\ref{fig:igrins_teff}). The surface temperature was locked to the value from Section~\ref{sec:star}. The measurements were the three transit depths and the global fit depth (which only constrains $D$). We wrapped this in an MCMC framework using \texttt{emcee}, running for 50,000 steps with 50 walkers. 

\begin{figure}[tbp]
    \centering
    \includegraphics[width=0.45\textwidth]{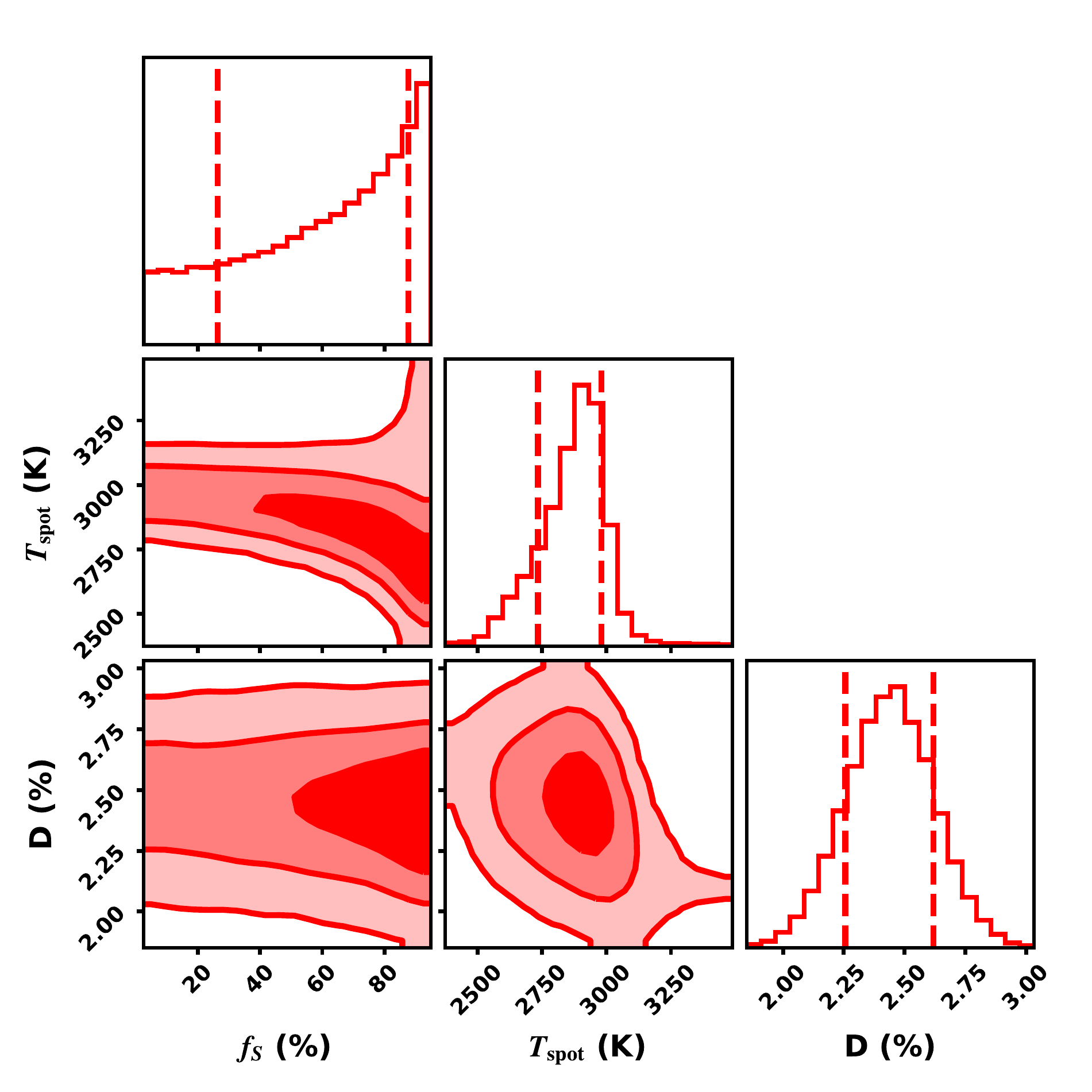}
    \caption{Corner plot of the posteriors from our transit+spot fit to the 2021 Apr 24 transit data for the spot parameters. The shaded regions correspond to 1-3$\sigma$ of the MCMC points and the dashed lines indicate the 1$\sigma$ interval. The constraints are the three transits ($g$, $Rc$, and $z_s$) as well as the overall depth from the global fit. The planet is assumed to be crossing a spotted region, which results in a weaker transit at bluer wavelengths. Although the fit favors a high spot fraction ($f_S>50\%$), this is unlikely given the stellar variability and wavelength-independent temperature. The region of the posterior with modest ($<30\%$) spot fractions and temperatures similar to the surface (3070\,K) can fully explain the anomalous $g$-band transit and are consistent with all other observables. 
    }
    \label{fig:spots}
\end{figure}

We show the resulting posterior in Figure~\ref{fig:spots}. The spots could not be much cooler than 2700\,K unless the spot coverage was $\gtrsim60\%$ because colder spots would change the $z_s$ transit. Spot coverage fractions $>$60\% are rare even in young stars \citep{2020ApJ...893...67M}, although not unheard of \citep[e.g.,][]{Gully-Santiago2017}. Such high fractions are also unlikely based on the stellar variability and spectrum. On the high end, the spot temperatures were limited by the stellar surface temperature. However, modest (5--30\%) spot fractions consistent with expectations for young stars can explain all observations. Since spot temperatures close to the surface temperature are allowed, even moderate ($\gtrsim$30\%) spot factions might not impact the spectroscopic analysis significantly (see Section~\ref{sec:star} and Figure~\ref{fig:igrins_teff}). We conclude that spots could easily explain the anomalous transit. 

As an additional test, we re-ran all $g$' transit data through \texttt{MISTTBORN}. This was effectively a repeat of our analysis from Section~\ref{sec:fpp} including the 2021 Apr 24 data; as in previous fits, a single GP was used to model stellar variability. The goal was to see if including this data could change our interpretation. The resulting transit depth was $2.27^{+0.33}_{-0.26}\%$, consistent with our earlier fits as well as the other $g$-band data (Figure~\ref{fig:transitcolor}). The depth change was insignificant because the SOAR $g$'-band data are far more precise, and significant stellar variability can explain a lot of the apparent discrepancy (which the GP can fit; see Figure~\ref{fig:gtransit}). The updated depth was still sufficient to rule out all possible binaries from the \texttt{MOLUSC} analysis.

\section{Why is TOI1227\,b so big?}\label{sec:model}

\begin{figure}[tbp]
    \centering
    \includegraphics[width=0.45\textwidth]{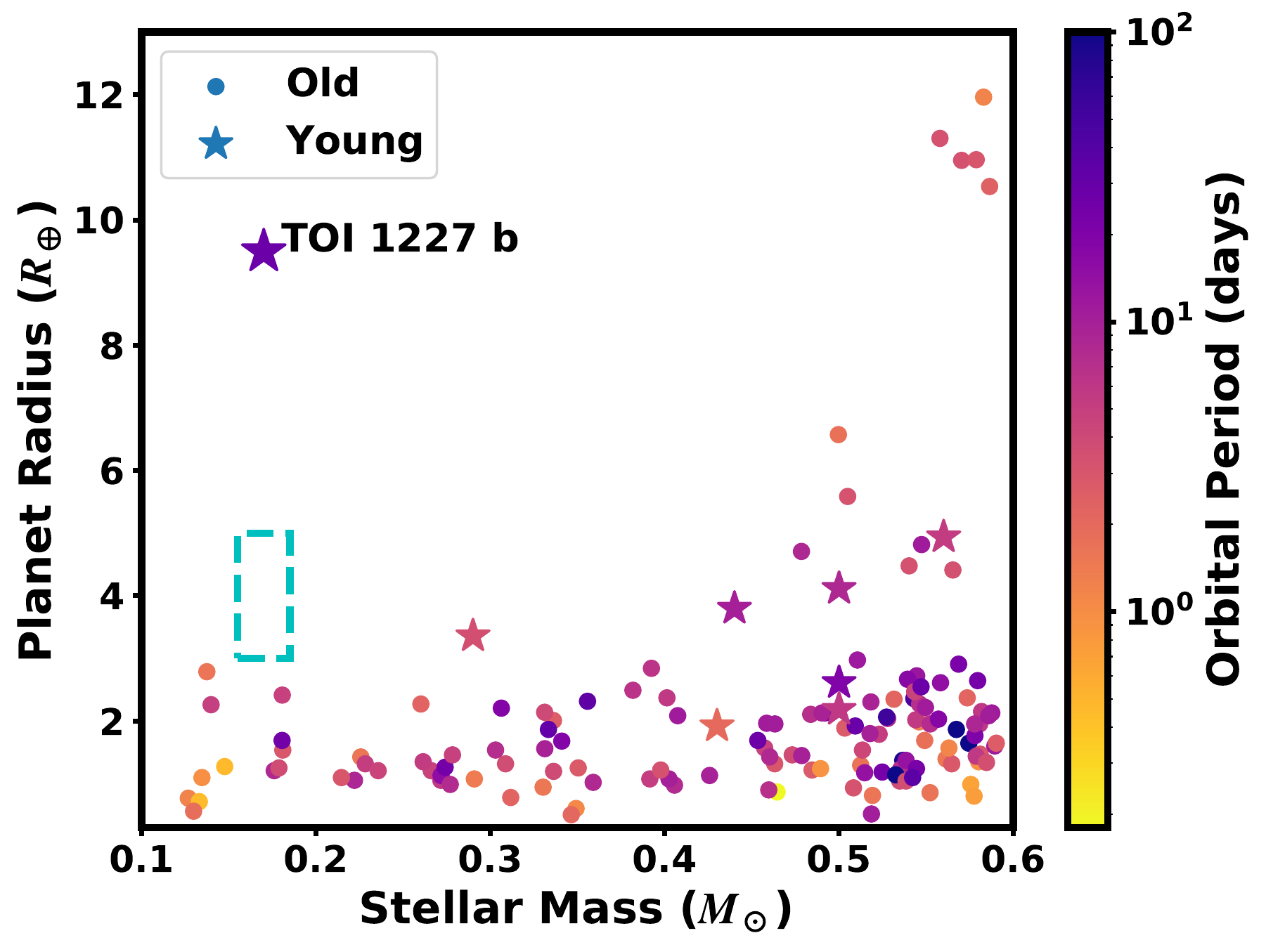}
    \caption{Planet radius as a function of host-star mass for transiting planets from the NASA exoplanet archive \citep{NASAexplanetarchive}. We denote systems around young stars ($<800$\, Myr) with stars, as many planets in Hyades/Praesepe have abnormally large radii \citep{Obermeier2016, Mann2017a}. All points are colored by their orbital period. The cyan region shows the likely final position of \planetname\ based on our MESA evolutionary models.
    }
    \label{fig:comparison}
\end{figure}

No Jovian-sized planets have been discovered around mid-to-late M dwarfs (Figure~\ref{fig:comparison}). Greater than 4$\,R_\oplus$ planets would have been easily detected in the (mid-)M dwarf samples covered by \kepler\ \citep{Dressing2015, 2016MNRAS.457.2877G, 2019AJ....158...75H} \ktwo\ \citep{2017AJ....154..207D, 2020ApJS..247...28H}, and ground-based surveys \citep[e.g., MEarth;][]{Berta2013}, but such programs found only upper limits in this regime. The planet experiences stellar irradiance compared to most known planets, and should not be inflated. Instead, by $>$100\,Myr we expect \planetname\ to more closely resemble the common sub-Neptunes seen around older stars. We explore this further below. 

\subsection{Evolutionary Modeling}\label{sec:evolution}

\begin{figure}[htbp]
    \centering
    \includegraphics[width=\columnwidth]{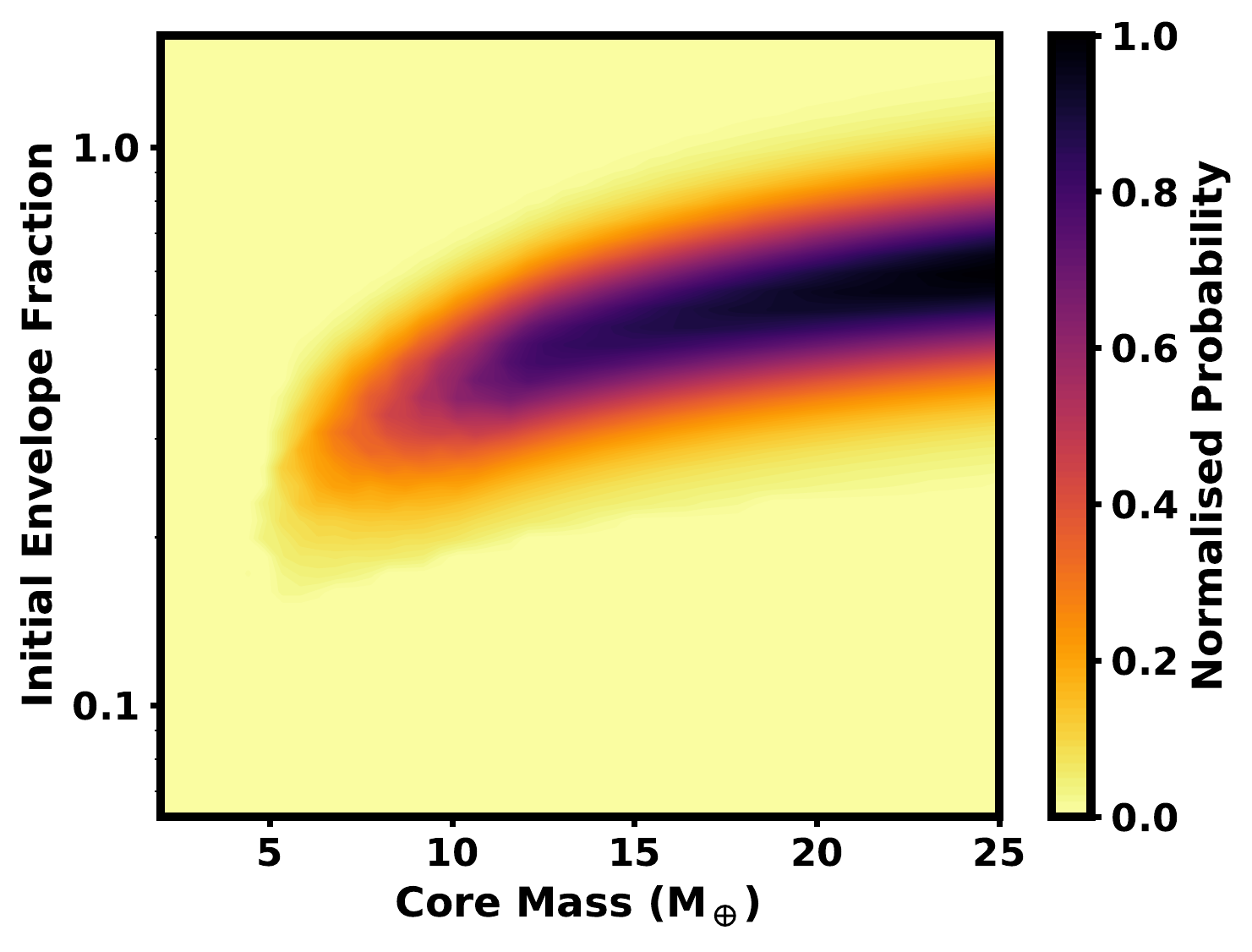}
    \caption{The marginalized probability distribution for \planetname's core mass and initial hydrogen dominated envelope mass fraction that explains its current properties. }
    \label{fig:XvsMcore}
\end{figure}

To determine the possible current internal structure and future evolution of \planetname\ we compared its current properties to evolutionary models. We followed the framework of \citet{2020MNRAS.498.5030O} where we constrained \planetname's core mass, initial hydrogen dominated atmosphere mass fraction and initial (Kelvin-Helmholtz) cooling timescale to match its current age and radius. We considered planets with core masses $<$25\,$M_\oplus$ and restricted ourselves to initial cooling timescales $<$1\,Gyr and applied uniform priors in core-mass, log initial envelope mass fraction and log initial cooling timescale. The evolutionary tracks were computed using {\sc mesa} \citep{Paxton2011, Paxton2013} and included mass-loss due to photoevaporation and irradiation from an evolving host star \citep{2013ApJ...775..105O, OwenMorton2016}. The methodology also accounted for uncertainty in the protoplanetary disk dispersal timescale. Figure~\ref{fig:XvsMcore} shows the resulting marginalized probability distribution for core-mass and initial envelope mass fraction.

With initial envelope mass fractions $<$1, our results confirmed that \planetname\ is likely a young inflated planet that is actively cooling, eventually becoming one of the ubiquitous sub-Neptunes. We note the slight preference for higher core mass is likely artificial because the models with higher core masses are strongly biased towards longer initial cooling times. While we selected an upper limit of 1\,Gyr in our modeling, this was likely an overestimate as the longest initial cooling timescales expected predicted by formation models that include an early ``boil-off'' phase are of order a few 100~Myr. However, our modeling indicated that core masses $\lesssim$5\,$M_\oplus$ are strongly ruled out. 

\begin{figure}[ht]
    \centering
    \includegraphics[width=\columnwidth]{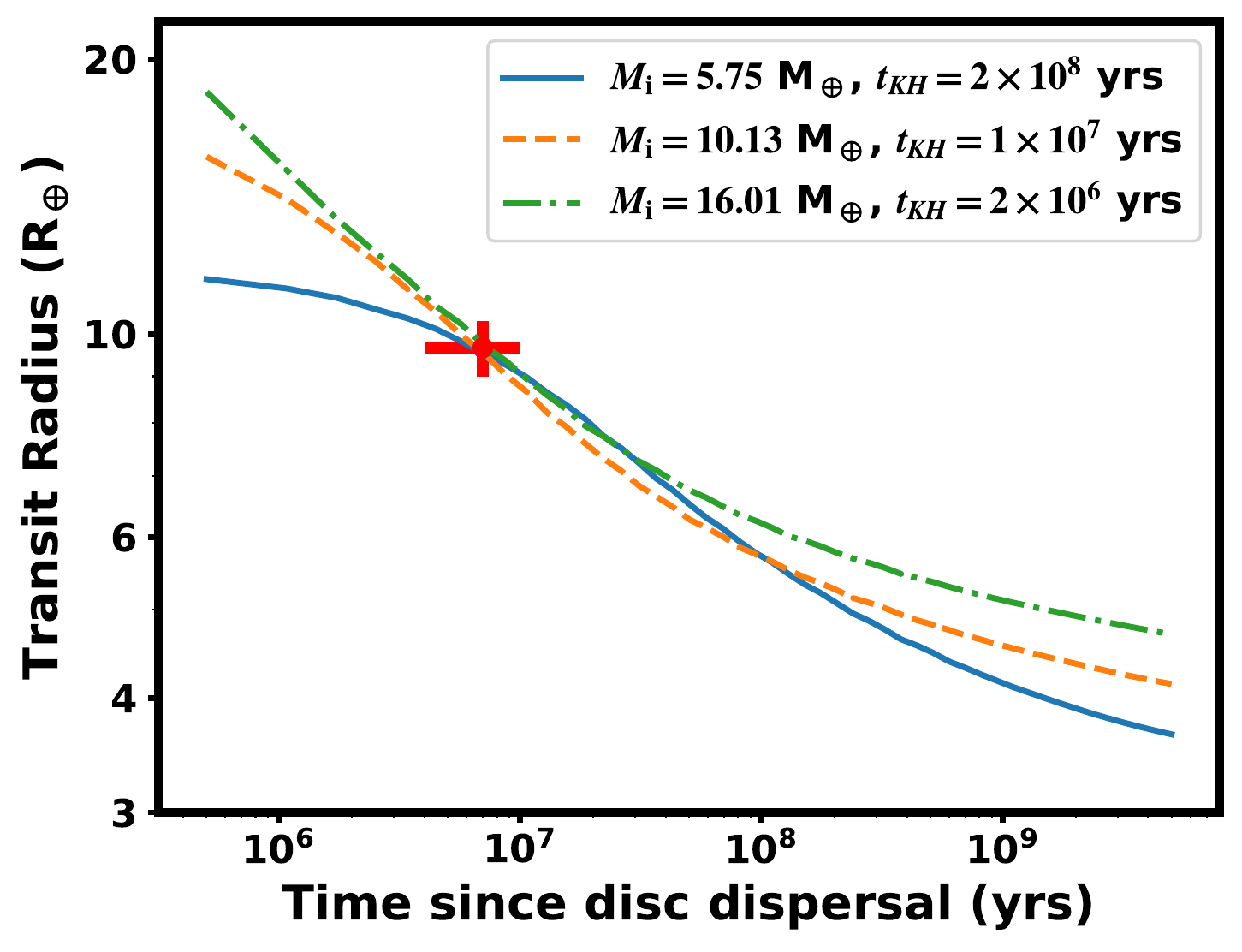}
    \caption{Possible evolutionary tracks for \planetname\ for a range of plausible initial masses ($M_i$) and cooling times ($t_{KH}$). These correspond to high (green) medium (orange) and low (blue) entropy. In choosing these evolutionary curves we have assumed the disk dispersed at a stellar age of 3~Myr. The red cross indicates the age and radius of \planetname, which are used as the observables on the model fit. }
    \label{fig:evol_tracks}
\end{figure}

We followed the future evolutionary pathways of a subset of the model planets by selecting three representative cooling tracks and appropriate initial cooling timescales. These results are presented in Figure~\ref{fig:evol_tracks}, showing that for plausible initial planet masses and initial cooling timescales they evolve into a sub-Neptune with a size 3--5\,$R_\oplus$ at billion-year ages. Such planets are still rare around M dwarfs, but much more consistent with the locus of detections (Figure~\ref{fig:comparison}). 

\section{Summary and Discussion}\label{sec:discussion}

\planetname\ is an $\simeq$11\,Myr giant (0.85\,$R_J$) planet orbiting a low-mass (0.17\,$M_\odot$) M star in the LCC region of the Scorpius-Centarus OB association. Recent work on LCC has identified substructures, i.e. smaller populations, each with slightly different Galactic motion, position, and age. \planetname\ was flagged as a member of the A0 group by \citet{2018ApJ...868...32G} and LCC-B by \citet{2021arXiv210509338K}, which are effectively the same group. Because of the contradictory and easily forgotten naming, we denote this group \population\ after the constellation containing most of the members.

As part of our effort to better measure the age of \planetname, we used rotation, lithium, and stellar evolution models to determine an age of 11$\pm$2\,Myr for \population, and hence \planetname. The young age placed \planetname\ as one of only two systems with transiting planets younger than 15\,Myr \citep[K2-33\,b;][]{Mann2016b, David2016b}, and one of only eleven transiting planets younger than 100\,Myr \citep[V1298\,Tau, DS\,Tuc, AU\,Mic, TOI\,837, HIP\,67522, and TOI\,942][]{2019ApJ...885L..12D, Newton2019, Benatti_dstuc, 2020Natur.582..497P, Rizzuto2020,  Bouma_2020, Zhou_2020, Carleo_2021}. Even including radial velocity detections only adds a few \citep{2016ApJ...826..206J, 2017MNRAS.467.1342Y, Donati:2017aa}, some of which are still considered candidates \citep{2020MNRAS.491.5660D, 2020A&A...642A.133D}.  

Like K2-33\,b, \planetname\ is inflated, with a radius much larger than known transiting planets orbiting similar-mass stars (Figure~\ref{fig:comparison}). \planetname's radius is closer to Jupiter than to the more common 1--3\,$R_\oplus$ planets seen around mid-to-late M dwarfs \citep{Berta2013, 2019AJ....158...75H}. Such a large planet would be obvious in surveys of older M dwarfs, especially given that the equivalent mature M dwarfs are both less variable and a factor of a few smaller than their pre-main-sequence counterparts. Thus, \planetname\ is evidence of significant radius evolution, especially when combined with similar earlier discoveries.

Interestingly, evolutionary models favor a final radius for \planetname\ of $>$3\,$R_\oplus$, which would still be abnormally large (Figure~\ref{fig:comparison}) for the host star's mass. However, that discrepancy is easier to explain as observational bias. Furthermore, the evolutionary tracks are poorly constrained by the existing young planet population, especially given that AU\,Mic\,b is the only $<$50\,Myr planet with both mass and radius measurements \citep{2021MNRAS.502..188K, 2021arXiv210913996C}. The lack of masses remains a significant challenge for these kinds of comparisons \citep{2020MNRAS.498.5030O}. \planetname\ is too faint for radial velocity follow-up in the optical, but might be within the reach of high-precision NIR monitoring. Our IGRINS monitoring achieved 40--50\,\mps\ precision, which is below the level from the stellar jitter. More dedicated monitoring and careful fitting of stellar and planetary signals should be able to detect a planet below $\simeq$40\,$M_\oplus$.

A less challenging follow-up would be spin-orbit alignment through Rossiter-Mclaughlin, where the signal is expected to be $>$200\,\mps\ and the transit timescale ($\simeq$5\,h) is much shorter than the rotational jitter (1.65\,days). Recent measurements of stellar obliquity in young planetary systems have found that they are generally aligned with their host stars \citep[e.g.,][]{2019AJ....158..197R, Zhou_2020, wirth2021toi942b}, suggesting that these planets formed in situ or migrated through their protoplanetary disk. However, the number of misaligned systems even around older stars is $\lesssim$10\% \citep[varying with spectral type and planet mass;][]{2009ApJ...696.1230F, MortonWinn2014, Campante_2016}. We likely need a larger sample of obliquity measurements in young systems for a significant statistical comparison. 

Follow-up and characterization of \planetname\ highlight the difficulties of characterizing and validating such young planets. We initially mischaracterized both the host star and planet, and ground-based transits yielded some contradictory results. The host star (\starname) was initially thought to be part of the $\simeq$5\,Myr \epscha\ cluster, based on the \texttt{BANYAN} code. At that age, the stellar host would be even larger, yielding an even larger radius for the planet and creating a more complex set of false-positive scenarios (e.g., the planet radius would be more consistent with a young brown dwarf). Without \gaia, it is unlikely that \population\ would have been recognized, and assigning \starname\ to a given population without a parallax would have been extremely challenging. 

Similarly, the THYME team originally flagged \planetname\ as a likely eclipsing binary based on its V-shaped transit and large depth. Only with the SALT/HRS spectra and a consistent transit depth from Goodman/SOAR did we pursue further follow-up. This motivated both spectroscopic monitoring with IGRINS and the suite of multi-band transit photometry observed over $\simeq$1.5 years to better characterize the planet. While the majority of the transit photometry painted a clear picture that the signal was from a planet, a single transit showed an inconsistent transit depth, suggesting a false positive. We argued in Section~\ref{sec:oddtransit} that regular flaring, the planet occulting spots, or simply low-precision photometry, could easily explain the unusual transit. Spots could also explain why our other $g$' transits were deeper than the $i'$ and $z_s$ data (Figure~\ref{fig:transitcolor}); if the transit cord has fewer spots than the rest of the stellar surface, the bluest transits appear deeper \citep[this was seen in K2-25;][]{Thao2020}. 

While the case for \planetname\ as a planet is strong, the initially confusing results raise concerns about the future follow-up of young planets. Radial velocity detection remains challenging. As with K2-25\,b, the discovery transit for \planetname\ was V-shaped, making it hard to classify on transit depth and shape alone. Transits from young planets are likely to show some chromaticity even if the planet is real given the presence of spots, and smaller planets are not as amenable to the large suite of ground-based follow-up used here. Earlier studies often made use of {\it Spitzer}, which provided the advantage of a wide wavelength range while operating in a regime where the effects of spots and flares are significantly smaller. With the end of {\it Spitzer}, we may need to focus on improving the NIR photometric precision from the ground. For now, we encourage caution when rejecting young planets based on metrics tuned for older systems. 

\acknowledgments
The authors thank the anonymous referee for providing insightful comments and suggestions on the paper. AWM would like to thank Bandit, who sat directly on AWM's keyboard whenever \texttt{MISTTBORN} was running, preventing him from working on this manuscript too much. The THYME collaboration also wants to thank Halee, Wally, Maizie, Dudley, Charlie, Marley, and Edmund for their thoughtful discussions and emotional support. 
AWM was supported through NASA’s Astrophysics Data Analysis Program (80NSSC19K0583), a grant from the Heising-Simons Foundation (2019-1490), and the \tess\ GI program (80NSSC21K1054). MLW was supported by a grant through NASA's \ktwo\ GO program (80NSSC19K0097). MJF was supported by NASA's Exoplanet Research Program (XRP; 80NSSC21K0393). This material is based upon work supported by the National Science Foundation Graduate Research Fellowship Program under Grant No. DGE-1650116 to PCT. MJF and RPM were supported by the NC Space Grant Graduate Research program. MGB and SPS were supported by the NC Space Grant Undergraduate Research program. MGB was also supported by funding from the Chancellor’s Science Scholars Program at the University of North Carolina at Chapel Hill. JEO is supported by a Royal Society University Research Fellowship and this project has received funding from the European Research Council (ERC) under the European Union’s Horizon 2020 research and innovation programme (Grant agreement No. 853022, PEVAP). ACR was supported as a 51 Pegasi b Fellow through the Heising-Simons Foundation. ERN acknowledges support from the \tess\ GI program (program G03141). 

Based on observations collected at the European Organisation for Astronomical Research in the Southern Hemisphere under ESO programmes 0101.A-9012(A), 0101.C-0527(A), 0101.C-0902(A), 081.C-0779(A), 082.C-0390(A), 094.C-0805(A), 098.C-0739(A), and 60.A-9022(C). This work used the Immersion Grating Infrared Spectrometer (IGRINS) that was developed under a collaboration between the University of Texas at Austin and the Korea Astronomy and Space Science Institute (KASI) with the financial support of the Mt. Cuba Astronomical Foundation, of the US National Science Foundation under grants AST-1229522 and AST-1702267, of the McDonald Observatory of the University of Texas at Austin, of the Korean GMT Project of KASI, and Gemini Observatory. This paper includes data collected by the TESS mission, which are publicly available from the Mikulski Archive for Space Telescopes (MAST). Funding for the TESS mission is provided by NASA’s Science Mission Directorate. This research has made use of the Exoplanet Follow-up Observation Program website, which is operated by the California Institute of Technology, under contract with the National Aeronautics and Space Administration under the Exoplanet Exploration Program. This work has made use of data from the European Space Agency (ESA) mission \emph{Gaia}\footnote{\url{https://www.cosmos.esa.int/gaia}}, processed by the \emph{Gaia} Data Processing and Analysis Consortium (DPAC)\footnote{\url{https://www.cosmos.esa.int/web/gaia/dpac/consortium}}. Funding for the DPAC has been provided by national institutions, in particular the institutions participating in the \emph{Gaia} Multilateral Agreement.  This work makes use of observations from the LCOGT network. Part of the LCOGT telescope time was granted by NOIRLab through the Mid-Scale Innovations Program (MSIP). MSIP is funded by NSF. This research received funding from the European Research Council (ERC) under the European Union's Horizon 2020 research and innovation programme (grant agreement n$^\circ$ 803193/BEBOP), and from the Science and Technology Facilities Council (STFC; grant n$^\circ$ ST/S00193X/1). This research has made use of the VizieR catalogue access tool, CDS, Strasbourg, France. The original description of the VizieR service was published in A\&AS 143, 23. Resources supporting this work were provided by the NASA High-End Computing (HEC) Program through the NASA Advanced Supercomputing (NAS) Division at Ames Research Center for the production of the SPOC data products. We acknowledge the use of public TOI Release data from pipelines at the TESS Science Office and at the TESS Science Processing Operations Center. We acknowledge the use of public TESS data from pipelines at the TESS Science Office and at the TESS Science Processing Operations Center. Part of this research was carried out at the Jet Propulsion Laboratory, California Institute of Technology, under a contract with the National Aeronautics and Space Administration (NASA). Part of this work was performed using resources provided by the Cambridge Service for Data Driven Discovery (CSD3) operated by the University of Cambridge Research Computing Service (\url{www.csd3.cam.ac.uk}), provided by Dell EMC and Intel using Tier-2 funding from the Engineering and Physical Sciences Research Council (capital grant EP/P020259/1), and DiRAC funding from the Science and Technology Facilities Council (\url{www.dirac.ac.uk}). Based on observations obtained at the international Gemini Observatory, a program of NSF’s NOIRLab, which is managed by the Association of Universities for Research in Astronomy (AURA) under a cooperative agreement with the National Science Foundation. on behalf of the Gemini Observatory partnership: the National Science Foundation (United States), National Research Council (Canada), Agencia Nacional de Investigaci\'{o}n y Desarrollo (Chile), Ministerio de Ciencia, Tecnolog\'{i}a e Innovaci\'{o}n (Argentina), Minist\'{e}rio da Ci\^{e}ncia, Tecnologia, Inova\c{c}\~{o}es e Comunica\c{c}\~{o}es (Brazil), and Korea Astronomy and Space Science Institute (Republic of Korea).

\facilities{Exoplanet Archive, LCOGT 1m (Sinistro), SOAR (Goodman), SALT (HRS), Gemini (IGRINS), \tess, ASTEP}

\software{ \texttt{MISTTBORN} \citep{MISTTBORN}, \texttt{IGRINS RV} \citep{2021AJ....161..283S}, \textit{emcee} \citep{Foreman-Mackey2013}, \textit{batman} \citep{Kreidberg2015}, matplotlib \citep{hunter2007matplotlib}, \texttt{corner.py} \citep{foreman2016corner}, \texttt{AstroImageJ} \citep{Collins17}, BANZAI \citep{McCully18}, \texttt{TAPIR} \citep{Jensen:2013}, \texttt{saphires} \citep{Tofflemireetal2019}, \texttt{TelFit} \citep{Gullikson2014}, BANYAN-$\Sigma$ \citep{BanyanSigma}, \texttt{unpopular} \citep{2021arXiv210615063H}, \texttt{synphot}
\citep{pey_lian_lim_2020_3971036} 
\texttt{MOLUSC} \citep{2021arXiv210609040W}
\texttt{FriendFinder} \citep{THYMEV}
}

\startlongtable
\begin{deluxetable*}{ccccccccccccccccccccc}
\centering
\tabletypesize{\scriptsize}
\tablewidth{0pt}
\tablecaption{Friends of \starname}\label{tab:population}
\tablehead{
\colhead{TIC} & \colhead{$\alpha$} & \colhead{$\delta$} & 
\colhead{RV}& \colhead{$\sigma_{\rm{RV}}$}& \colhead{RV}&  \colhead{EqW Li}& \colhead{Li} & \colhead{$P_{\rm{rot}}$} & \colhead{Qual}  & \colhead{CMG} & \colhead{SpyGlass} & \colhead{FF} \\
\colhead{} & \colhead{(J2016.0)} & \colhead{(J2016.0)} & 
\colhead{(\kms)} &   \colhead{(\kms)}&   \colhead{Source} & \colhead{mA} & \colhead{Source} & \colhead{(days)} & \colhead{}
}
\startdata
360156606 & 186.76734 & -72.45185 & 13.30 & 0.30 & IGRINS & 513 & Goodman & 1.65 & Q0 & Y & Y & Y \\
360003096 & 186.40895 & -72.82357 & $\ldots$ & $\ldots$ &  & 598 & Goodman & 0.58 & Q0 & Y & Y & Y \\
360260204 & 187.87099 & -72.83042 & $\ldots$ & $\ldots$ &  & $\ldots$ &  & 10.22 & Q3 & Y & Y & Y \\
359697676 & 184.44271 & -72.37392 & $\ldots$ & $\ldots$ &  & 382 & Goodman & 2.09 & Q0 & Y & Y & Y \\
360259773 & 187.57311 & -72.98529 & $\ldots$ & $\ldots$ &  & $\ldots$ &  & 6.53 & Q1 & Y & Y & Y \\
360331566 & 188.12874 & -72.91861 & $\ldots$ & $\ldots$ &  & 479 & Goodman & 5.52 & Q0 & Y & Y & Y \\
360625930 & 189.85218 & -72.73502 & $\ldots$ & $\ldots$ &  & $\ldots$ &  & 6.60 & Q0 & Y & Y & Y \\
360259534 & 187.70690 & -73.07898 & $\ldots$ & $\ldots$ &  & 563 & Goodman & 8.64 & Q2 & Y & Y & Y \\
326657935 & 188.45791 & -71.42464 & $\ldots$ & $\ldots$ &  & $\ldots$ &  & 1.61 & Q0 & N & Y & Y \\
359997243 & 186.21572 & -72.60394 & 7.90 & 1.20 & Kharchenko2007 & 23 & ESO & 2.99 & Q1 & Y & Y & Y \\
360906004 & 190.90566 & -71.99760 & $\ldots$ & $\ldots$ &  & $\ldots$ &  & 1.82 & Q0 & Y & Y & Y \\
360212499 & 187.32060 & -73.91157 & $\ldots$ & $\ldots$ &  & $\ldots$ &  & 0.47 & Q3 & N & Y & Y \\
359766629 & 185.15013 & -73.88416 & $\ldots$ & $\ldots$ &  & 428 & Goodman & 2.91 & Q0 & Y & Y & Y \\
447994659 & 185.27027 & -71.28040 & 14.30 & 0.45 & Schneider2019 & 517 & ESO & 6.81 & Q0 & Y & Y & Y \\
360626787 & 189.65123 & -73.02830 & $\ldots$ & $\ldots$ &  & $\ldots$ &  & 4.92 & Q2 & N & N & Y \\
359851737 & 185.87389 & -73.17399 & $\ldots$ & $\ldots$ &  & $\ldots$ &  & 11.03 & Q2 & N & N & Y \\
359766954 & 184.93144 & -74.06594 & 14.97 & 0.84 & Schneider2019 & 558 & ESO & 6.75 & Q0 & Y & Y & Y \\
360627462 & 189.62496 & -73.26625 & $\ldots$ & $\ldots$ &  & 439 & Goodman & 1.97 & Q0 & Y & Y & Y \\
359288962 & 182.61217 & -72.11260 & $\ldots$ & $\ldots$ &  & 222 & Goodman & 2.38 & Q0 & Y & Y & Y \\
314231280 & 184.37701 & -71.00238 & $\ldots$ & $\ldots$ &  & $\ldots$ &  & 3.22 & Q0 & Y & Y & Y \\
359767456 & 184.97269 & -74.33597 & $\ldots$ & $\ldots$ &  & 458 & Goodman & 10.22 & Q0 & Y & Y & Y \\
313853780 & 184.01031 & -71.05102 & $\ldots$ & $\ldots$ &  & $\ldots$ &  & 18.23 & Q2 & Y & Y & Y \\
454975214 & 179.87431 & -72.64049 & $\ldots$ & $\ldots$ &  & $\ldots$ &  & 0.47 & Q0 & N & Y & Y \\
360771943 & 190.48124 & -73.18481 & $\ldots$ & $\ldots$ &  & $\ldots$ &  & 1.26 & Q0 & N & Y & Y \\
328480578 & 190.73212 & -70.57249 & $\ldots$ & $\ldots$ &  & $\ldots$ &  & 1.01 & Q0 & N & Y & Y \\
313754471 & 184.10687 & -71.39463 & $\ldots$ & $\ldots$ &  & $\ldots$ &  & 1.51 & Q0 & Y & Y & Y \\
329538372 & 191.89997 & -70.52056 & $\ldots$ & $\ldots$ &  & $\ldots$ &  & 4.16 & Q0 & N & Y & Y \\
958526152 & 185.50216 & -70.01776 & $\ldots$ & $\ldots$ &  & $\ldots$ &  & 1.50 & Q0 & N & Y & Y \\
334999132 & 194.30115 & -71.32633 & $\ldots$ & $\ldots$ &  & $\ldots$ &  & 1.68 & Q2 & Y & Y & Y \\
360261300 & 187.60514 & -72.43166 & $\ldots$ & $\ldots$ &  & $\ldots$ &  & 1.03 & Q0 & Y & Y & Y \\
327147189 & 189.14024 & -69.91816 & $\ldots$ & $\ldots$ &  & $\ldots$ &  & 8.20 & Q0 & N & Y & Y \\
359357695 & 182.72617 & -75.13193 & $\ldots$ & $\ldots$ &  & 515 & Goodman & 3.43 & Q0 & Y & Y & Y \\
360631514 & 189.83799 & -75.04427 & 12.85 & 2.63 & RAVEDR5 & 413 & ESO & 3.99 & Q0 & Y & Y & Y \\
328477573 & 190.55432 & -69.73018 & $\ldots$ & $\ldots$ &  & $\ldots$ &  & 0.72 & Q0 & N & Y & Y \\
327656671 & 189.69631 & -71.66432 & $\ldots$ & $\ldots$ &  & $\ldots$ &  & 6.29 & Q0 & Y & Y & Y \\
312803013 & 182.90841 & -71.17671 & 13.50 & 0.10 & Desidera2015 & 272 & ESO & 5.24 & Q0 & N & Y & Y \\
326542774 & 188.20581 & -69.66814 & $\ldots$ & $\ldots$ &  & $\ldots$ &  & 3.42 & Q0 & N & Y & Y \\
454980196 & 180.01050 & -74.73522 & $\ldots$ & $\ldots$ &  & $\ldots$ &  & 1.64 & Q0 & N & Y & Y \\
410981161 & 178.04938 & -70.69887 & $\ldots$ & $\ldots$ &  & $\ldots$ &  & 1.40 & Q0 & Y & Y & Y \\
361112047 & 191.97708 & -75.42557 & $\ldots$ & $\ldots$ &  & $\ldots$ &  & 0.33 & Q0 & Y & Y & Y \\
312803129 & 182.98363 & -71.13739 & $\ldots$ & $\ldots$ &  & $\ldots$ &  & 4.03 & Q0 & Y & Y & Y \\
359892714 & 185.70196 & -74.17238 & $\ldots$ & $\ldots$ &  & $\ldots$ &  & 0.47 & Q0 & Y & Y & Y \\
327665414 & 189.89608 & -69.13804 & $\ldots$ & $\ldots$ &  & 445 & Goodman & 3.61 & Q0 & N & Y & Y \\
361392250 & 193.58588 & -72.66762 & $\ldots$ & $\ldots$ &  & $\ldots$ &  & 1.82 & Q0 & Y & Y & Y \\
405040121 & 185.59257 & -71.61785 & $\ldots$ & $\ldots$ &  & $\ldots$ &  & 0.83 & Q0 & N & Y & Y \\
327147732 & 189.28324 & -69.77014 & $\ldots$ & $\ldots$ &  & $\ldots$ &  & 1.42 & Q3 & N & N & Y \\
359573863 & 183.64167 & -74.77805 & $\ldots$ & $\ldots$ &  & $\ldots$ &  & 1.65 & Q0 & Y & Y & Y \\
360086059 & 186.49203 & -75.85329 & $\ldots$ & $\ldots$ &  & $\ldots$ &  & 1.28 & Q0 & N & Y & Y \\
454757744 & 175.87292 & -74.31052 & $\ldots$ & $\ldots$ &  & $\ldots$ &  & 1.92 & Q0 & Y & Y & Y \\
327549481 & 189.39036 & -69.00110 & $\ldots$ & $\ldots$ &  & $\ldots$ &  & 0.94 & Q0 & N & N & Y \\
328234948 & 190.21623 & -68.94963 & 34.05 & 19.38 & \gaia\ DR2 & $\ldots$ &  & 3.00 & Q0 & N & Y & Y \\
326538701 & 188.35305 & -68.81545 & $\ldots$ & $\ldots$ &  & $\ldots$ &  & 1.66 & Q0 & N & Y & Y \\
311731608 & 181.28789 & -70.07076 & $\ldots$ & $\ldots$ &  & $\ldots$ &  & 9.57 & Q0 & N & Y & Y \\
334930665 & 194.03409 & -69.44842 & 13.80 & 0.30 & Torres2006 & $\ldots$ &  & 4.59 & Q0 & N & N & Y \\
361113174 & 191.98801 & -76.10258 & $\ldots$ & $\ldots$ &  & $\ldots$ &  & 0.58 & Q0 & N & Y & Y \\
455000299 & 179.92537 & -76.02397 & 12.63 & 1.40 & \gaia\ DR2 & 460 & ESO & 7.90 & Q0 & N & Y & Y \\
297549461 & 180.65657 & -69.19232 & 7.60 & 3.70 & Kharchenko2007 & 25 & ESO & 0.30 & Q1 & Y & Y & Y \\
359000352 & 180.59159 & -73.05367 & $\ldots$ & $\ldots$ &  & 358 & Goodman & 2.04 & Q0 & Y & Y & Y \\
454636797 & 173.52469 & -72.87915 & $\ldots$ & $\ldots$ &  & $\ldots$ &  & 4.38 & Q1 & N & N & Y \\
359065340 & 181.02908 & -72.26877 & $\ldots$ & $\ldots$ &  & $\ldots$ &  & 0.78 & Q0 & Y & Y & Y \\
454851844 & 177.68682 & -74.18711 & 15.00 & 1.20 & Murphy2013 & 404 & ESO & 1.06 & Q0 & N & N & Y \\
360899642 & 190.93459 & -74.20175 & $\ldots$ & $\ldots$ &  & $\ldots$ &  & 1.05 & Q0 & N & N & Y \\
410986249 & 178.32330 & -71.14346 & $\ldots$ & $\ldots$ &  & $\ldots$ &  & 1.75 & Q0 & Y & Y & Y \\
313865023 & 184.11613 & -68.96561 & $\ldots$ & $\ldots$ &  & $\ldots$ &  & 1.68 & Q3 & N & N & Y \\
359139846 & 181.37133 & -76.01453 & $\ldots$ & $\ldots$ &  & $\ldots$ &  & 0.42 & Q3 & Y & Y & Y \\
335376063 & 194.60605 & -70.48039 & 9.60 & 0.90 & Desidera2015 & $\ldots$ &  & 2.00 & Q0 & N & N & Y \\
454541357 & 171.77052 & -72.32937 & $\ldots$ & $\ldots$ &  & $\ldots$ &  & 1.19 & Q0 & Y & Y & Y \\
327667179 & 189.67291 & -68.76646 & $\ldots$ & $\ldots$ &  & $\ldots$ &  & 0.79 & Q0 & Y & N & Y \\
361289085 & 193.06188 & -76.47441 & $\ldots$ & $\ldots$ &  & $\ldots$ &  & 1.87 & Q0 & N & Y & Y \\
334660676 & 193.77151 & -68.75523 & $\ldots$ & $\ldots$ &  & $\ldots$ &  & 1.70 & Q0 & N & N & Y \\
359571841 & 183.85674 & -76.05843 & $\ldots$ & $\ldots$ &  & $\ldots$ &  & 0.80 & Q0 & N & N & Y \\
959134031 & 191.49579 & -68.14649 & $\ldots$ & $\ldots$ &  & $\ldots$ &  & 4.75 & Q2 & N & N & Y \\
454718471 & 175.20625 & -74.99425 & 10.30 & 1.00 & Murphy2013 & $\ldots$ &  & 0.50 & Q0 & Y & Y & Y \\
313174402 & 183.28939 & -69.19817 & $\ldots$ & $\ldots$ &  & $\ldots$ &  & 0.66 & Q2 & N & Y & Y \\
311374236 & 181.00944 & -69.61571 & $\ldots$ & $\ldots$ &  & $\ldots$ &  & 2.27 & Q2 & N & N & Y \\
327039106 & 188.81350 & -70.71889 & $\ldots$ & $\ldots$ &  & $\ldots$ &  & 24.56 & Q2 & N & N & Y \\
329454577 & 191.84103 & -68.14453 & 3.03 & 2.49 & RAVEDR5 & $\ldots$ &  & 3.66 & Q0 & N & N & Y \\
394908401 & 203.37566 & -73.69372 & $\ldots$ & $\ldots$ &  & $\ldots$ &  & 0.67 & Q0 & Y & Y & Y \\
335367096 & 194.54401 & -68.41302 & $\ldots$ & $\ldots$ &  & $\ldots$ &  & 0.73 & Q0 & N & N & Y \\
359144755 & 181.46321 & -73.17935 & $\ldots$ & $\ldots$ &  & $\ldots$ &  & 1.29 & Q0 & Y & Y & Y \\
281757097 & 186.29977 & -67.36249 & $\ldots$ & $\ldots$ &  & $\ldots$ &  & 1.05 & Q0 & N & N & Y \\
312197475 & 181.90714 & -69.20425 & $\ldots$ & $\ldots$ &  & $\ldots$ &  & 0.66 & Q1 & N & Y & Y \\
311974232 & 181.72372 & -67.55967 & $\ldots$ & $\ldots$ &  & $\ldots$ &  & 1.70 & Q0 & N & N & Y \\
312204205 & 181.90170 & -70.92725 & $\ldots$ & $\ldots$ &  & $\ldots$ &  & 0.39 & Q1 & Y & N & Y \\
314759973 & 184.94741 & -68.39638 & $\ldots$ & $\ldots$ &  & $\ldots$ &  & 2.81 & Q0 & N & N & Y \\
327667195 & 189.67779 & -68.76370 & 7.20 & 1.20 & Kharchenko2007 & 67 & ESO & 0.79 & Q2 & Y & N & Y \\
360154672 & 187.07901 & -73.10969 & 14.55 & 1.15 & \gaia\ DR2 & $\ldots$ &  & 6.16 & Q0 & N & N & Y \\
405089321 & 185.49499 & -67.91525 & $\ldots$ & $\ldots$ &  & $\ldots$ &  & 4.69 & Q0 & N & N & Y \\
454611617 & 173.34613 & -76.36925 & $\ldots$ & $\ldots$ &  & $\ldots$ &  & 1.57 & Q0 & N & N & Y \\
358994080 & 180.72684 & -77.31060 & 14.40 & 0.60 & Malo2014 & 256 & ESO & 2.50 & Q0 & N & N & Y \\
326660816 & 188.69208 & -70.32378 & $\ldots$ & $\ldots$ &  & $\ldots$ &  & 1.51 & Q0 & N & N & Y \\
340221485 & 201.14607 & -70.34068 & $\ldots$ & $\ldots$ &  & $\ldots$ &  & 1.64 & Q1 & Y & N & Y \\
297460759 & 180.12239 & -70.84458 & $\ldots$ & $\ldots$ &  & $\ldots$ &  & 0.63 & Q0 & N & N & Y \\
313648837 & 183.72736 & -68.26724 & $\ldots$ & $\ldots$ &  & $\ldots$ &  & 1.31 & Q2 & N & N & Y \\
454335224 & 167.71655 & -72.92027 & $\ldots$ & $\ldots$ &  & $\ldots$ &  & 1.04 & Q0 & N & N & Y \\
907780433 & 178.85370 & -69.07816 & $\ldots$ & $\ldots$ &  & $\ldots$ &  & 1.05 & Q1 & Y & N & N \\
313857039 & 184.16726 & -70.12676 & 13.87 & 0.54 & \gaia\ DR2 & $\ldots$ &  & 4.11 & Q0 & Y & N & N \\
454826793 & 177.17699 & -73.62319 & $\ldots$ & $\ldots$ &  & $\ldots$ &  & 1.16 & Q0 & Y & N & N \\
328706525 & 190.98583 & -68.21051 & $\ldots$ & $\ldots$ &  & $\ldots$ &  & 1.41 & Q0 & Y & N & N \\
341448790 & 203.04825 & -71.79406 & $\ldots$ & $\ldots$ &  & $\ldots$ &  & 5.30 & Q0 & Y & N & N \\
329457630 & 191.73068 & -68.72506 & $\ldots$ & $\ldots$ &  & $\ldots$ &  & 2.58 & Q0 & N & Y & N \\
328235910 & 190.46474 & -68.74517 & $\ldots$ & $\ldots$ &  & $\ldots$ &  & 1.42 & Q0 & N & Y & N \\
329162367 & 191.64321 & -68.07945 & 11.27 & 0.79 & \gaia\ DR2 & $\ldots$ &  & 14.12 & Q2 & N & Y & N \\
328985850 & 191.19354 & -68.21358 & $\ldots$ & $\ldots$ &  & $\ldots$ &  & 0.91 & Q1 & N & Y & N \\
329162173 & 191.42380 & -68.11292 & 12.04 & 0.94 & \gaia\ DR2 & $\ldots$ &  & 5.30 & Q0 & N & Y & N \\
359137193 & 181.14973 & -77.52634 & 10.40 & 2.00 & LopezMarti2013 & 420 & ESO & 4.87 & Q0 & N & Y & N \\
359484011 & 183.57172 & -73.35947 & $\ldots$ & $\ldots$ &  & $\ldots$ &  & 1.68 & Q0 & N & Y & N \\
410983415 & 178.01551 & -71.54682 & $\ldots$ & $\ldots$ &  & $\ldots$ &  & 1.35 & Q2 & N & Y & N
\enddata
\tablecomments{Reference code: Torres2006 = \citet{Torres2006}, Kharchenko2007 = \citet{2007AN....328..889K}, Murphy2013 = \citet{2013MNRAS.435.1325M}, LopezMarti2013 = \citet{LopezMarti2013}, Malo2014 = \citet{Malo2014a}, RAVEDR5 = \citet{Kunder2017}, \gaia\ DR2 = DR2 = \citet{DR2_velocities}, Schneider2019 = \citet{Schneider2019}, Desidera2015 = \citet{Desidera2015}, IGRINS = this work. 
}
\end{deluxetable*}

\clearpage
\bibliography{fullbiblio}{}
\bibliographystyle{aasjournal}

\appendix
\section{Isochronal ages using mixture models}\label{sec:mixture}

Ages derived from direct comparison to theoretical evolutionary models are impacted by a wide range of systematics. Binaries, for example, make the association look younger; an effect that varies with stellar mass \citep{2021ApJ...912..137S}. The effect can be mitigated by explicitly including binaries in their model likelihood assuming an initial mass function and binary fraction \citep{2021arXiv210509338K}. This is less effective if the sample is biased towards or against binaries, which could easily arise from \gaia-based astrometry until later releases that account for binary motion. For young associations like those considered here, field interlopers (non-members) also bias the result towards older ages. Specific to populations like Musca is the presence of interlopers from other populations in LCC, which could bias the age in a direction that depends on the nearby populations.

Our solution is to fit the data with two populations simultaneously (a mixture model). The first population is the (ideally) single-age and single-star population of interest. The second is the `outlier' population, which may contain nonmembers (field or nearby populations), binaries, stars with inaccurate photometry or parallaxes, or simply regions of the CMD that are poorly captured by the model grid. Following \citet{HoggRecipes}, the likelihood would be a sum of the two populations weighted by an amplitude:
\begin{eqnarray}\label{eq:mixture}
\like &\propto& \prod_{i=1}^N\left[(1-\Pbad) P(\mathrm{interest}) + \Pbad\ P(\mathrm{outlier}) \right] 
\nonumber \\
\like &\propto&
 \prod_{i=1}^N \left[\frac{1-\Pbad}{\sqrt{2\,\pi\,\sigma_{M_{x,i}}^2}}
 \,\exp\left(-\frac{[M_{x,i}-f(age,E(B-V),col_{i})]^2}{2\,\sigma_{M_{x,i}}^2}\right)\right.
 \nonumber \\ & & \quad
 \left.+ \frac{\Pbad}{\sqrt{2\,\pi\,[\Vbad+\sigma_{M_{x,i}}^2]}}
 \,\exp\left(-\frac{[M_{x,i}-\Ybad]^2}{2\,[\Vbad+\sigma_{M_{x,i}}^2]}\right)\right],
\end{eqnarray}
where $M_{x,i}$ is the absolute magnitude of the $i$th star in band $x$, $f(age,E(B-V),col_{i})$ is the predicted absolute magnitude from the model isochrones for a given age, reddening ($E(B-V)$), and color ($col_{i}$), and $\sigma_{M_{x,i}}$ is the total error in the star's absolute magnitude and the propagated error from the star's color. The parameters $\Pbad$, $\Vbad$, and $\Ybad$ are free parameters that describe the second population: the amplitude of the outliers (or the fraction of the population that are outliers), the mean absolute magnitude of the outliers, and the variance in the absolute magnitude of the outliers, respectively.

The simple likelihood above works well when most outliers are field M dwarfs, which are well described by a single Gaussian distribution in absolute magnitude versus color. However, we have both field and young interlopers that may occupy a large range of the CMD. Fortunately, such stars are still (mostly) restricted to the physically allowed regions of the CMD. Thus, we change the second term in the likelihood from a simple distribution in absolute magnitude to an offset from the population of interest:
\begin{eqnarray}\label{eq:mixture2}\displaystyle
P(\mathrm{outlier}) = \frac{1}{\sqrt{2\,\pi\,[\Vbad+\sigma_{M_{x,i}}^2]}}
 \,\exp\left(-\frac{[M_{x,i}-f(age,E(B-V),col_{i})-\Ybad]^2}{2\,[\Vbad+\sigma_{M_{x,i}}^2]}\right).
\end{eqnarray}
A downside of using an offset from the CMD to model the outliers is that the three outlier parameters can be biased by targets in regions of parameter space that are not well described by a simple offset, such as sources with poor parallaxes or M-dwarf white-dwarf binaries. This can be fixed by applying initial quality cuts. A more complicated problem is that the offset from the expected CMD likely varies with luminosity/mass, while we assumed a single parameter.

\begin{figure*}[tbh]
    \centering
    \includegraphics[width=0.49\textwidth]{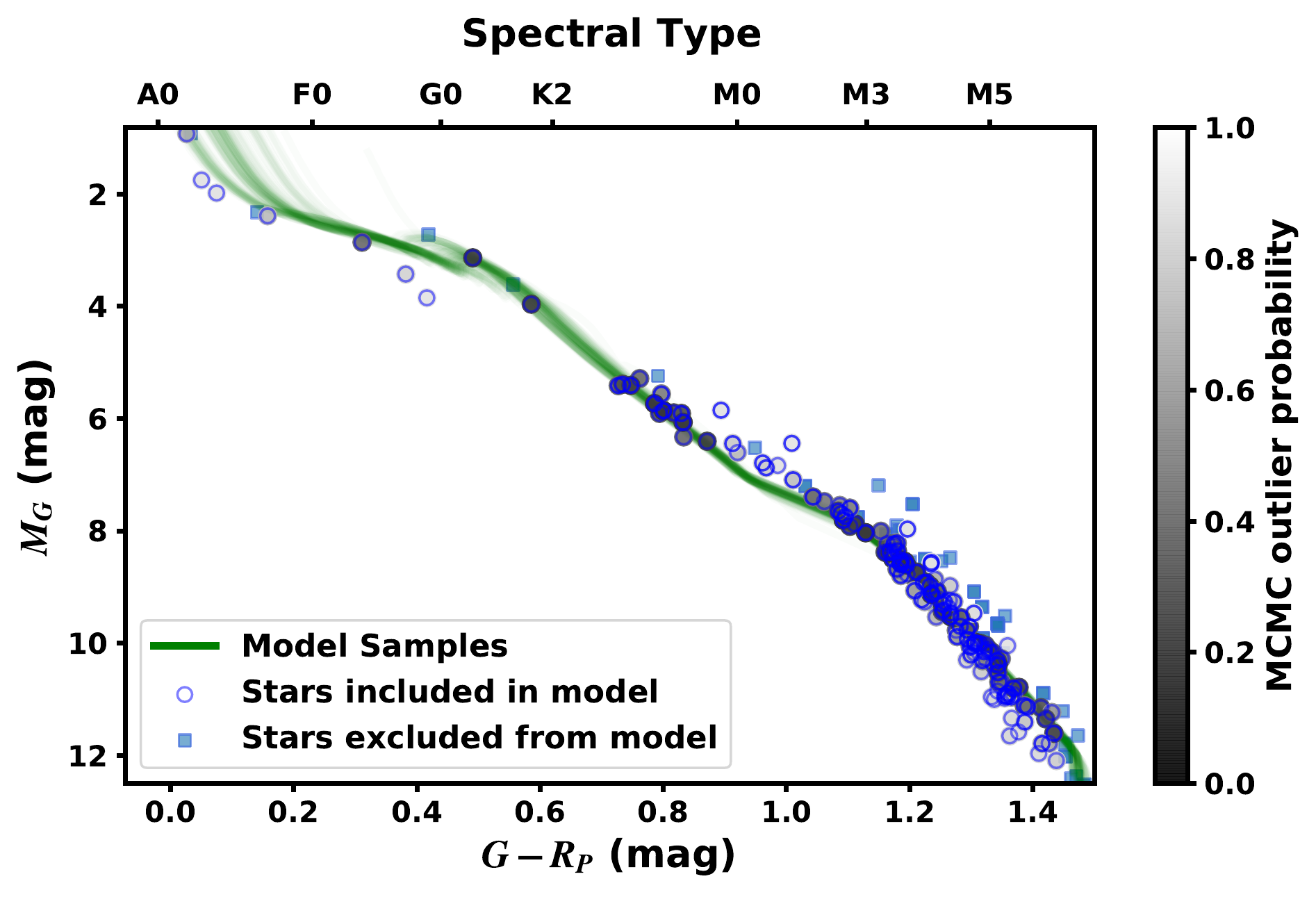}
    \includegraphics[width=0.49\textwidth]{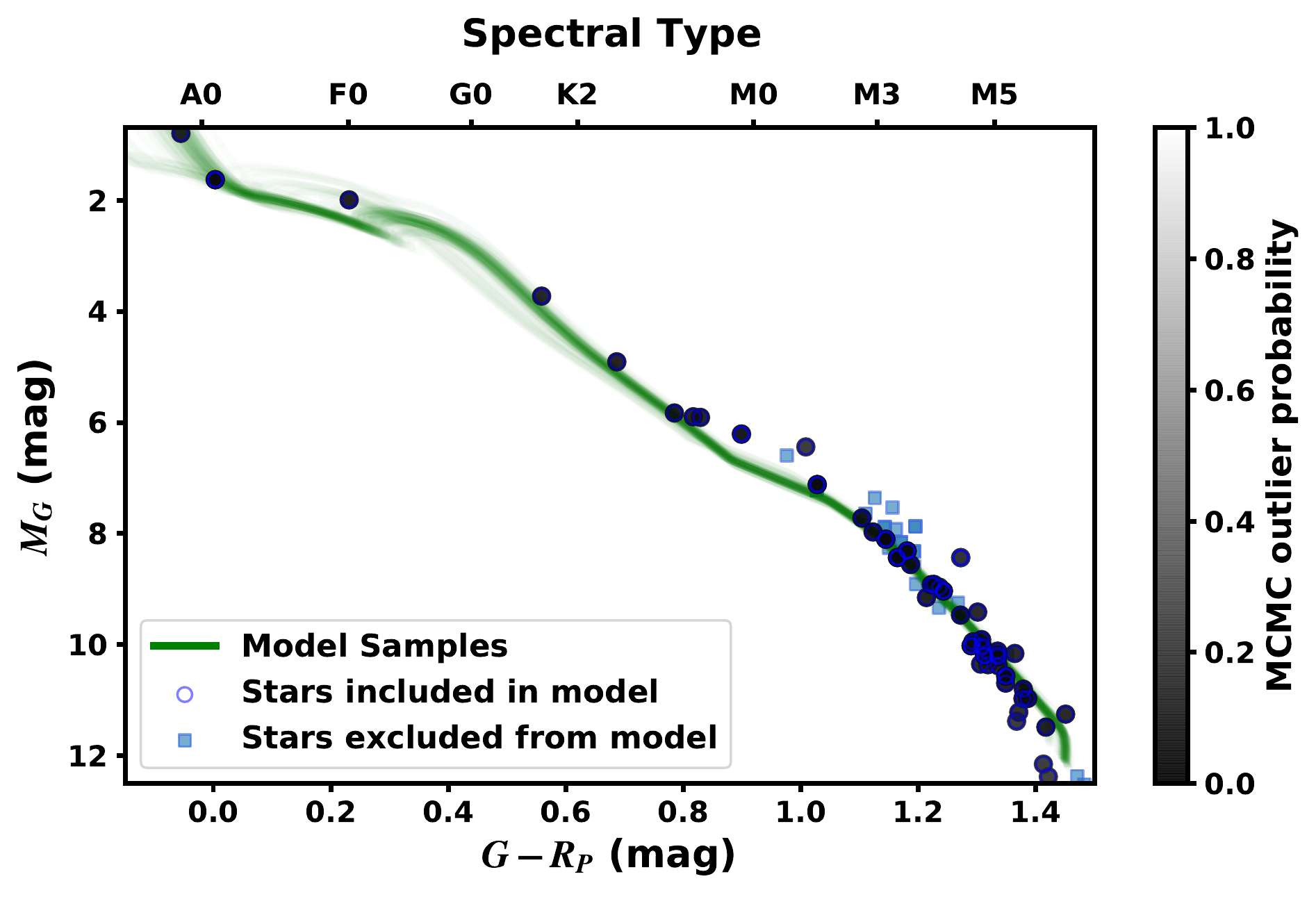}
    \caption{CMDs of the Crux-S (left) and LCC-B (right) groups from \citet{2021arXiv210509338K}. Points including in our fit are labeled as blue circles, with blue squares representing stars excluded (mostly because of a high RUWE). Each point is shaded by the probability of being an outlier (i.e., not well-described by the model). This may include non-members, nearby association interlopers, binaries, or simply stars with poor parallaxes/photometry. 
    }
    \label{fig:mixture_test}
\end{figure*}

As a test, we ran our mixture model on the LCC-C (Crux-S) group and LCC-B groups
from \citet{2021arXiv210509338K}, which are similar cases to \population. In both cases, we used the PARSECv1.2S isochrones \citep{PARSEC}, assumed Solar metallicity, and cut out stars with RUWE$>$1.2. We restricted the comparison to \gaia\ magnitudes, as those were available for all stars, but tests adding in $JHK$ magnitudes yielded similar results. We show the resulting CMD and fit in Figure~\ref{fig:mixture_test}. For Crux-S, we derived an age of 12.6$^{+1.6}_{-0.8}$\,Myr, consistent with the 14.6$\pm$0.8 Myr from \citet{2021arXiv210509338K}. For LCC-B, we estimated an age of 11.2$^{+1.1}_{-0.8}$\,Myr, consistent with 13.0$\pm$1.4 from \citet{2021arXiv210509338K}. We also performed similar tests with older associations, such as MELANGE-1 from \citet{THYMEV}, which gave us an age of 210$\pm$70\,Myr, an excellent match to the Li- and gyro-based ages ($250^{+50}_{-70}$\,Myr) from \citet{THYMEV}.

One advantage of the mixture model approach is the handling of regions where the models perform poorly. For example, in both test cases the PARSEC models fail to reproduce the population of M0-M2 dwarfs (Figure~\ref{fig:mixture_test}). Given that the magnetic models from DSEP better reproduce this region, it is likely that the offset is due to errors in the models as opposed to a high rate of binarity in that part of the selected population. In the case of LCC-C, there are few sources in that regime, so they have little effect on the overall age. For Crux-S, the mixture model weights these down in the fit. 

However, the ability of this method to discard points is also a problem with this method; for Crux-S, the fit identifies many of the high-mass stars as medium-probability outliers, likely because it is challenging to fit the low-mass stars simultaneous with the high-mass ones \citep{Feiden2016}. A solution to this would be to run some systems with independent age estimates \citep[such as from lithium depletion or traceback ages,][]{2014prpl.conf..219S} to identify such systematics and apply corrections or weights. However, this requires a much more extensive set of associations with non-isochronal ages. A less elegant approach is to run the mixture code with multiple model grids. 

Another drawback is when the population is poorly described by two models. However, the framework can be increased to explicitly model additional populations with sufficient terms. For example, one could use one term to handle the population of interest, a term to handle binaries with a prior or limits on the $\Ybad$-like term, and a term to model the (assumed) single-age population of young interlopers from other LCC groups. We found the simpler model to be sufficient for our purposes here, but hope to explore this in future iterations.

\end{document}